\documentclass[prd,a4paper, 
nofootinbib, 
]{revtex4}
\usepackage{graphicx}
\usepackage{amsfonts}
\usepackage{amssymb}
\usepackage{amsmath}
\usepackage{slashed} 
\usepackage{enumerate}
\usepackage{color}
\usepackage{comment}
\usepackage{tikz}
\usepackage{hyperref}
\hypersetup{
    colorlinks=true,% make the links colored
}
\usepackage[utf8]{inputenc}
\usepackage{erewhon}
\usepackage{subfigure}
\usepackage[normalem]{ulem}

\def\eq#1{{(\ref{#1})}}
\definecolor{roguered}{rgb}{0.72, 0.3, 0.2}
\definecolor{northernblue}{rgb}{0.21, 0.1, 0.83}
\definecolor{leafygreen}{rgb}{0.13,0.55,0.15}
 \hypersetup{colorlinks, citecolor=northernblue, linkcolor=roguered, urlcolor=black}
\newcommand{\be}{\begin{equation}}
\newcommand{\ee}{\end{equation}}
\newcommand{\bea}{\begin{eqnarray}}
\newcommand{\eea}{\end{eqnarray}}

\newcommand{\ic}{i}

\definecolor{gbcolor}{rgb}{.13,.22,.8}
 
\definecolor{gbcolor2}{rgb}{.6,.2,.6}

\makeatletter
\def\Dated@name{Dated: }
\makeatother

\makeatletter
\def\l@subsection#1#2{}
\def\l@subsubsection#1#2{}
\makeatother

\begin{document}

%%%%%%%%%%%%%%%%%%%%%%%%%%%%%%%%%%%%%%%%%%%%%%%%%%%%%%%%%%%  
\hfill{} IFT-UAM/CSIC-21-45

\hfill{} DESY 21-048

\hfill{} TUM-HEP-1334-21

\title{Revisiting isocurvature bounds in models unifying the axion with the inflaton}
\date{\today}
\author{Guillermo Ballesteros$^{a,b}$}
\email[\quad]{guillermo.ballesteros@uam.es}
\author{Andreas Ringwald$^{c}$}
\email[\quad]{andreas.ringwald@desy.de}
\author{Carlos Tamarit$^{d}$}
\email[\quad]{carlos.tamarit@tum.de}
\author{Yvette Welling$^{c}$}
\email[\quad]{yvette.welling@desy.de}
\affiliation{Instituto de Física Teórica UAM/CSIC, Calle Nicolás Cabrera 13-15, Cantoblanco E-28049 Madrid, Spain$^a$}
\affiliation{Departamento de Física Teórica, Universidad Autónoma de Madrid (UAM), Campus de Cantoblanco, 28049 Madrid, Spain$^b$}
\affiliation{ Deutsches Elektronen-Synchrotron DESY,
Notkestra\ss e 85, D-22607 Hamburg, Germany$^c$}
\affiliation{Physik-Department T70, Technische Universit\"at M\"unchen, James-Franck Stra\ss e 1, D-85748 Garching, Germany$^d$}

%%%%%%%%%%%%%%%%%%%%%%%%%%%%%%%%%%%%%%%%%%%%%%%%%%%%%%%%%%%%%%%%%%%%%
\begin{abstract}
\noindent

Axion scenarios in which the spontaneous breaking of the Peccei-Quinn symmetry takes place before or during inflation, and in which axion dark matter arises from the misalignment mechanism, can be constrained by {Cosmic Microwave Background isocurvature bounds}. {Dark matter isocurvature is} thought to be suppressed in  models {with} axion-inflaton interactions, for which {axion perturbations are assumed to} freeze at horizon {crossing} during inflation. {However,} this assumption can be an oversimplification {due to the} interactions {themselves}. {In particular,} non-perturbative effects during reheating {may} lead to a dramatic growth of axion perturbations. We {perform} lattice calculations in two models in which the Peccei-Quinn field participates in inflation. We find that  the growth of axion perturbations is such that the Peccei-Quinn symmetry is restored for an axion decay constant  $f_A\lesssim10^{16}$-$10^{17}$GeV, leading to an over-abundance of dark matter, {unless $f_A \lesssim 2 \times 10^{11}$~GeV}. 
For $f_A\gtrsim10^{16}$-$10^{17}$GeV we still find a large growth of axion perturbations at low momentum, such that a naive extrapolation to CMB scales {suggests a} violation of the isocurvature bounds.

 \end{abstract}
 
 \maketitle

{
  \hypersetup{linkcolor=black}
  \tableofcontents
} 
 
%%%%%%%%%%%%%%%%%%%%%%%%%%%%%%%%%%%%%%%%%%%%%%%%%%%%%%%%%%%%%%%%%%%

%%%%%%%%%%%%%%%%%%%%%%%%%%%%%%%%%%%%%%%%%%%%%%%%%%%%%%%%%%%%%%%%%%%
 %%%%%%%%%%%%%%%%%%%%%%%%%%%%%%%%%%%%%%%%%%%%%%%%%%%%%%%%%%%%%%%%%%%

 \section{The axion and isocurvature fluctuations}
 
In the Standard Model of particle physics (SM), the amount of violation of CP symmetry in the strong interactions
can be quantified by a single real number: the $\theta_{\rm QCD}$ parameter of Quantum Chromodynamics (QCD). This quantity is sourced by CP violating interactions between gluons and the CP violating phase of the quark mass mixing matrix. Measurements of the electric dipole moment of the neutron have set the very stringent bound $\theta_{\rm QCD} \lesssim 10^{-10}$ \cite{Afach:2015sja}, meaning an extremely low level of CP violation in the strong sector of the SM. Since there is a priori no reason for the strong interactions to be CP invariant (and the level of CP violation in weak interactions is much more significant) the origin of this bound --which has been termed {\it the strong CP problem}-- is considered one of the main puzzles of theoretical particle physics and a compelling motivation for extending the SM.\footnote{{The standard understanding of the strong CP problem has been recently questioned
in Ref.~\cite{Ai:2020mzh}.}}
The problem may be pushed away by simply positing that $\theta_{\rm QCD}$ is very small by sheer chance. However, in physics, and in quantum field theory in particular, the appearance of unexpected zeros or nearly vanishing quantities, such as $\theta_{\rm QCD}$, often signals the existence of an underlying symmetry or a dynamical mechanism at play. The most popular solution to the strong CP problem features both, as it is based on the spontaneous (i.e.\ dynamical) breaking of a new $U(1)$ global symmetry. This (Peccei-Quinn or PQ) symmetry \cite{Peccei:1977hh} effectively promotes $\theta_{\rm QCD}$ to a field (often denoted as $A$) which dynamically settles to $A=0$, thus solving the strong CP problem. This happens because the PQ symmetry is anomalous under QCD, in such a way that nonperturbative QCD effects generate a potential for $A$ with a minimum at $A=0$ \cite{Vafa:1984xg}. The pseudo-Goldstone boson arising from the spontaneous breaking of the PQ symmetry \cite{Weinberg:1977ma,Wilczek:1977pj} is called {\it the axion} and its mass $m_A$ is inversely proportional to the energy scale $f_A$ at which the PQ symmetry is broken. A calculation from chiral perturbation theory gives \cite{diCortona:2015ldu}:
\begin{align}\label{eq:ma}
m_A\simeq 5.691(51)\left(\frac{10^{12} {\rm GeV}}{f_A}\right) \mu{\rm eV}\,.
\end{align}

 When the expansion rate of the Universe becomes smaller than $m_A$ (we use $c=\hbar =1$), the axion field undergoes classical oscillations around the minimum of its (approximately quadratic) potential, behaving as a pressureless perfect fluid and contributing to the cold dark matter content of the Universe. This is known as ``misalignment mechanism''  \cite{Abbott:1982af, Dine:1982ah, Preskill:1982cy}. The cosmological energy density of this axion condensate depends on $f_A$ and the initial conditions for the axion field in the radiation era. The latter depend on how the PQ symmetry breaks in the early Universe, and more concretely during primordial inflation and the subsequent reheating process. The details of this breaking {also} determine whether the decay of topological defects formed due to the breaking of the PQ symmetry contribute to the relic axion density. If $f_A$ lies in the adequate range of values (which depends on the PQ breaking history), the axion can account for the totality of the (cold) dark matter in the Universe. This enhances the appeal of the axion solution to the strong CP problem. 

{{If} the PQ
  symmetry is broken after inflation --the so-called \emph{post-inflationary} scenario-- the axion field {takes}
 random initial conditions in different patches of the Universe, with an expected average of  $A/f_A\sim\mathcal{O}(1)$. In this scenario,
axions sourced by the decay of cosmological defects (axion strings) at the intersection of patches with different initial values of $A/f_A$ {do contribute to the dark matter abundance.} Neglecting {these contributions}  gives a lower bound on the axion dark matter abundance that only depends on $f_A$ (see e.g.\ \cite{Borsanyi:2016ksw}), 
\begin{align} \label{abundance}
\Omega_A\,h^2\gtrsim 0.12 \left(\frac{f_A}{2\times 10^{11} \rm{GeV}}\right)^{7/6}\,,
\end{align}
where, {by convention}, the $\mathcal{O}(1)$ parameter $h$ is defined through the value of the current Hubble constant $H_0=100\,h\, \rm{km/s/Mpc}$, 
 and
} we have chosen to normalize the expression so that the prefactor agrees with the observational measurements of the dark matter abundance \cite{Aghanim:2018eyx}. {This leads 
to  models with
$f_A\lesssim 2\times10^{11}$~{GeV}, and axion masses $m_A\gtrsim28$-$100\,\mu{\rm eV}$. 
{While the upper bound in $f_A$ (lower bound in $m_A$) is known precisely from the results in Ref.~\cite{Borsanyi:2016ksw}, there remains a  sizable theoretical uncertainty} coming from the difficulty of estimating the contributions of decaying strings 
\cite{Klaer:2019fxc,Gorghetto:2018myk,Hindmarsh:2019csc,Gorghetto:2020qws}, which can shift the preferred values of $m_A$ and $f_A$ by up to an order of magnitude with respect to those saturating the bound of Eq.~\eqref{abundance}~\cite{Gorghetto:2020qws}.

Equation~\eqref{abundance} implies that {$f_A \gg 10^{11}$GeV is} incompatible with the post-inflationary scenario. {Such} high values of $f_A$ are particularly motivated  by grand unified theories in which the axion decay constant may be tied to the unification scale \cite{Wise:1981ry,Reiss:1981nd,Ernst:2018bib,DiLuzio:2018gqe,Ernst:2018rod,Boucenna:2018wjc,Babu:2018qca,Chakrabortty:2019fov,Ballesteros:2019tvf,FileviezPerez:2019ssf,DiLuzio:2020qio,Lazarides:2020frf}. Hence, for large axion decay constants one has to consider the} possibility that the PQ symmetry is spontaneously broken {before or} during inflation and never restored afterwards. This scenario (commonly dubbed {\it pre-inflationary}), will be the focus of this paper, with the aim of assessing whether minimal high-scale axion models can have consistent cosmological histories. The pre-inflationary scenario requires $f_A$ to be larger than the Hubble scale during inflation, $H_{\rm inf}$ and that the PQ symmetry is not restored during reheating, either thermally or non-thermally. Any topological defects produced after the symmetry is broken are diluted away by inflation and do not contribute to the axion energy density. {In this case, since no large axion perturbations are generated during and after inflation, one can assume that  our current Hubble patch comes from a region with a common initial $A/f_A$, which is denoted as $\theta_i$ and is often called {\it misalignment angle}. {In the pre-inflationary scenario} the axion {abundance} depends 
on both $f_A$ and $\theta_i$ as \cite{Borsanyi:2016ksw},
 \begin{align}\label{eq:Omegah2preinf}
  \Omega_A\,h^2\simeq 0.12 \left(\frac{\theta_i}{0.004}\right)^2\left(\frac{f_A}{10^{16} \rm{GeV}}\right)^{7/6}\,.
 \end{align}

For $f_A$ near the typical unification scale of $10^{16}$ GeV the axion mass of Eq.~\eqref{eq:ma} is of the order of neV.
 } 
 A large number of experiments (ALPS \cite{Bahre:2013ywa}, CAST \cite{Anastassopoulos:2017ftl}, 
ABRACADABRA \cite{Ouellet:2018beu}, 
ADMX \cite{Braine:2019fqb}, CULTASK \cite{Chung:2016ysi}, SHAFT \cite{Gramolin:2020ict}, CAPP \cite{Kwon:2020sav}, HAYSTAC \cite{Brubaker:2016ktl}, ORGAN \cite{McAllister:2017lkb}, QUAX \cite{Alesini:2020vny}) {search} for axions across the range $10^8 {\rm GeV} \lesssim f_A \lesssim 10^{18} {\rm GeV}$ and several other are proposed to join the search in the near future (ARIADNE \cite{Geraci:2017bmq}, BRASS \cite{brass01}, CASPEr \cite{Budker:2013hfa}, KLASH \cite{Alesini:2017ifp}, MADMAX \cite{Brun:2019lyf}, IAXO \cite{Armengaud:2014gea}). Most experimental efforts exploit the generically unavoidable coupling between axions and photons; see \cite{Irastorza:2018dyq} for a review. {ABRACADABRA (and the future 
DM Radio Cubic Meter experiment) as well as CASPEr are} of particular interest for {models in which $f_A$ is $\mathcal{O}(10^{16})\,{\rm GeV}$} as they could probe the axion masses {at (DM Radio Cubic Meter) or near (CASPEr: $m_A < {\rm neV}$)} 
{the range of masses associated with values of $f_A$ tied to the scale of grand unification. }  
In general, values of $f_A$ below $\sim {\rm few}\times 10^8\, {\rm GeV}$ are excluded by neutrino data from the supernova SN1987A 
\cite{Raffelt:2006cw,Carenza:2020cis}.
Values of $f_A$ above $\sim 10^{18} {\rm GeV}$ have been argued to be ruled out due to the non-observation of gravitational waves from axion induced superradiance on black holes~\cite{Arvanitaki:2010sy}. Both limits suffer from astrophysical uncertainties. {For instance, the supernova bound has been questioned in \cite{Bar:2019ifz}.}

{Since the axion fluctuations generated during inflation in the pre-inflationary scenario are not erased by later processes, they can affect the Cosmic Microwave Background (CMB)}
{as} {\emph{isocurvature} fluctuations in the cosmological plasma during the radiation era. {These are fluctuations  in the energy densities $\rho_i$ of different particle species in the plasma satisfying:}
\begin{equation}\label{eq:isodef}
 \frac{\delta\rho_i}{\rho_i+p_i}\neq\frac{\delta\rho_j}{\rho_j+p_j},\text{    \,\,\, for\,\,\,  }i\neq j ,
\end{equation}
(see Appendix \ref{app:iso} for more details). If all species enter thermal equilibrium in the radiation era {before decoupling}, such isocurvature fluctuations are absent {from the CMB} since $\delta\rho_i/(\rho_i+p_i)=(3/4)\delta\rho_i/\rho_i=3\delta T/T$. In a pre-inflationary scenario, however, the axions {do not reach thermal equilibrium with the plasma.} The axion interactions are suppressed by inverse powers of $f_A$, and for their rates to become large enough to achieve equilibration $T>f_A$ {is needed}, which in general would lead to a thermal restoration of the PQ symmetry. Hence, one expects isocurvature fluctations which are stringently constrained by the} CMB data from Planck \cite{Akrami:2018odb}. {{ Treating the axion as massless, assuming its fluctuations freeze at horizon crossing during inflation and allowing for the value of $f_A$ during inflation {($f_{A,\rm inf}$)} to be different from that corresponding to the minimum of the {axion} potential, the {CMB} isocurvature constraint can be written as follows (see the discussion in Section~\ref{sec:isobound})
\begin{align}\label{eq:ISO_bound_massless}
 \left(\frac{f_A}{10^{16}{\,\rm GeV}}\right)^{7/6}\left(\frac{H_{\rm inf}}{10^{9}{\,\rm GeV}}\right)^2\left(\frac{f_{A,\rm inf}}{10^{16}{\,\rm GeV}}\right)^{-2}\lesssim 1.
\end{align}
In general, it is assumed that $f_{A,\rm inf}=f_A$, so that the isocurvature bound can be avoided {if} the inflationary Hubble scale $H_{\rm inf}$ is sufficiently small. This requires $H_{\rm inf}<10^9$ GeV, for $f_A>10^{16}$ GeV. Such {a value} of $H_{\rm inf}$ {is} very small. {In particular,  it is at odds with simple large-field
models, e.g.\ such as those featuring} an inflaton with {a flattened potential due to a coupling to the Ricci scalar (see e.g.\ \cite{Bezrukov:2007ep} as an example of such a non-minimal quartic chaotic inflation model),}
for which typical values of $H_{\rm inf}$ are above $10^{13}$ GeV.
\footnote{For single-field inflationary models with $H_{\rm inf}<10^9$~{GeV} the amount of B-modes from primordial gravitational waves that would be produced at CMB scales would be unobservable with any conceivable future probe.}

{At face value, the} bound of Eq.~\eqref{eq:ISO_bound_massless} {implies that}
pre-inflationary axion dark matter scenarios, including those inspired by grand unification, are {in}compatible with standard {(high-scale)} inflation. Possible ways to avoid {this} stringent isocurvature {bound} have been 
proposed in the literature. For example, as is clear from Eq.~\eqref{eq:ISO_bound_massless}, taking $f_{A,\rm inf}\gg f_A$ can help suppressing isocurvature fluctuations, and this can be achieved if the axion is embedded into {a} complex scalar whose modulus drives inflation {with}
$f_{A,\rm inf}\sim M_P$ \cite{Fairbairn:2014zta,Ballesteros:2016xej,Boucenna:2017fna}.   In this case, the isocurvature bound can be satisfied with  $H_{\rm inf}$ and $f_A$ around $10^{13}$~GeV. The SMASH model 
\cite{Ballesteros:2016euj,Ballesteros:2016xej} is a particular example of an axion embedded into the inflaton. {It was shown in \cite{Ballesteros:2016xej}  that CMB} isocurvature fluctuations are suppressed {in SMASH} if $f_A\lesssim 10^{14}$ GeV. However, even for these values of $f_A$ 
the PQ symmetry is restored {in this model} during preheating, {and} it was concluded in \cite{Ballesteros:2016xej} that there was no room for {viable PQ breaking before or during inflation}.
In fact, in Ref.~\cite{Ballesteros:2016xej} it was {argued}
that the PQ restoration would not take place for $f_A\gtrsim4\times10^{16}$ GeV, {although} this was not confirmed with dedicated reheating simulations.

Other possible mechanisms to avoid the axion isocurvature bound rely on {violating} some of the assumptions leading to Eq.~\eqref{eq:ISO_bound_massless}. A possibility is that the 
axion was not massless during inflation {but heavier than} 
$H_{\rm inf}$. Then, its fluctuations during inflation would become suppressed. This 
was considered in Ref.~\cite{Nakayama:2015pba}, where the extra contributions to the axion mass {arise} from interactions of the complex field containing the axion with a complex inflaton.

The analysis of the fate of isocurvature fluctuations in both of the above types of scenarios {can however be improved}.
First, in the models of Refs.~\cite{Fairbairn:2014zta,Ballesteros:2016euj,Boucenna:2017fna,Ballesteros:2016xej} 
{it} was not {taken into account that} as long as the fields are not at the minimum of the potential the axion field is not massless (note that Goldstone's theorem only applies at the potential minimum). It thus remains to analyze the impact of this mass in the axion perturbations during inflation. Furthermore, the calculations always assumed that the axion perturbations froze after horizon crossing and stopped evolving afterwards. Such assumption was also made in the model of Ref.~\cite{Nakayama:2015pba}. However, the freezout at horizon crossing  is not guaranteed for fields with nonzero masses, such as the axion during the oscillations of the background throughout the reheating process. In particular, the axion modes can be tachyonic, and this can lead to an explosive growth of perturbations.

The aim of this paper is to reconsider the previous calculations in the literature concerning the predicted spectrum of axion isocurvature fluctuations, by abandoning the assumptions of masslessness during inflation and the freezing of perturbations at horizon crossing. We will
follow the evolution of the axion perturbations both in the linear regime during inflation --solving uncoupled differential equations for the different momentum modes-- and in the nonlinear regime of reheating --carrying out lattice simulations for the evolution of the axion and other fields like the inflaton and the Higgs, using initial conditions obtained from the results in the linear regime. We consider as examples the scalar sector of SMASH and a two-field model inspired by Ref.~\cite{Nakayama:2015pba}. With respect to SMASH, a secondary aim of our calculation will be  to extend the reheating simulations of Ref.~\cite{Ballesteros:2016xej} 
revisiting the conclusion that the PQ restoration is expected to  be avoided for $f_A\gtrsim4\times10^{16}$~{GeV},
which was based on extrapolations of 
simulations for much smaller values of $f_A$.
{
Our results show that during reheating the axion perturbations in both models go through a phase of exponential amplification, after which the isocurvature power spectrum decays with the inverse square of the scale factor of the universe. The initial amplification can lead to a restoration of the PQ symmetry for axion decay constants {smaller than} $10^{16}$-$10^{17}$ GeV, while for higher values we find that a naive extrapolation of the isocurvature power spectrum to CMB scales would exceed the {Planck
bound.}}

The paper is organized as follows. In Section~\ref{sec:model} we describe the models analyzed in the paper. Section \ref{sec:isobound} is devoted to recovering the usual estimates of the axion isocurvature bounds, and {to} describe the improvements carried out in the paper. The results of our calculations of the evolution of the power spectrum of axion perturbations during and after inflation are given in Section \ref{sec:results}. We conclude in Section~\ref{sec:conclusions}. {Additional details are given in appendices. Appendix~\ref{app:iso} summarizes the definition and relevant properties of isocurvature fluctuations, while Appendix~\ref{app:isoax} focuses then on those sourced by axions. {The axion mass controlling this source is discussed in Appendix~\ref{app:isomass}.} Details on the treatment of Higgs decays are given in Appendix~\ref{app:Higgsdecays}, while consistency checks of the lattice computations using a mean-field approximation are provided in Appendix~\ref{app:meanfieldapproximation}.}

\section{\label{sec:model}Model setup}

{In this section we describe the two models to be analyzed in the paper, which correspond to the two ways to relax the axion isocurvature bound mentioned in the introduction. Aside from the Higgs scalar, {Model 1 features an axion with large $f_{A,\rm inf}$ coming from a complex scalar whose modulus plays the role of the inflaton.} 
Model 2 features two new complex scalars, one responsible for driving inflation and one containing the phase associated with the 
axion, which is rendered very massive during inflation.}

\subsection{\label{sec:model1}Model 1: axion embedded into a complex inflaton}

We consider a complex scalar $\sigma$, non-minimally coupled to gravity, with a global $U(1)_\text{PQ}$ symmetry  {and a portal coupling to the SM Higgs $H$. {This is the scalar field content of the}
SMASH model \cite{Ballesteros:2016euj,Ballesteros:2016xej}. The {action contains the following terms:}
\begin{equation}
S \supset  \int d^4 x \sqrt{-g_J} \left[-\left(\frac{M_p^2  + \xi ({2}|\sigma|^2 - f_A^2)}{2}\right)R_J + |\partial \sigma |^2 -V_J(|\sigma|,H) \right]\ .
\end{equation}
Here $R_J$ is the Ricci scalar of spacetime {curvature (in the so-called {\it Jordan frame})} and {$\xi \lesssim \mathcal{O}(1)$} is a dimensionless coupling.
{We} assume the usual scalar potential for the PQ field $\sigma$ {supplemented with a portal coupling to the Higgs,}
\begin{equation}\label{eq:Vmodel1}
V_J = 
 V_{\rm SM}+\frac{1}{4}\lambda_\sigma\left(\rho^2 -f_A^2\right)^2{+\lambda_{H\sigma}\left(H^\dagger H-\frac{v^2}{2}\right) \rho^2}\,,
\end{equation}
where $V_{\rm SM}$ is the Standard Model Higgs potential, and 
\begin{equation}\label{eq:Adef}
\sigma = \frac{\rho}{\sqrt{2}}e^{i \theta} =\frac{\rho}{\sqrt{2}}e^{\ic {A/\rho}}\,.
\end{equation}
{The field $A=\rho\theta$ corresponds to the axion, with a canonical normalization in the Jordan frame}. {In Eq.~\eqref{eq:Vmodel1} $v$ denotes the usual Higgs vacuum expectation value (VEV) at the electroweak scale. In our calculations, $v$ will be much below the physical scales  relevant for the dynamics of the axion fluctuations, and its effect can be ignored. }
For our considerations the most important feature of the potential is that it has its global minimum {at a nonzero value of $\rho$:} {$f_A=\langle\rho\rangle$.}
Our findings should, however, also apply to more general setups featuring a global $U(1)$ symmetry. For instance, when the minimum of the potential is set by another scale.\footnote{In particular, when the quadratic and quartic terms are related by a single scale it should be straightforward to generalize our findings.} Moreover, we expect that
{Planck-suppressed} operators {in $V_J$} may modify 
the inflationary dynamics, but not substantially the reheating dynamics, {which is the most relevant phase for our analysis.}
{Focusing on inflationary backgrounds {where the Higgs satisfies $H=0$,} and carrying out} a {Weyl} transformation {of the metric}, the corresponding {\it Einstein frame} action for $\sigma$ reads
\begin{equation}\label{eq:Einstein_frame_action}
S(\sigma,H=0) =  \int d^4 x \sqrt{-g} \left[-\frac{M_p^2 }{2}R +G_{ij} \partial_\mu \phi^i \partial_\mu \phi^j-V(\rho) \right]\,,
\end{equation}
with {${\phi} = (\rho, \theta)$.} The Weyl transformation is characterized by 
\begin{equation}
{\Omega^2 = 1 + \xi\,\frac{\rho^2 - f_A^2}{M_p^2}\,,}
\end{equation} 
such that the resulting scalar potential is given by $V(\rho) = V_J(\rho)/\Omega^4(\rho)$, and the nonzero components of the field metric are
{\begin{align}\label{eq:metric}
G_{\rho\rho} =&\,\frac{1}{\Omega^2}\left(1+6\,\xi^2\frac{\rho^2}{M_p^2\,\Omega^2}\right), & G_{AA}=&\,\frac{G_{\theta\theta}}{{\rho}^2}={\frac{1}{\Omega^2}}\,.
\end{align}}
In this model the CMB constraints \cite{Akrami:2018odb} are satisfied for {$\xi \gtrsim 0.003$ (or $\xi \gtrsim 0.007$ if the Universe enters radiation domination immediately after inflation, as in SMASH), see e.g.\ \cite{Ballesteros:2016euj} and the more recent analysis in \cite{Ringwald:2020vei}. 
{Perturbative unitarity requires $\xi<1$ \cite{Barbon:2009ya,Burgess:2009ea} and we will use $\xi=0.1$ for our calculations.}\footnote{The issue of unitarity in Higgs inflation is still being explored in the literatture, \cite{Ema:2017rqn,Gorbunov:2018llf,Ema:2019fdd}.}

\begin{figure}
\centering     
\subfigure[]{\label{fig:a}\includegraphics[width=0.49\textwidth]{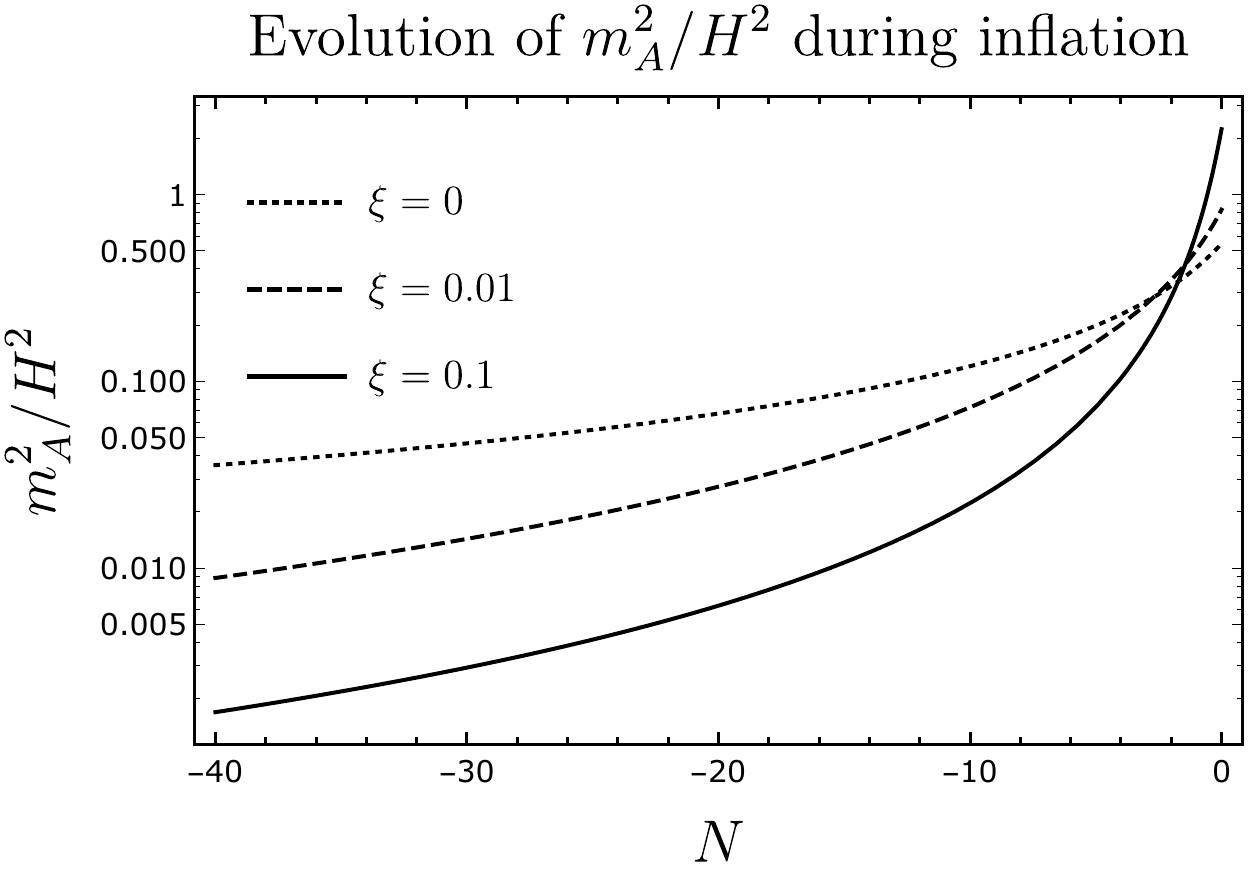}}
\hfill
\subfigure[]{\label{fig:b}\includegraphics[width=0.49\textwidth]{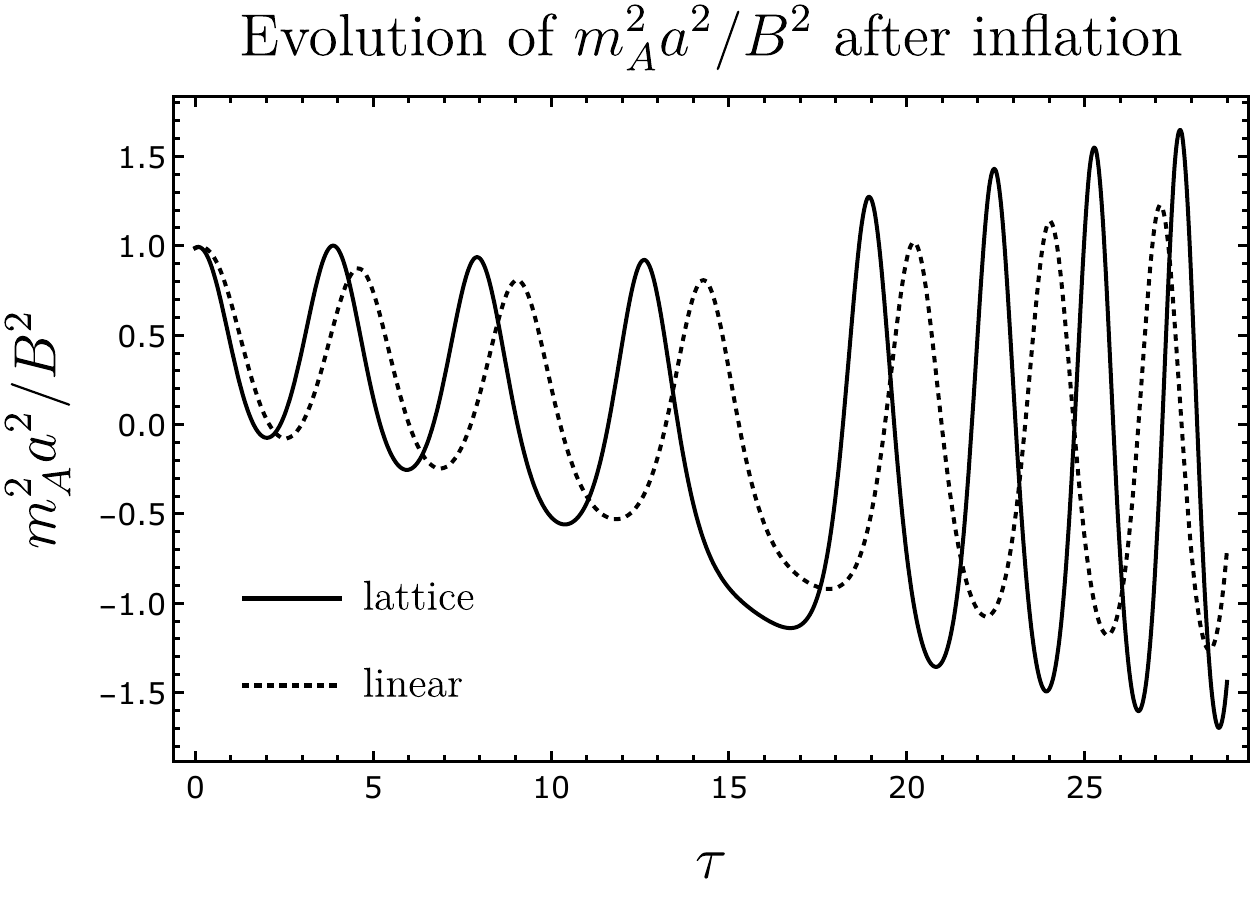}}
\caption{(a) The evolution of the isocurvature mass  {squared}, 
Eq. \eqref{eq:ma1}, during inflation as a function of {the number of e-folds} 
in units of the Hubble scale $H$ {for three values} of the non-minimal coupling, {$\xi$.}
(b) The evolution of the isocurvature mass {squared} after inflation as function of (rescaled) conformal time of Eq.~\eqref{eq:tauBdef}.  We evaluate the isocurvature mass on the {averaged background obtained from a lattice simulation} 
with $\xi=0$, but with 
initial conditions of the fields given by their evolution during inflation with $\xi = 0.1$ consistent with section \ref{sec:results}. At the end of inflation the $\xi$ corrections to the isocurvature mass are almost negligible. For comparison we also show the linear evolution where backreaction on the background has been neglected (black, dotted) with the same initial field amplitudes and the $\xi$-dependent corrections to the mass neglected. }
\label{fig:isomassevolution}
\end{figure}

{The (first order and canonically normalized) gauge invariant perturbation corresponding to the field direction orthogonal to the inflationary trajectory is (see e.g.\ \cite{Cespedes:2012hu}),
\begin{align}\label{eq:calS}
{\cal F}=\sqrt{G_{\theta\theta}}\, \delta \theta.
\end{align}
The superhorizon modes of ${\cal F}$
evolve according to the equation 
\begin{equation}\label{eq:iso_orthogonal_evolution}
 \frac{\partial^2 {\mathcal{F}}}{\partial N^2} +  (3-\epsilon) \frac{\partial {\mathcal{F}}}{\partial N} + \frac{m^2_\text{iso}}{H^2}\mathcal{F} = 0\ ,
\end{equation}
where $N$ is the number of e-folds ($dN = H\, dt$) and $\epsilon = - d \log H /dN$. Gauge invariance requires the mass squared $m^2_\text{iso}$ of these modes to be defined including the curvature of the field manifold, described by the metric whose components are given in \eq{eq:metric}. Its expression is (approximately) \begin{align}\label{eq:ma1}
 m^2_{{A}}=\frac{ 1-\xi f_A^2/M_p^2}{\Omega^4(\Omega^2+6\xi\rho^2/M_p^2)}
\lambda_\sigma  \left(\rho^2 - f_A^2\right).
\end{align} 
It may be somewhat surprising to have a mass associated to the axion direction, given that the potential $V(\rho)$ is formally independent of it. However, the usual definition of masses in terms of second-order partial derivatives of the potential is not invariant under field redefinitions, and is not the most adequate for describing gauge invariant fluctuations. The usual definition only agrees with the one invariant under field redefinitions if the kinetic terms are diagonal and canonically normalized. This happens in our case if we use the basis $\sigma=1/\sqrt{2}(\sigma_r+{\rm i}\sigma_\theta)$, in the Jordan frame. The details of the construction of the field-redefinition-invariant masses are given in Appendix~\ref{app:isomass}. 

During inflation, with $\rho\gg f_A$, the axion {perturbation} mass is {not zero}, and it can have an effect in {their evolution for superhorizon scales} before the end of inflation, which {(to the best of our knowledge)} has so far not been taken into account in the literature. After inflation, with the background $\rho$ oscillating around the minimum of the potential and passing through values with $\rho<f_A$, the {mass of the axion perturbations} can become temporarily tachyonic. This is expected to lead to an exponential growth of axion perturbations. As the latter can evolve to become $\mathcal{O}(1)$, the linear treatment of the growth of breaks down and so does eventually any perturbative analysis. For this reason we will resort to nonperturbative lattice simulations.}

{In Figure \ref{fig:isomassevolution} we show the evolution of the isocurvature mass during (left figure) and after inflation (right figure). During inflation the natural time scale is the Hubble time $1/H$, so we plot the isocurvature mass as a function of {the number of} e-folds 
and in units of the Hubble rate. We choose $N=0$ to be the end of inflation. After inflation, the natural time scale of oscillations in a quartic potential is captured by the following dimensionless rescaled conformal time variable,
\begin{align}\label{eq:tauBdef}
\tau = \int \frac{B dt}{a}, \quad\text{  with   }\quad B \equiv \sqrt{\lambda_\sigma} \rho_\text{end},
\end{align}
where $ \rho_\text{end}$ is the value of the background radial field at the end of inflation. Therefore, we                                                                                                                                                       
plot the isocurvature mass as a function of $\tau$ and in units of $B/a$. We choose $\tau =0$ at the end                                                                                                                                                                                    
of inflation. Deep inside the epoch of inflation the isocurvature mass is non-zero but positive and {in} the slow-roll approximation is given by
\begin{equation}                                                                                                                                                                                                             
\frac{m_A^2}{H^2} \approx \frac{3}{-2N}\frac{1+6\xi}{1-8\xi N}\ .
\end{equation}
For small $\xi \ll 1/|N|$ the mass scales like {$1/N$}, whereas for larger $\xi$ it is suppressed as $1/N^2$, though {it rises faster}  at the end of inflation. This results into a slightly enhanced suppression of the amplitude of isocurvature perturbations for small values of $\xi$, see section \ref{sec:evolutionisoinflation}. After inflation we neglect the $\xi$-dependent contributions to the isocurvature mass but use $\xi=0.1$ to set the initial field value $\rho_\text{end}$.  In the figure 
$f_A = 5\times 10^{17}~\text{GeV}$, which sets $\tau \sim 15$ for the time at which the radial field starts to oscillate around the true minimum. From the right figure it is clear that the isocurvature perturbations experience multiple periods of tachyonic instability. This suggests that lattice simulations will be required to capture the strongly non-linear evolution of all fields.

\subsection{\label{sec:model2}Model 2: axion interacting with a complex inflaton}

{Inspired by Ref.~\cite{Nakayama:2015pba},} Model 2 includes two complex scalar fields $\sigma$ and $\phi$ with opposite PQ charges. {At late times, only $\sigma$ is assume{d} to develop a 
{VEV} that breaks the PQ symmetry, so that the axion is contained in the phase of $\sigma$. The field $\phi$ is assumed to be the main driver of inflation. 
In order to have a pre-inflationary axion scenario with a well defined initial misalignment angle,  the inflationary trajectory should have 
$|\sigma|\neq 0$, which can be achieved {provided that a specific combination of quartic couplings is negative (see below).}
The PQ symmetry is compatible with}  an interaction term $\sim \sigma^2 \phi^2 + \text{h.c.}$ that yields a mass for the imaginary component of $\sigma$ if $\phi$ takes a non-zero expectation value during inflation. This could drastically suppress the amplitude of isocurvature perturbations during inflation and relax the axion isocurvature problem \cite{Nakayama:2015pba}. 

The scalar potential {of the model} in the Jordan frame is given by
\begin{align}
V_J = &\,V_H+\lambda_\sigma\left(|\sigma|^2 - \frac{f_A^2}{2}\right)^2 + m_\phi^2 |\phi|^2 +\lambda_\phi |\phi|^4 + 2 \lambda_{\phi\sigma}|\phi|^2\left(|\sigma|^2 - \frac{f_A^2}{2}\right) + 2 \hat \lambda_{\phi\sigma}\left(\phi^2 \sigma^2 +\text{h.c.} \right)\\
%%%%
&{+2\lambda_{H\phi}\left(H^\dagger H-\frac{v^2_h}{2}\right)\phi^\dagger\phi
 %%%%
 +2\lambda_{H\sigma}\left(H^\dagger H-\frac{v^2_h}{2}\right)\left(\sigma^\dagger\sigma-\frac{v^2_\sigma}{2}\right)\,,}
\end{align}
where an interaction term $\sim \phi \sigma$ is forbidden if we endow $\sigma$ with an extra $\mathbb{Z}_2$ symmetry. {As in Model 1, we consider nonminimal gravitational couplings
\begin{align}
 {\cal L}\supset -\frac{1}{2} \left(\xi_\sigma(2|\sigma|^2-f_A^2)+2\xi_\phi |\phi|^2\right)R_J .
\end{align}
For large field values and negative values of $2\hat{\lambda}_{\phi\sigma}+\lambda_{\phi\sigma}$, an inflationary valley arises in the direction
\begin{align}\label{eq:valley}
 \frac{|\sigma|^2}{|\phi|^2}\approx-\frac{(2\hat{\lambda}_{\phi\sigma}+\lambda_{\phi\sigma})}{\lambda_\sigma}\equiv \tan\omega,
\end{align}
{along {which} the potential can be captured by an effective quartic interaction with coupling
\begin{align}\label{eq:lambdaeff}
 \lambda_{\rm inf}=\cos^4\omega\left(\lambda_\phi-\frac{4\hat\lambda^2_{\phi\sigma}}{\lambda_\sigma}\right).
\end{align}

}
}

{We will further assume that $|2\hat{\lambda}_{\phi\sigma}+\lambda_{\phi\sigma}|\ll\lambda_\sigma$, such that the inflationary trajectory is mainly aligned with $\phi$. For achieving a flat potential in the Einstein frame compatible with the CMB constraints we consider} a non-zero non-minimal coupling $\xi_\phi$, whereas we may set $\xi_\sigma =0$ for simplicity. Parametrizing 
\begin{equation}\label{eq:AdefModel2}
\sigma = \frac{{\rho_\sigma}}{\sqrt{2}}e^{{\ic A/\rho_\sigma}}, \quad \text{and} \quad \phi  = \frac{{\rho_\phi}}{\sqrt{2}}e^{i b},
\end{equation}
 and considering the limit $\omega \approx 0,\lambda_{\phi\sigma}\ll\hat\lambda_{\phi\sigma}$ one can estimate the isocurvature mass during inflation {with the covariant formalism of  Appendix \ref{app:isomass}. In the limit $\rho_\phi\gg f_A$ this gives}
\begin{equation}\label{eq:ma2}
m_{A,{\rm inf}}^2 \sim {-\frac{4 \rho_\phi ^2 \hat{\lambda }_{\phi \sigma }}{\Omega_\phi ^2}-\frac{\xi  \rho_\phi ^4 \lambda _{\phi }}{\Omega_\phi ^4 \left(\xi  (6 \xi +1) \rho_\phi ^2+1\right)}},
\end{equation}
{where
\begin{align}
 \Omega^2_\phi=1+\frac{\xi_\phi \rho_\phi^2}{M_P^2}.
\end{align}
}
{In Eq.~\eqref{eq:ma2} we included a subscript ``inf'' for the axion mass to emphasize that the expression is only valid along the inflationary valley of Eq.~\eqref{eq:valley}. For large field values during inflation, with $\xi\rho_\phi^2/M_p^2>1$, one has $m^2_A\sim -4\hat\lambda_{\phi\sigma}M_P^2/\xi_\phi$. On the other hand, for such field values}
the Hubble rate scales like
\begin{equation}
H^2 \sim \frac{V_E}{3M_p^2} \sim {\frac{\lambda_\phi M_P^2}{12\xi^2_\phi}.}
\end{equation}
{Fitting the temperature power spectrum of the CMB requires $\xi_\phi\sim 2\times10^{5}\sqrt{\lambda_\phi}$ \cite{Bezrukov:2007ep}, which leads to  
\begin{align}
\frac{m^2_A}{H^2}\sim -\frac{10^{6}\hat{\lambda}_{\phi\sigma}}{\sqrt{\lambda_\phi}}, \quad \frac{\xi_\phi \rho_\phi^2}{M_P^2}>1.
\end{align}
Hence, even for $|\hat\lambda_{\phi\sigma}|\sim\sqrt{ \lambda_\phi}$ the axion becomes very massive during inflation, and its power spectrum suppressed. As before, we will focus on $\xi_\phi=0.1$, so that the CMB constraint can be satisfied for $\lambda_{\rm inf}\sim 10^{-11}$. Taking ${\hat\lambda}_{\phi\sigma} = -10^{-7}$,  ${\hat\lambda}_{\sigma} = 10^{-5}$ ensures then that the axion remains very heavy during inflation and that {the angle} $\omega$ stays small ({see Eq.\ \eq{eq:lambdaeff})}. After inflation the isocurvature perturbation may nevertheless be enhanced {due to nonperturbative effects during preheating, which were not {considered} in Ref.~\cite{Nakayama:2015pba}. We will include such effects by performing lattice simulations.}

\section{\label{sec:isobound}Revisiting the computation of the isocurvature bound}

In order to connect the primordial isocurvature perturbation to the dark matter - photon isocurvature component, that has been constrained by the {CMB} \cite{Akrami:2018odb}, 
we assume that all species, except for the axions, reach thermodynamic equilibrium after reheating has completed. {In SMASH for example, the quantitative estimates in Ref.~\cite{Ballesteros:2016xej} showed that the number densities of SM particles produced during reheating where high enough to achieve interaction rates above the Hubble scale, which is expected to lead to the thermalization of the SM particles {and of other particles  with sizable interactions with the former.} In the scenarios with $f_A\sim10^{11}$ GeV that were the focus of Ref.~\cite{Ballesteros:2016xej}, all particles including the axion itself were expected to reach thermal equilibrium. {Here we focus instead} on models with larger values of $f_A$ where the axion interactions are further suppressed, preventing equilibration}. 

{The former assumption}  implies that the energy densities of all {species other than the axion} were once determined by a common temperature, and hence they are adiabatic with respect to each other. {We recall that} in a relativistic thermal plasma, with $p=\rho/3$ for all species in the plasma , $\delta\rho_i/(\rho_i+p_i)=3{\delta T}/{T}$ and the adiabaticity condition is always satisfied, see Eq.~\eqref{eq:isodef}.
As reviewed in appendix \ref{app:iso}, the observable isocurvature perturbation {between axions and photons} is then given by 
\begin{equation}
\mathcal{S}_{A\gamma} \approx\frac{\theta^2-\langle \theta^2\rangle}{\langle\theta^2\rangle}.
\label{eq:isofraction}
\end{equation}
{In a pre-inflationary scenario, {the misalignment angle is $\theta(t,x)=\theta_i+\delta\theta(t,x)$ with $\langle\theta\rangle\approx \theta_i$.} Defining the power spectrum  $\Delta_X(t,|\bf{k}|)$ for an observable $X(t,\bf{k})$ as
\begin{align}\label{eq:Deltadef}
 \langle X(t,{\bf k}) X(t,{\bf k}')\rangle\equiv \frac{16\pi^5}{|{\bf k}|^3}\delta({\bf{k+k}'})\Delta_X(t,|{\bf k}|).
\end{align}
Then, under the assumptions above, the  power spectrum $\Delta_{S_{A\gamma}}(k_\star)$ of $S_{A\gamma}$  at a reference scale $k_\star$ can be approximated as 
\begin{align}\label{eq:isopower}
 \Delta_{S_{A\gamma}}(k_\star)\approx\frac{4}{\theta_i^2}\,\Delta_{\delta\theta}(k_\star).
\end{align}
{This follows directly from Eqs.~\eqref{eq:isofraction}, \eqref{eq:Deltadef} keeping the leading terms in $\delta\theta$.
This power} spectrum is to be contrasted with that of the {comoving} curvature perturbations, $\Delta_{\cal R}(k_\star)$, which is 
dominated by the contribution from adiabatic modes. $\Delta_{\cal R}(k_\star)$ is constrained by CMB and baryon acoustic oscillations (BAO) measurements giving \cite{Aghanim:2018eyx}
\begin{align}\label{eq:As}
\Delta_{\cal R}(0.05\,{\rm Mpc}^{-1})=(2.5\pm0.3)\times10^{-9}.
\end{align}
Taking all dark matter to be of axionic origin,\footnote{The final prediction can be easily generalized assuming a smaller energy density of axionic dark matter $\Omega_A < \Omega_\text{dm}$ (see e.g. \cite{Beltran:2006sq}).} created from the misalignment mechanism, our {scenario}
is captured by the {so-called} `Cold Dark matter Isocurvature' (CDI) model \cite{Linde:1985yf,Stompor:1995py}, which is commonly parametrized by the fraction 
\begin{equation}
\beta_\text{iso}(k_\star) \equiv\frac{\Delta_{S_{A\gamma}}(k_\star)}{\Delta_{S_{A\gamma}}(k_\star)+\Delta_{{\cal R}}(k_\star)} \approx\frac{4}{\theta_i^2}\frac{\Delta_{\delta\theta}(k_\star)}{\Delta_{\cal R}(k_\star)}.
\label{eq:beta}
\end{equation}}
In the last step we {used} $\beta_\text{iso}  \ll1$ and plugged in Eq. \eqref{eq:isopower}. 
Currently the strongest constraint on the CDI fraction is given by \cite{Akrami:2018odb}
\begin{align}\label{eq:isobound}
\beta_\text{iso}(0.002\,{\rm Mpc}^{-1})\lesssim 0.035,                                                           
\end{align}
considering that in our scenario the adiabatic and CDI modes are uncorrelated.
At first sight this seems to force 
a rather low scale of inflation, or else the model is ruled out \cite{Beltran:2006sq, Hertzberg:2008wr,Wantz:2009it}. {Indeed, if we assume}
that the axion perturbations acquire the variance of a massless scalar field during inflation, {which freezes at horizon crossing:\footnote{As mentioned in the introduction, we allow for a value of $f_A$ during inflation different to the one at late times.}
\begin{align}\label{eq:poweraxionmassless}
 \Delta_{\delta\theta,m_A=0}(k_\star)= \left.\left(\frac{H}{2\pi f_{A,\rm inf}}\right)^2\right|_{k_\star=a H}\equiv \left(\frac{H_{\rm inf}(k_\star)}{2\pi f_{A,\rm inf}}\right)^2,
\end{align}
then, substituting Eq.~\eqref{eq:poweraxionmassless} into Eq.~\eqref{eq:beta}, imposing Eq.~\eqref{eq:As} and the relic abundance constraint of Eq.~\eqref{eq:Omegah2preinf} leads to
\begin{equation}\label{eq:betaisomassless}
 \beta_{\text{iso},m_A=0}(k_\star) \sim 0.03 \left(\frac{f_A}{10^{16} \text{GeV}}\right)^{7/6}\left(\frac{f_{A,\rm inf}}{10^{16} \text{GeV}}\right)^{-2}\left(\frac{H_{\rm inf}(k_\star)}{10^9 \text{GeV}}\right)^{2}.
\end{equation}
As mentioned in the introduction, 
for  models with $f_{A,\rm inf}=f_A$ this is in tension with observations  unless the Hubble parameter is suppressed with respect to its {typical} values ($H \gtrsim 10^{13}$ GeV) in simple viable inflationary models involving scalar fields with non-minimal gravitational couplings.}
In \cite{Fairbairn:2014zta} it was however argued that, if the radial part of the PQ field drives inflation, the effective decay constant changes during inflation \cite{Linde:1991km},  {allowing $f_{A,\rm inf}$ to take Planckian values,} and opening up the range $10^{12} \text{GeV} \lesssim f_A \lesssim 10^{15} \text{GeV}$ to be compatible with observations for inflation driven by PQ field with non-minimal gravitational couplings featuring $H\gtrsim 10^{13}$. The upper bound in the previous window of $f_A$ was amended in Ref.~\cite{Ballesteros:2016xej} to $f_A\lesssim10^{14}$ GeV, with the difference arising from the fact that $f_{A,\rm inf}$ is not exactly equal to the value of the canonically normalized real inflaton field.

{Eq.~\eqref{eq:betaisomassless}} is based on two assumptions {that are not necessarily valid: first,} that the axions are massless during and after inflation, {and second, that the power spectrum remains frozen after horizon crossing during inflation.
As seen in the previous section, if the axion field {is} dynamical during inflation, Goldstone's theorem does not apply and the axion can develop a mass. This invalidates both of the above assumptions. Previous calculations in the literature have not accounted for both effects simultaneously. With respect to models analogous to our Model 1, the estimates of Refs.~\cite{Fairbairn:2014zta,Ballesteros:2016xej} did not account for {massive axion perturbations} during inflation. In Ref.~\cite{Ballesteros:2016xej} it was emphasized how axion perturbations grew during reheating, and that axionic perturbations were expected to be resonantly amplified in a window of momenta including zero, even when neglecting $f_A$, {leading to} the possibility of tachyonic axion masses at the origin (see Eq.~\eqref{eq:ma1}). This implies a violation of the second assumption above.   For small enough $f_A$, the axion fluctuations seen in Ref.~\cite{Ballesteros:2016xej} during reheating actually lead to  a non-thermal restoration of the PQ symmetry that remains incompatible with pre-inflationary scenarios. {Indeed, it is not clear whether the late time values of the time-dependent isocurvature perturbations at the scales probed by the CMB can remain below the Planck bound. Besides, although the growth of fluctuations was not explored in detail} for large $f_A$, it was hypothesized that the PQ restoration could fail for $f_A\gtrsim 10^{16}$~GeV: the perturbations grow exponentially when the background oscillates around the origin of field space, and it was estimated that for $f_A\gtrsim 10^{16}$ GeV the field would settle into a potential well before many oscillations could be completed. This hypothesis will be tested in Section~\ref{sec:results}.  In regards to analogues of Model 2, the analysis of Ref.~\cite{Nakayama:2015pba} accounted for a large axion mass during inflation, suppressing the primordial isocurvature perturbations, but still assumed their freezout at horizon crossing, and did not account for the possible growth of perturbations during reheating. 

In the next sections we will perform computations for models 1 and 2 that do not rely on the 
assumptions spelled above. In section  \ref{sec:evolutionisoinflation} we will study the evolution of the isocurvature power spectrum during inflation, accounting for the nonzero axion mass. In this regime perturbations are small and one can use a linear analysis, solving for the evolution of the Fourier modes with a proper normalization that that allows to estimate directly the power spectra of quantum fluctuations from the mode amplitudes. Once the power spectra are obtained up until the end of inflation, we will use the results as initial conditions for a lattice simulation, which {aims to} capture the nonlinear effects expected when the background oscillates after the end of inflation. This will be done in Section \ref{sec:results}.}

\section{Evolution of isocurvature perturbations during inflation}\label{sec:evolutionisoinflation}

{In this section we study the evolution of the power spectrum of {cosmological} perturbations, including axion isocurvature perturbations, during inflation. The power spectra are related to 2-point correlators of observables dependent on spatial momenta $\bf{k}$. For an observable $X(t,\bf{k})$, its power spectrum $\Delta_X(t,|\bf{k}|)$ is defined as in Eq.~\eqref{eq:Deltadef}.
When choosing $X$ as an operator corresponding to the fluctuation $\delta\hat\varphi_n$ of a real scalar field, then,  as is familiar from the expansion of quantum fields in terms of creation and annihilation operators multiplied by mode functions that solve the classical equations of motion, one can recover the 
quantum averages for fields $\varphi_n$ in terms of the mode functions. This requires solving the evolution equations for the coefficients, and implementing the appropriate normalization such that for modes well inside the horizon one recovers the usual Minkowski mode functions. That is, expressing
\begin{align}
 \delta\hat\varphi_n(t,{\bf k})=a_{n,\bf k} \delta\varphi_{n,{\bf k}}(t)+a_{n,{\bf k}}^\dagger \delta\varphi^\star_{n,\bf k}(t)
\end{align}
with 
\begin{align}
 [a_{n,\bf k},a^\dagger_{n,\bf{k}'}]=(2\pi)^3\delta^{(3)}({\bf k}-{\bf k}')
\end{align}
one has
\begin{align}\label{eq:power_spectra_from_modes}
\Delta_{\delta\varphi_n}(t,|{\bf k}|)=\frac{|\bf k|^3}{2\pi^2}\,|\delta\varphi_{n,\bf k}(t)|^2.
\end{align}
Let us consider modes $\delta\varphi_{n,\bf k}$ {corresponding to multi-field} fluctuations in a Friedmann-Robertson Walker (FRW) background metric $ds^2=dt^2-a(t)^2d{\bf x}^2$ (with $H=\dot a/a$).}
In the Einstein frame, neglecting deviations {metric in field space} with respect to a flat metric (which holds for small $\xi$) the equations of motion for the modes $\varphi_{n,\bf{k}}$ {are}
\cite{Gordon:2000hv}{
\begin{align}\label{eq:eoms}
 \delta\ddot\varphi_{n,\bf k}(t)+3H\dot\delta\varphi_{n,\bf k}(t)+\left(\frac{|{\bf k}|^2}{a^2}\delta_{nm}+\partial_m\partial_n V(\bar\varphi_p)-\frac{1}{M_P^2 a^3}\frac{d}{dt}\left(\frac{a^3}{H}\dot{\bar\varphi}_n\dot{\bar\varphi}_m\right)\right)\delta\varphi_{m,\bf k}(t)=0.
\end{align}
In a field basis where the mixing between fields is absent, the boundary condition that recovers the Minkowski mode functions well inside the horizon is
\begin{align}\label{eq:bc}
 \varphi_{n,\bf k}(t)\rightarrow\frac{e^{-i|\bf  k|\int_{t_0}^t dt'/a(t')}}{2^{1/2}\,(|{\bf {k}}|^2+\partial_n^2 V(\bar\varphi_m))^{1/4}} \quad \text{for $|{\bf k}|\gg a(t) H(t)$}.
\end{align}
}
In our models the axion field is contained in the phase $\theta$ of a complex field $\sigma$, with $\theta=\arg\sigma$. The inflationary background is assumed to be aligned with a direction of fixed $\bar\theta=\theta_i$, where $\theta_i$ is the misalignment angle entering the dark-matter abundance constraint of Eq.~\eqref{eq:Omegah2preinf}. As follows from the latter equation, for large $f_A$ one has $\theta_i\ll1$ {(assuming all dark matter is in the form of axions)}, so that the inflationary trajectory is mostly aligned with the real part of $\sigma$. {For convenience we rotate our basis to be aligned with $\theta_i$. {In the new basis with a zero value of $\arg\sigma$, the minimum of the axion potential will be displaced from zero, so there is no conflict with the misalignment mechanism, in which $\theta_i$ should be understood as the initial deviation of $\theta$ from the minimum of the axion potential.} Writing
\begin{align}\label{eq:sigmari}
 \sigma\equiv\frac{1}{\sqrt{2}}(\sigma_r+\ic \sigma_\theta),
\end{align}
then the axion isocurvature perturbation of Eq.~\eqref{eq:isofraction} can be approximated {to linear order} as
\begin{align}\label{eq:Sagammaapprox}
 S_{A\gamma}\approx \frac{2\delta \sigma_\theta}{\theta_i\bar\sigma_r}.
\end{align}
Thus we can estimate $S_{A\gamma}$ during inflation from the background value $\bar\sigma_r$ and the real scalar fluctuation $\delta\sigma_\theta$, which can be computed by solving Eq.~\eqref{eq:eoms} with the appropriate mass and the boundary condition of Eq.~\eqref{eq:bc}. We note that in the literature of inflation with multiple scalar fields, it is customary to define  isocurvature perturbations during inflation {--see e.g.\ \cite{Cespedes:2012hu,Achucarro:2012sm}), where they are defined to be orthogonal to the inflationary trajectory--} which in principle are not straightforwardly related to the  definition of isocurvature fluctuations in the plasma during the radiation era (reviewed in Appendix \ref{app:isomass} and which lead to Eq.~\eqref{eq:isodef}). 
In our {models} the inflationary trajectory will be  aligned with $\sigma_r$. Thus $\sigma_\theta$ corresponds to {the } orthogonal direction, and 
our $S_{A\gamma}$ of Eq.~\eqref{eq:Sagammaapprox} is directly related to one of the  isocurvature modes {often} studied in multi-field inflationary models.

}

\subsection{Model 1}

Model 1 admits inflationary trajectories with a constant misalignment angle $\arg\sigma=\theta_i$, which we take to be very small so as to satisfy the axion relic abundance constraint {Eq.\  \eqref{eq:Omegah2preinf}} for large $f_A$. {As mentioned before, we choose {a field basis ($\sigma_r, \sigma_\theta$) such that} the {inflaton background field}
is aligned with $\sigma_r$ {and the}
mass matrix is  diagonal.
During inflation 
${f_A}\ll\rho$ and thus the results for the power spectra are essentially independent of $f_A$. We choose 
$\xi=0.1$, which requires {$\lambda\simeq1.27\times10^{-11}$} in order to fit the {CMB constraint}
Eq.~\eqref{eq:As}. We further choose $\lambda_{H\sigma}=10^{-6}$, motivated by the fact that the SMASH {model} favours similar values in order to guarantee stability of the Higgs potential at large fields with respect to quantum fluctuations of the top quark \cite{Ballesteros:2016xej}. The  power spectra of the perturbations $\delta\sigma_r$, $\delta\sigma_\theta$, $S_{A\gamma}$ can be obtained by using  Eqs.\ \eqref{eq:power_spectra_from_modes} and \eqref{eq:Sagammaapprox}. In spatially flat gauge the power spectrum of $\phi_r$ is related to that of the dimensionless curvature perturbation $\cal R$ via
\begin{align}
 \Delta_{\cal R} = \frac{H^2}{(\dot{\bar\sigma}_r)^2}\,\Delta_{\delta\sigma_r}.
\end{align}
The results for the power spectra at the end of inflation are illustrated in Fig.~\ref{fig:power_spectra_model_1}. As expected from the previous discussion, the power spectra of $\sigma_i$, $S_{A\gamma}$ are suppressed with respect to their values assuming masslessness, and there  is no freezing of the spectrum at horizon crossing. At the end of inflation, however, the suppression for $\xi=0.1$ is mild (just a factor of $\sim 2$ for superhorizon modes). Note that the behaviour of the power spectra changes between superhorizon ($k< a_{\rm end} H_{\rm end}$) and subhorizon ($k> a_{\rm end} H_{\rm end}$), where $a_{\rm end}$ and $H_{\rm end}$ denote the scale factor and Hubble constant at the end of inflation; we have used units with $a_{\rm end}=1$. For subhorizon modes the power spectra approach the Minkowski result following from Eq.~\eqref{eq:bc}. On the right panel of Fig.~\ref{fig:power_spectra_model_1} we show both the curvature and isocurvature perturbations, the latter for different choices of the misalignment angle $\theta_i$ corresponding to $f_A=10^{14}$ GeV and $f_A=5\times10^{17}$ GeV {(for all dark matter in axions)}. For the isocurvature power spectra on the right plot of Fig.~\ref{fig:power_spectra_model_1}, at each value of $f_A$ we show not only the result obtained as detailed before (given by solid lines), but also when one neglects the isocurvature mass (dashed lines), and when one not only neglects the isocurvature mass but further assumes that the power spectrum freezes at horizon crossing (dotted lines). The latter case corresponds to the previous estimates in the literature \cite{Ballesteros:2016xej}, according to which $f_A<10^{14}$ GeV was thought to satisfy the isocurvature bounds. This can be seen from the fact that the blue dotted line in Fig.~\ref{fig:power_spectra_model_1}, corresponding to the isocurvature power spectrum for $f_A=10^{14}$ GeV in the approximations of Ref.~\cite{Ballesteros:2016xej},  remains safely below the extrapolation of the curvature power spectrum (solid orange line) to small CMB scales. However, for the improved estimation of the isocurvature power spectrum at $f_A=10^{14}$~GeV (solid blue line), one would infer a violation of the isocurvature bound if the spectra were to remain constant after inflation. Assuming masslessness and freezout at horizon crossing underestimates the isocurvature power spectrum by a factor that grows with decreasing $k$, and which is already of the order of 20 for the smallest scales in Fig.~\ref{fig:power_spectra_model_1}. Since the better estimate of the isocurvature power spectrum flattens out at small scales, one can extrapolate the deviation from the usual calculations to CMB scales, and again a factor around 25 is expected. Assuming masslessness but no freezout at horizon crossing overpredicts the power spectrum by a factor of approximately
3.4 
for the smaller scales in the figure, which gives a factor around 4.5 when extrapolating to CMB scales. Note that if one assumes the fluctuation $\delta_\sigma$ to be massless, its power spectrum freezes at horizon crossing as for any massless {scalar} in {de} Sitter. However, the power spectrum of $S_{A\gamma}$ does not freeze out due to the additional suppression by 
$\bar\sigma_r$, as in Eq.~\eqref{eq:Sagammaapprox}.

\begin{figure}[h]
\centering     
\includegraphics[width=.45\textwidth]{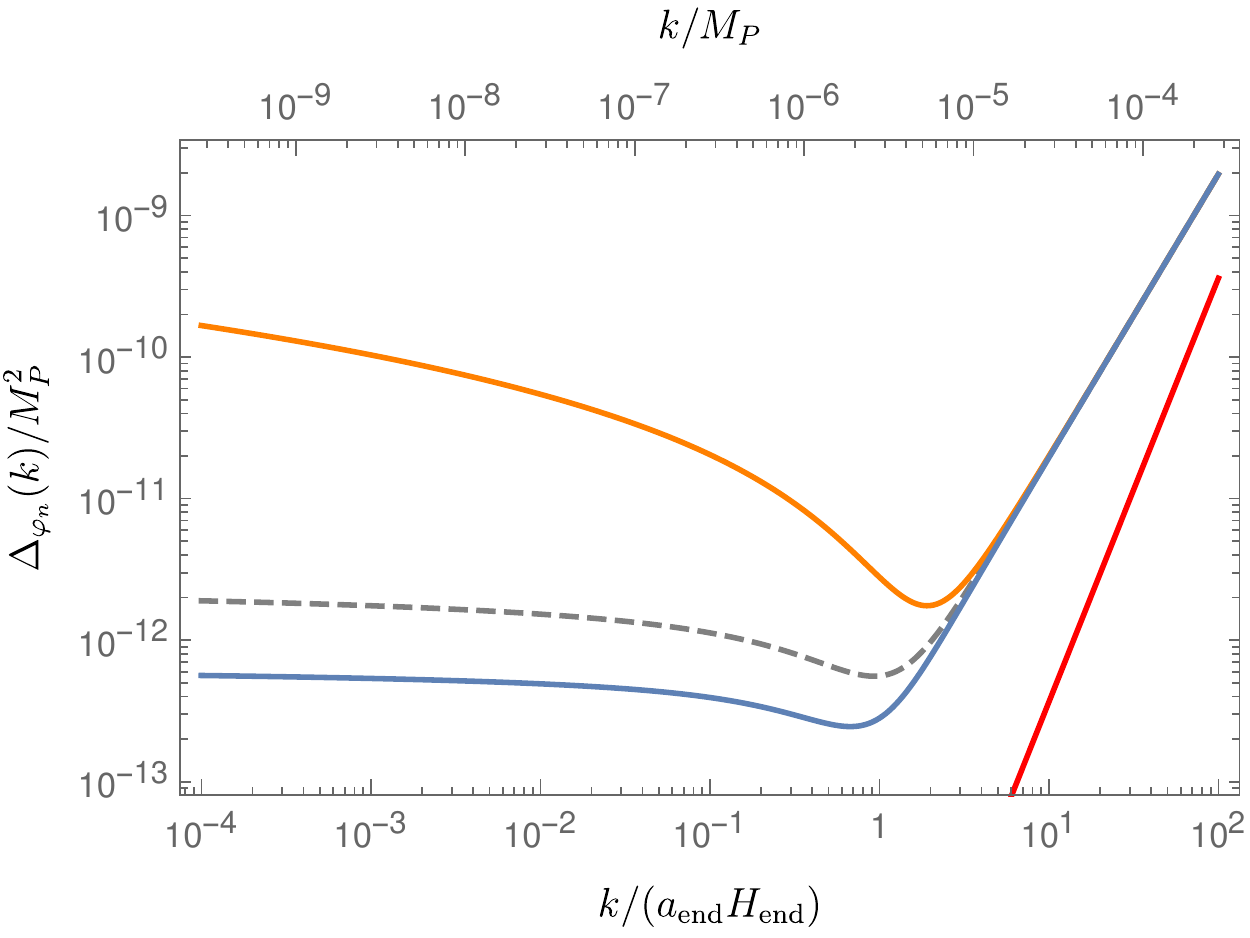}
\includegraphics[width=.45\textwidth]{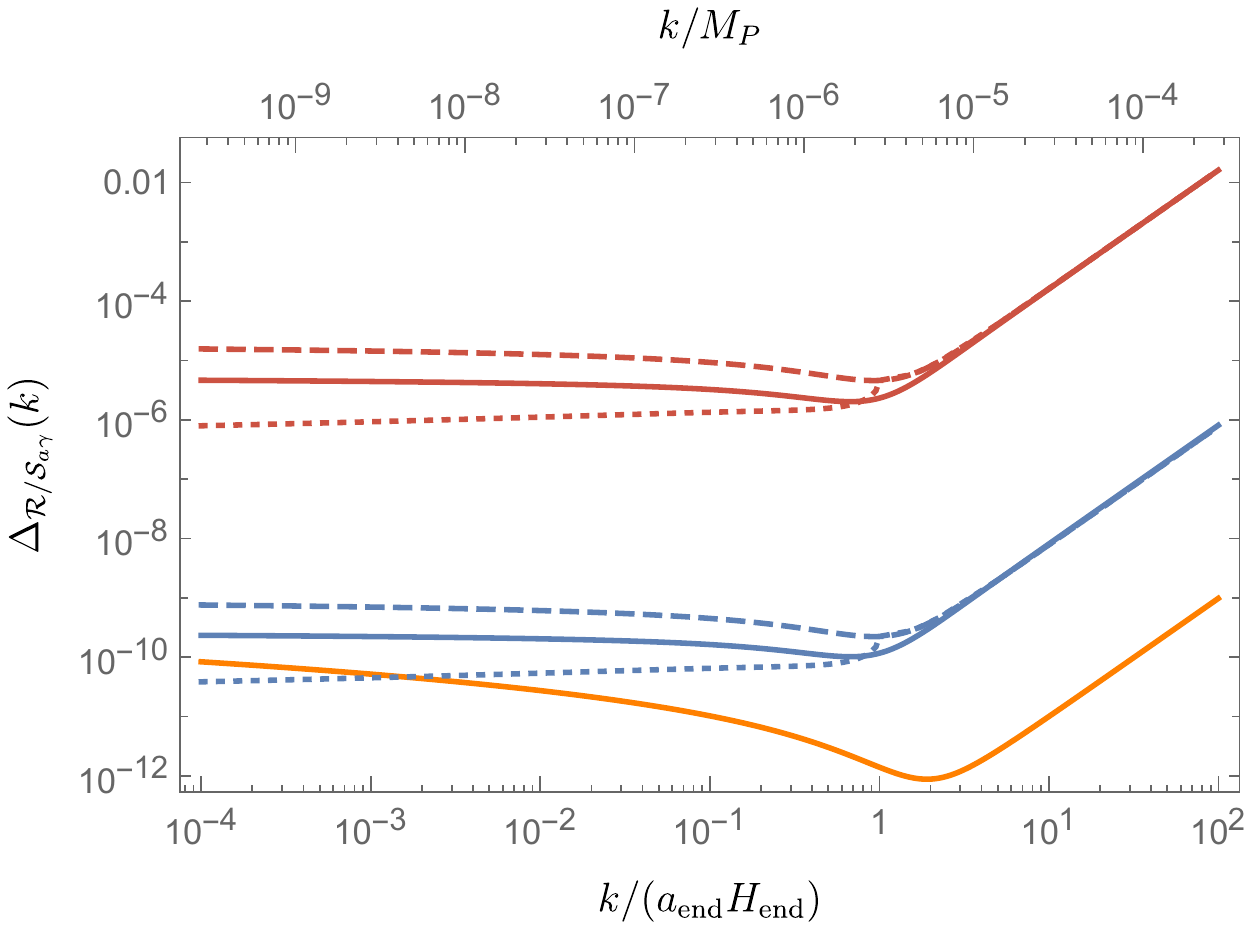}
\caption{{Power spectra at the end of inflation in Model 1 for $\xi=0.1, \lambda=1.27\times 10^{-11}$,  $\lambda_{H\sigma}=10^{-6}$. The left plot gives the spectra for $\sigma_r$ (orange), $\sigma_\theta$ including mass effects (blue) or without them (dashed gray), and $h$ (red). The right plot gives the dimensionless power spectra for the curvature perturbation ${\cal R}$ (orange), and the axion isocurvature perturbation ${\cal S}_{A\gamma}$ for $f_A=5\times10^{17}$ GeV (red) and $f_A=10^{14}$ GeV (blue). The dashed lines correspond to the result when mass effects are ignored, and the dotted lines give the result when assuming masslessness and freezout of the power spectrum at horizon crossing, as in previous estimates in the literature. The latter assumption underestimates the isocurvature spectrum at the end of inflation by a factor of the order of 20 for small scales.} } 
\label{fig:power_spectra_model_1}
\end{figure}

{We can {also} use Eq.~\eqref{eq:iso_orthogonal_evolution} to study the suppression of the super-horizon isocurvature spectrum for more general inflationary scenarios with different values of $\xi$. We expect that similar results will apply for $S_{A\gamma}$, given that due to Eq.~\eqref{eq:Sagammaapprox}, one has $S_{A\gamma}\approx {\cal F}/\bar\sigma_r$.}
We define a decay factor $d(\xi, f_A)$ {that quantifies} how much the isocurvature power spectrum has decayed with respect to the massless case by the end of inflation, that is, 
\begin{align}\label{eq:decayfactor}
 \Delta_{\mathcal{S}}(t_\text{end},k_\star)=e^{-d(\xi, f_A)} \frac{H_{\rm inf}^2}{4\pi^2}.
\end{align}
We evaluate the decay factor numerically, and find $d(\xi = 0.001) \sim 3.8$, $d(\xi = 0.01) \sim 2.5$, $d(\xi = 0.1) \sim 1.3$ and $d(\xi = 1) \sim 0.8$. Moreover, the decay factors decrease with increasing $f_A$, but this becomes only relevant for $f_A \gtrsim 0.1 M_p$. As we will see in \ref{sec:results}, the enhancement factor after inflation is extremely large in comparison, rendering the precise value of the decay factor irrelevant, {and for practical purposes it suffices to know it is $\mathcal{O}(1)$.}

\subsection{Model 2\label{subsec:model2:inflation}}

{For this model} we consider an inflationary background as in Eq.~\eqref{eq:valley} {and, similarly to Eq.~\eqref{eq:sigmari}, we define
\begin{align}
 \phi=\frac{1}{\sqrt{2}}\,(\phi_r+\ic\phi_\theta).
\end{align}} 
Again, one can consider a small misalignment angle $\theta_i={\rm arg} \sigma$, {and rotate the basis} such that the background trajectory is  aligned with the fields $\sigma_r$ and $\phi_r$. With $\lambda_{\phi\sigma}=0, \xi_\sigma=0$,  the effective inflationary quartic coupling along the background trajectory is $\lambda_{\rm inf}=\cos^4\omega(\lambda_\phi-{4\hat\lambda^2_{\phi\sigma}}/{\lambda_\sigma})$, and the effective nonminimal coupling is $\xi_{\rm inf}=\cos^2\omega \,\xi_\phi$, where $\omega$ was defined in Eq.~\eqref{eq:valley}.  A choice of parameters ensuring the same effective inflationary parameters as in the {example of Model 1}, while {also} yielding very heavy axion fluctuations during inflation, is $\xi_\phi=0.1$, $\lambda_\phi=4.01\times10^{-9}$, $\lambda_\sigma=10^{-5}$, $-\hat\lambda_{\phi\sigma}=\lambda_{H\phi}=\lambda_{H\sigma}=10^{-7}$. The resulting power spectra at the end of inflation for the fluctuations $\delta\phi_{r/\theta}, \delta\sigma_{r/\theta}$ are given in  Fig.~\ref{fig:power_spectra_model_1}. Note how the fluctuations in $\sigma_{r/i}$ are now heavily suppressed, yielding negligible axion isocurvature fluctuations during inflation. We find that the power spectra for the massive fields can be recovered by substituting in Eq.~\eqref{eq:power_spectra_from_modes} the Minkowski mode functions of Eq.~\eqref{eq:bc}, even for superhorizon modes.

\begin{figure}[h]
\centering     
\includegraphics[width=.45\textwidth]{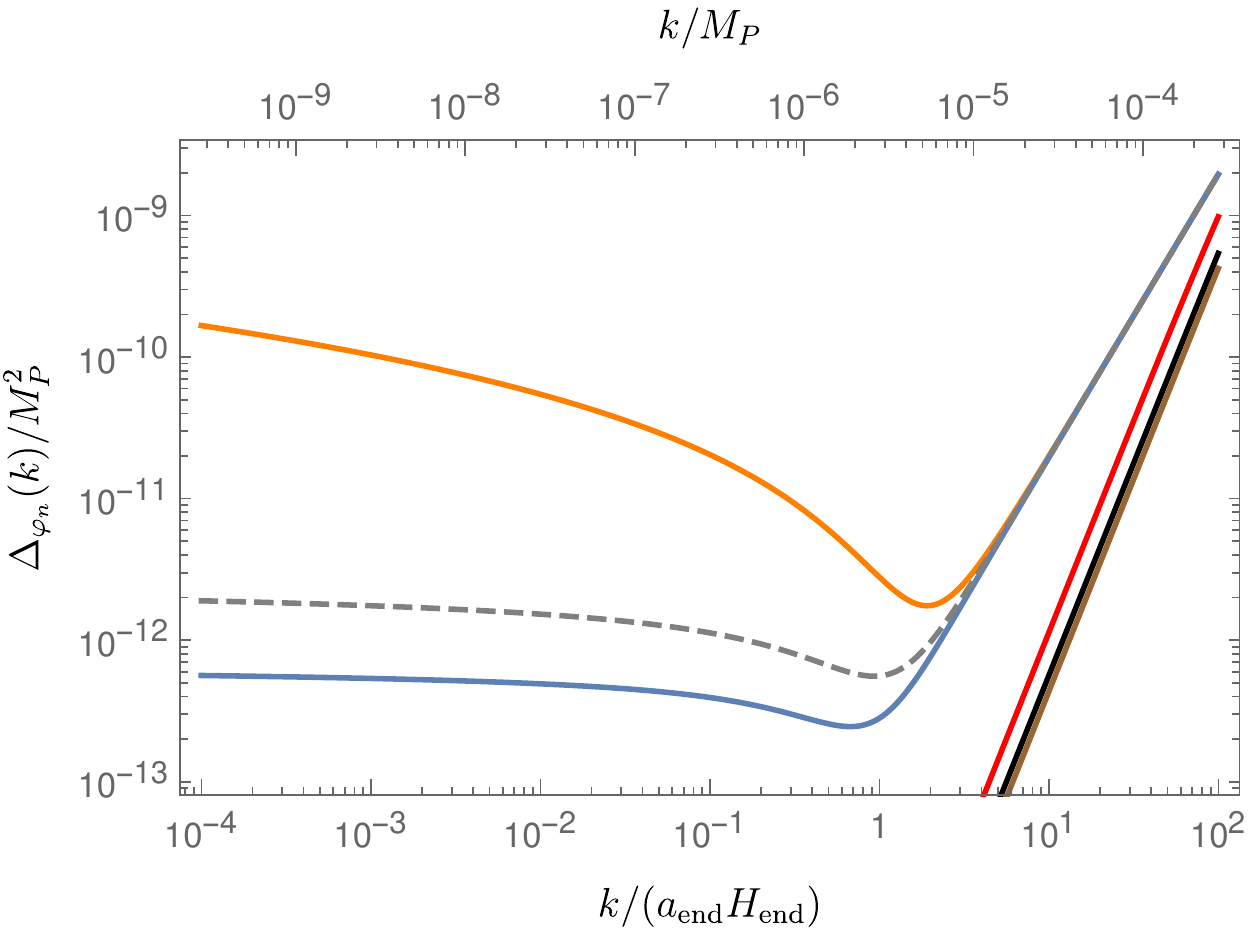}
\caption{{Power spectra at the end of inflation in Model 2 for $\xi_\phi=0.1, \xi_\sigma=0, \lambda_\phi=4.01\times 10^{-9},\lambda_\sigma=10^{-5},\lambda_{\phi\sigma}=0$,  $-\hat\lambda_{\phi\sigma}=\lambda_{H\phi}=\lambda_{H\sigma }=10^{-7}$. The spectra correspond to $\phi_r$ (orange), $\phi_i$ including mass effects (blue) or without them (dashed gray), $h$ (red), $\sigma_r$ (brown), $\sigma_i$ (black).} } 
\label{fig:power_spectra_model_2}
\end{figure}

\section{\label{sec:results}Evolution of isocurvature perturbations after inflation}

In this section we study the impact of non-perturbative effects during reheating on the power spectrum of the isocurvature perturbation $S_{A\gamma}$. For this we will take the power spectra at the end of inflation computed in the previous section, and use them to set initial conditions for lattice simulations of the evolution of the scalar fields. Here we make use of the fact that the quantum dynamics in high-occupancy states can be approximated by averaging over classical dynamics with random initial conditions sampled from an initial quantum wave-function {see \cite{Hertzberg:2016tal} for a discussion in the case of bosonic fields).}  The power spectra $\Delta_{\delta\varphi_n}(|\bf k|)$ at the end of inflation are variances of the probability distribution for the momentum modes of the fluctuations $\delta\varphi_n(|\bf k|)$. Assuming Gau{\ss}ianity 
we can 
generate a random sample of initial fluctuations in Fourier space for each field. Once an initial condition is fixed, a classical evolution is performed. Even  a single classical simulation can capture the quantum dynamics, because within a fixed window of $|\bf k|$ there can be many discrete lattice momenta whose initial conditions were sampled from the quantum probability distribution. Hence, a single lattice simulation is in effect evolving in parallel many modes within a given momentum shell, and performing averages over the momenta in the shell will capture the effect of quantum fluctuations. While the initial conditions and the power spectra are then computed in Fourier space, the fields are evolved in configuration space including nonlinear effects.

\subsection{Model 1}

\subsubsection{Required suppression factor of isocurvature perturbations} 
\label{subsec:suppression} 
{Before embarking on the details of the lattice simulations,}
let us estimate the suppression factor $\kappa$ of the power spectrum of isocurvature perturbations compared to its value at the end of inflation, required to match the {Planck bound} \cite{Akrami:2018odb}, that is
\begin{equation}\label{eq:kappadef}
\kappa \equiv \frac{\Delta_{\mathcal{S}_{A\gamma}}(t_\text{CMB},k_\star)}{\Delta_{\mathcal{S}_{A\gamma}}(t_\text{end},k_\star)} \quad \longrightarrow \quad \beta_\text{iso} \approx \kappa \frac{\Delta_{\mathcal{S}_{A\gamma}}(t_\text{end},k_\star)}{\Delta_{\mathcal{R}}(k_\star)}< 0.035 \ .
\end{equation}
{In the previous equation, $t_{\rm end}$ and $t_{\rm CMB}$ denote the time at the end of inflation and at {photon} decoupling, respectively. {In the second expression, we have omitted a time dependence in the curvature power spectrum assuming it freezes at horizon crossing.} 
We can then parametrize the isocurvature power spectrum at the end of inflation as (see Eqs.~\eqref{eq:isopower}, \eqref{eq:calS}, \eqref{eq:decayfactor})}
\begin{equation}
 \Delta_{\mathcal{S}_{A\gamma}}(t_\text{end},k_\star) =
 \frac{4 e^{-d(\xi, f_A)}}{G_{\theta\theta} \theta_\text{in}^2} \left(\frac{H_{\rm inf}(k_\star)}{2\pi}\right)^2.
\end{equation}
{Computing $G_{\theta\theta}$ from Eq.~\eqref{eq:metric}, we find that at the end of inflation}
it takes values {$0.4M^2_p \lesssim G_{\theta\theta} \lesssim 5 M^2_p$}  as the non-minimal coupling ranges between $1 > \xi > 10^{-3}$ . Meanwhile, the Hubble parameter varies between {$10^{13}~\text{GeV} < H_{\rm inf} < 10^{14}~\text{GeV}$} \cite{Ballesteros:2016euj}. 
Moreover, for sufficiently large $f_A$, the misalignment angle {can be determined {assuming all the DM in in the form of axions.}
Taking all this together we find that the maximal allowed value for the suppression factor {must be in the range} %$5\cdot 10^{-26} (f/\text{GeV})^{7/6} \lesssim\langle \mathcal{S}^2_{A\gamma}(t_\text{end})\rangle \lesssim 3 
{\begin{equation}\label{eq:kappamax}
10^{-5}\lesssim \kappa_\text{max}  \left(\frac{f_A}{5\cdot 10^{17}~\text{GeV}}\right)^{7/6} \lesssim 10^{-6}\ .
\end{equation}
}

\subsubsection{Lattice simulations\label{subsec:lattice_simulations}}

In this section we focus on the evolution of the isocurvature perturbations after inflation, taking as initial conditions the results of the previous sections for $\xi=0.1$. For the post-inflationary evolution of the fields after inflation we assume a flat FLRW background metric and  neglect the effects of the nonminimal gravitational coupling {whose effect is small for small enough field values}. 
{We solve 
the evolution of the fields $\phi_i,\phi_r,h$ and the scale factor of the metric, neglecting metric perturbations. We describe the decay of the Higgs into SM particles through a decay term controlled by a rate $\Gamma_h$ (described in Appendix \ref{app:Higgsdecays}). } 
Analogous terms are not necessary for the fields $\phi_i,\phi_r$, as they  only interact directly among themselves and with the Higgs, and all these interactions are already included in the equations of motion.
As the Higgs {excitations} will decay into relativistic SM particles, we model the decay products in terms of a homogeneous relativistic fluid with energy density $\rho_{\rm SM}(t)$ and pressure $p_{\rm SM}(t)=\rho_{\rm SM}(t)/3$. We include the feedback of $\rho_{\rm SM}$ into the evolution of the scale factor $a(t)$, and we compute the time evolution of $\rho_{\rm SM}$ by imposing covariant conservation of the  total stress-energy momentum tensor up to effects from spatial gradients, as these are neglected for $\rho_{\rm SM}(t)$ itself. In summary, we {solve the }
 following 
 equations for 
 $\phi_r(t,\vec{x}),\phi_i(t,\vec{x}),h(t,\vec{x})$,  
 $a(t)$ 
 and 
 $\rho_{\rm SM}(t)$:
\begin{align}\label{eq:eomlattice}\begin{aligned}
&\ddot\varphi_n+3\frac{\dot a}{a}\dot\varphi_n-\frac{1}{a^2}\vec{\nabla}^2\varphi_n+\frac{\partial V(\varphi_m)}{\partial\varphi_n}+\Gamma_n\dot{\varphi}_n=0,\\
%%%%
&\dot{\rho}_{\rm SM}+4 {\frac{\dot a}{a}} \rho_{\rm SM} - \Gamma_h \dot{h}^2=0,\\
%%%%
&3M^2_P\left(\frac{\dot a}{a}\right)^2=\rho_{\rm SM}+V_J+\frac{1}{2}\sum_n\dot{\varphi}_n^2+\frac{1}{2a^2}\sum_n({\bf\nabla}\phi_n)^2\,.
\end{aligned}\end{align}
{Where,}
$n,m=1,2,3$, and 
$\varphi_1=\sigma_r$, $\varphi_2=\sigma_i$, $\varphi_3=h$. We take $\Gamma_1=\Gamma_2=0$, while for the decay rate $\Gamma_3=\Gamma_h$ we take the sum of the SM partial widths, collected in Appendix \ref{app:Higgsdecays}, and substitute VEV insertions like $\langle h^2\rangle$ by averages of $h^2$ over the lattice. For the Higgs mass squared we use the average of $\partial^2 V/\partial h^2$, and require a positive result; otherwise $\Gamma_h$ is treated as zero. We note that the SM decay rates into massive gauge bosons diverge for $\langle h^2\rangle\rightarrow0$, as the assumption of massive bosons with 3 polarizations breaks down. In practice, a nonzero value of  $\langle h^2\rangle$ quickly develops as the fluctuations start growing, but for numerical stability we only consider these decay channels for 
$\langle h^2\rangle$ above a certain cutoff, see Appendix \ref{app:Higgsdecays} for more details.

The lattice calculations are implemented using a modified version of the {\tt CLUSTEREASY} software \cite{Felder:2000hq,Felder:2008zz}. The modifications amount to the following:
\begin{itemize}
 \item Implementation of Higgs decay terms. 
 
 \item Implementation of the evolution of the SM radiation density $\rho_{\rm SM}$.
 
 \item Modification of the evolution of the scale factor to account for $\rho_{\rm SM}$.
  
 \item Modification of the 
 initial conditions in order to use as input the power spectra at the end of inflation derived in Section~\ref{sec:evolutionisoinflation}.
\end{itemize}

With respect to the last point, we note that by default  {\tt CLUSTEREASY} assigns initial conditions for the fluctuations in momentum space by sampling with a Gau{\ss}ian probability distribution whose standard deviation is given by the modulus of the Minkowski solution of Eq.~\eqref{eq:bc}.   Comparing with the results for the power spectra at the end of inflation displayed in Fig.~\ref{fig:power_spectra_model_1}, {\tt CLUSTEREASY}'s initial conditions are correct for subhorizon modes ($k>a H$), but not for superhorizon fluctuations. 

As a technical aside, we shall mention that {\tt CLUSTEREASY} uses a staggered leapfrog method for solving the differential equations, in which, if the values of the variables $y$ and their second time derivatives are known for a time $t$, the values of the first time derivatives are only known at $t-dt/2$, where $dt$ is the discrete timestep of the numerical method. This is not well adapted to equations featuring single time derivatives, as the Higgs equation with a nonzero decay rate, or the equation for $\rho_{\rm SM}$. We opt for a simple workaround in which we estimate a first derivative $\dot y(t)$ of a variable $y$ as $\dot y(t)= \dot y(t-dt/2)+dt/2 \,\ddot y(t)+{\cal O}(dt^2)$.

Once the code provides the time-dependent solutions for the fields across the lattice, we can compute isocurvature power spectra by determining the angular variable $\theta$ and correspondingly the isocurvature fluctuation in Eq.~\eqref{eq:isofraction} at each lattice site. By performing a discrete Fourier transform and interpreting the fluctuations in Fourier space as samples following a probability distribution corresponding to the mode functions of a quantum field, we can estimate the power spectra at a scale $|\bf k|$ by using the analogous of Eq.~\eqref{eq:power_spectra_from_modes} and averaging over a {(spherical)} momentum shell centered on $|\bf k|$:
\begin{align}\label{eq:Sagammalattice}
 \Delta_{S_{A\gamma}}(|{\bf k}|)\approx\frac{|{\bf k}|^3}{2\pi^2}\, \frac{1}{N_{\rm shell}(|{\bf k}|,\Delta_k)}\sum_{\rm shell(|\bf k|,\Delta_k)} |S_{A\gamma}(\bf k)|^2,
\end{align}
where {``$_{\rm{shell}}$($|\bf k|,\Delta_k$)''} denotes a collection of discrete Fourier momenta  fitting inside a momentum shell with radius $|\bf k|$ and width $\Delta_k$, and $N_{\rm shell}(|{\bf k}|,\Delta_k)$ denotes the number of discrete momenta in the shell.\\

The {\tt CLUSTEREASY} implementation  makes use of dimensionless units for the spacetime coordinates and fields {as follows:}
\begin{align}\label{eq:lattice_dimless_units}
  \tau=&\,\int\frac{Bdt}{a},& \hat{\bf x}=&\, B\,{\bf x}, & B=&\,\sqrt{\lambda_{\rm inf}}\,\rho_{\rm inf, end}, & \lambda_{\rm inf}=\left\{\begin{array}{cr}\lambda_\sigma, &\text{Model 1}\phantom{,}\\                                                                                                                                                                                               %%%
\cos^2\omega\left(\lambda_\phi-\frac{4\hat{\lambda}^2_{\phi\sigma}}{\lambda_\sigma}\right), &\text{Model 2},
\end{array}\right.
\end{align}
where $\rho_{\rm inf}$ is the canonically normalized field 
along the inflationary trajectory. The lattice simulations are performed in finite spatial cubes with sides of length $L$ in dimensionless units {${\bf \hat x}$}. The number of discrete points per edge will be taken as a power of 2 and denoted as $N$. The minimum distance  between points will be denoted as $\Delta\hat{x}$, so that $L=(N-1)\Delta\hat{x}$. Similarly, we take $\Delta\tau$  to be the discrete timestep for the simulations.

The lattice simulations capture the dynamics of the fluctuations with $N^3$ discrete spatial comoving Fourier momenta which in lattice units go as
\begin{align}
 \hat{\bf k}=\frac{\bf k}{B}=\frac{2\pi}{L}(i_1,i_2,i_3), \quad -\frac{N}{2}\leq i_m\leq \frac{N}{2}-1.
\end{align}
The smallest nonzero comoving momentum  resolved in the simulation is thus ${k}_{\rm min}=2\pi B/L$, while the maximum momentum is $k_{\rm max}=\sqrt{3}\pi N/L$. As eventually we want to extrapolate the isocurvature power spectrum to CMB scales, which correspond to modes that were superhorizon at the end of inflation, we are interested in  capturing modes with associated lengths above the horizon scale at the end of inflation. Including these low energy modes is also important to capture possible instabilities due to the axion mass acquiring negative values. 
At the end of inflation, where effects due to the nonminimal gravitational couplings are subleading, Friedmann equations imply that the Hubble constant goes as
\begin{align}
 H\approx\frac{\sqrt{\lambda_{\rm inf}}\rho_{\rm inf}^2}{\sqrt{12} M_P}.
\end{align}
For our choices of parameters, we have $\rho_{\rm inf,end}\approx M_P$, so that the Hubble scale at the end of inflation is around {$B$}. Then one can capture physical momenta below the Hubble scale at the end of inflation by having $k_{\rm min}=B  \hat{k}_{\rm min}<B$, which can be achieved if $L > 2\pi$.

In principle 
particle production {should occur} at scales related to the frequency of oscillation of the background {fields} at the end of inflation, which in a quartic potential also goes as the scale $B$. To capture such modes one needs $k_{\rm max}> B$, which requires $L<\sqrt{3}\pi N$. {If long wavelength and short wavelength modes evolved independently, the behaviour of fluctuations well above the horizon would not be affected by the fluctuations with frequencies around $B$. However, modes of different wavelength are coupled through the non-linear evolution. Nevertheless, modes at scales separated by many orders of magnitude might be expected to evolve separately, at least to some extent. For this reason, and for practical computational purposes, we will mostly focus on large values of $L$ that capture superhorizon dynamics, without necessarily requiring that our simulations include the dynamics at the scale $B$. Nevertheless, we will perform consistency checks by computing isocurvature power spectra for different boxes capturing different intervals of $|\bf k|$, checking if the results are compatible for overlapping values of momenta.}

In order to identify the comoving momentum $|\bf k|$ at the end of inflation that corresponds to the CMB pivot scale of $0.002 {\rm Mpc}^{-1}$,  we have to match physical momentum scales $|{\bf k}|/a$, accounting for differences in normalization of the scale factor. {The evolution of the latter, as well as the associated Hubble rate, can be obtained from the results of our simulations up to their maximum reach in $\tau$. In Model 1, reheating is efficient and takes place within the simulated time, so that the onset of radiation domination is captured; extrapolating to large times and matching with the observed value of the Hubble constant at the present time leads to
\begin{align}\label{eq:a0end}
 \log\frac{a_0}{a_{\rm end}}\approx 63.
\end{align}
The same value can be obtained by assuming that radiation domination starts immediately after inflation. For large $f_A$  there can be periods of matter domination before the end of reheating; however, these periods are brief and they do not affect the estimate of Eq.~\eqref{eq:a0end}. For Model 2 reheating is less efficient and our simulations do not reach the time in which the SM radiation dominates the energy density. Thus, there remains an uncertainty in the number of efolds after inflation; nevertheless, we will still use Eq.~\eqref{eq:a0end} as a reasonable estimate. }

In our choice of units $a_{\rm end}=1$, the comoving momentum corresponding to the CMB pivot scale is obtained by demanding $k_\star/a_0\approx k_\star/\exp(63)=0.002 {\rm Mpc}^{-1}$, leading to
\begin{align}
 k_\star\approx1.2\times10^{-32} M_P.
\end{align}
For the choice of inflationary quartic $\lambda_{\rm inf}=1.27\times10^{-11}$ compatible with a nonminimal gravitational coupling $\xi_{\rm inf}=0.1$, giving $\rho_{\rm inf,end}=1.7 M_P$, one has $B\approx 6\times10^{-6}M_P$. Thus the CMB pivot scale in lattice units is
\begin{align}
 {\hat k}_\star = \frac{ k_\star }{B} \approx2\times10^{-27}.
\end{align}
Capturing this scale in the simulations would require extraordinarily large boxes with $L>2\pi/ { \hat k}|_\star= {\cal O}(10^{27})$. This cannot be done while maintaining precision in the presence of
other dimensionless couplings of order one in lattice units. Hence we 
restrict our simulations to much smaller values of $L$, and {show}
a regular pattern 
for the isocurvature power spectrum at increasingly small momenta, {which we use to extrapolate} 
to the CMB pivot scale. {This extrapolation spans} a huge number of orders of magnitude, and thus the extrapolated results should be taken with due care. Some consistency checks of our lattice calculations are given in Appendix \ref{app:meanfieldapproximation}, where we compare the lattice results against those coming from solving linearized equations for the Fourier modes, improved by using the lattice averages and variances of the fields to determine background quantities.\\

\renewcommand{\theparagraph}{\S\arabic{paragraph}}
\setcounter{secnumdepth}{4}
\setcounter{tocdepth}{4}

\paragraph{\bf{\boldmath$f_A$ and PQ symmetry restoration.}}\mbox{}\vspace{-0.3cm}\\

As a first application of our lattice simulations, we {determine} the values of $f_A$ for which the PQ symmetry is not restored by nonperturbative effects, as needed for a valid pre-inflationary scenario. As mentioned before, Ref.~\cite{Ballesteros:2016xej} hypothesized that PQ restoration should be expected for $f_A\gtrsim 4\times 10^{16}$ GeV. To assess this we can simply run simulations for different values of $f_A$ and plot the variance of the angular variable $\theta$ in configuration space, obtained by performing averages over the lattice. 
We use the same choices of coupling as in Section~\ref{sec:evolutionisoinflation}: $\lambda=1.27\times10^{-11}$, $\lambda_{H\sigma}=10^{-6}$. The results for simulations with $L=14848$, $N=64$, a timestep $\Delta\tau=0.0005$ and 9 different values of $f_A$ between $10^9$~GeV and $5\times10^{17}$ GeV are shown in the upper plot in Fig.~\ref{fig:variance}. For 
$f_A\leq 10^{17}$ GeV the results for the variance of $\theta$ 
fall on top of each other, and the variance grows quickly to order one values,  {implying}
restoration of the PQ symmetry. 
For $f_A\gtrsim2\times10^{17}$ GeV
the growth of the variance is thwarted and is minimized for $f_A\sim 5\times 10^{17}$ GeV.  The slight growth in the perturbations for $f_A=4\times10^{17}$~GeV is due to 
the Higgs field, which can act as a source for perturbations in $\sigma_\theta$ 
before the Higgs fluctuations decay. 
The lower plot in Fig.~\ref{fig:variance} illustrates the growth in the relative energy density of the SM bath, $\Omega_{\rm SM}=\rho_{\rm SM}/\rho$, at early times. For $f_A<2\times10^{17}$ GeV the production of SM particles is 
blocked and $\rho_{\rm SM}$ {remains} essentially zero.
The SM particle production opens up {for higher values of $f_A$} and reheating {becomes} much more efficient. The results can be understood in terms of the average {fields} setting into a minimum of the potential at early times. Shortly after reheating, for large background field values one can neglect the quadratic terms, and  the average field can be understood as oscillating in a quartic potential {(see the discussion around Eq.\ \eq{eq:a0end}}). In this background, $\sigma_r$ and $\sigma_\theta$ have oscillating masses, and their fluctuations follow a Lam\'{e} equation which predicts exponential growth in wide bands of momenta \cite{Greene:1997fu}. For the $\sigma_\theta$ fluctuations the resonance band in the lattice units of Eq.~\eqref{eq:lattice_dimless_units} is $0<|{\bf \hat{k}}|<1/2$, which is partly captured by our choice of $L,N$, {which has}
$\hat{k}_{\rm max}\approx0.02$. We have checked that {these} results {about PQ restoration and reheating} do not change when capturing {moderately} larger  momenta by 
decreasing $L$ or increasing $N$. The large growth in the perturbations in the components of $\sigma$ leads to a large effective mass of the Higgs 
\begin{align}\label{eq:m2Heff}
 m^2_{H,\rm eff}\,{\simeq}\,\lambda_{H\sigma}(\langle|\sigma|^2\rangle-f_A^2),
\end{align}
 which blocks the production of Higgs perturbations and their decay into SM radiation. When the background {fields} settle into a potential well, the potential can be captured by quadratic terms, and in this case when neglecting the expansion of the universe the equations for the fluctuations can be {described in terms of a} 
 Mathieu equation \cite{Kofman:1997yn}. In our case, assuming that a $\bar\sigma_r$ background oscillates around  $f_A$ with an amplitude $x f_A$ and a frequency equal to the mass $m_r=\sqrt{2\lambda} f_A$ of $\sigma_r$ around the minimum,
\begin{align}
 \bar\sigma_r\approx f_A(1+ x \cos m_r t),
\end{align}
taking $a=1$ and assuming $x\ll1$ one gets a Mathieu equation for $\sigma_\theta$, 
\begin{align}
 \frac{d^2}{dz^2}\sigma_\theta(z)+\left(\frac{4{|\bf k|}^2}{m_r^2}-4x \cos 2z\right)\sigma_\theta(z)=0, \quad z=\frac{m_r t}{2}.
\end{align}
{This} equation has narrow resonance bands, the wider {of which corresponds} to an exponential growth {$\sigma_\theta\sim\exp(x m_r t/2)$}. The expansion of the universe, which was ignored in the previous arguments, causes a redshifting of the amplitude of oscillation $x$, and the resonance will only be effective if the growth rate $x m_r$ is above the Hubble constant \cite{Kofman:1997yn}. Hence, if the field gets trapped in the  minimum early on, with a small amplitude of oscillation, there will be no resonant growth of perturbations in $\sigma_\theta$. This effect seems to be indeed behind the failure of the PQ restoration for $f_A>2\times 10^{17}$ GeV seen in Fig.~\ref{fig:variance}. To illustrate this, in Fig.~\ref{fig:background_fa} we show the lattice average of $\sigma_r$ 
at early times for three values of $f_A$. For $f_A=10^{15}$ GeV the field keeps going through oscillations that cross the origin, and remains 
in the quartic regime with exponential growth of perturbations. For $f_A=2\times 10^{17}$ GeV the field settles into a minimum after around 10 crossings of the origin, while for $f_A=5\times 10^{17}$ GeV the background undergoes just 4 crossings before quickly settling into a minimum. We have checked that $x m_r/H$ drops below 1 in the latter simulation for $\tau\gtrsim150$, {whereas it stays much larger than 1} for $f_A=10^{17}$ GeV. Finally, let us note that when the field settles into a minimum and the perturbations decay the effective Higgs mass of Eq.~\eqref{eq:m2Heff} goes to zero, so that Higgs production and its subsequent decay into SM radiation become allowed. This explains the growth of $\Omega_{\rm SM}$ seen in the lower plot of Fig.~\ref{fig:variance} for $f_A>2\times10^{17}$ GeV, and why the SM particle production is faster for larger $f_A$ for which the background settles faster around the minimum.

\begin{figure}[t]
\centering     
\includegraphics[width=.57\textwidth]{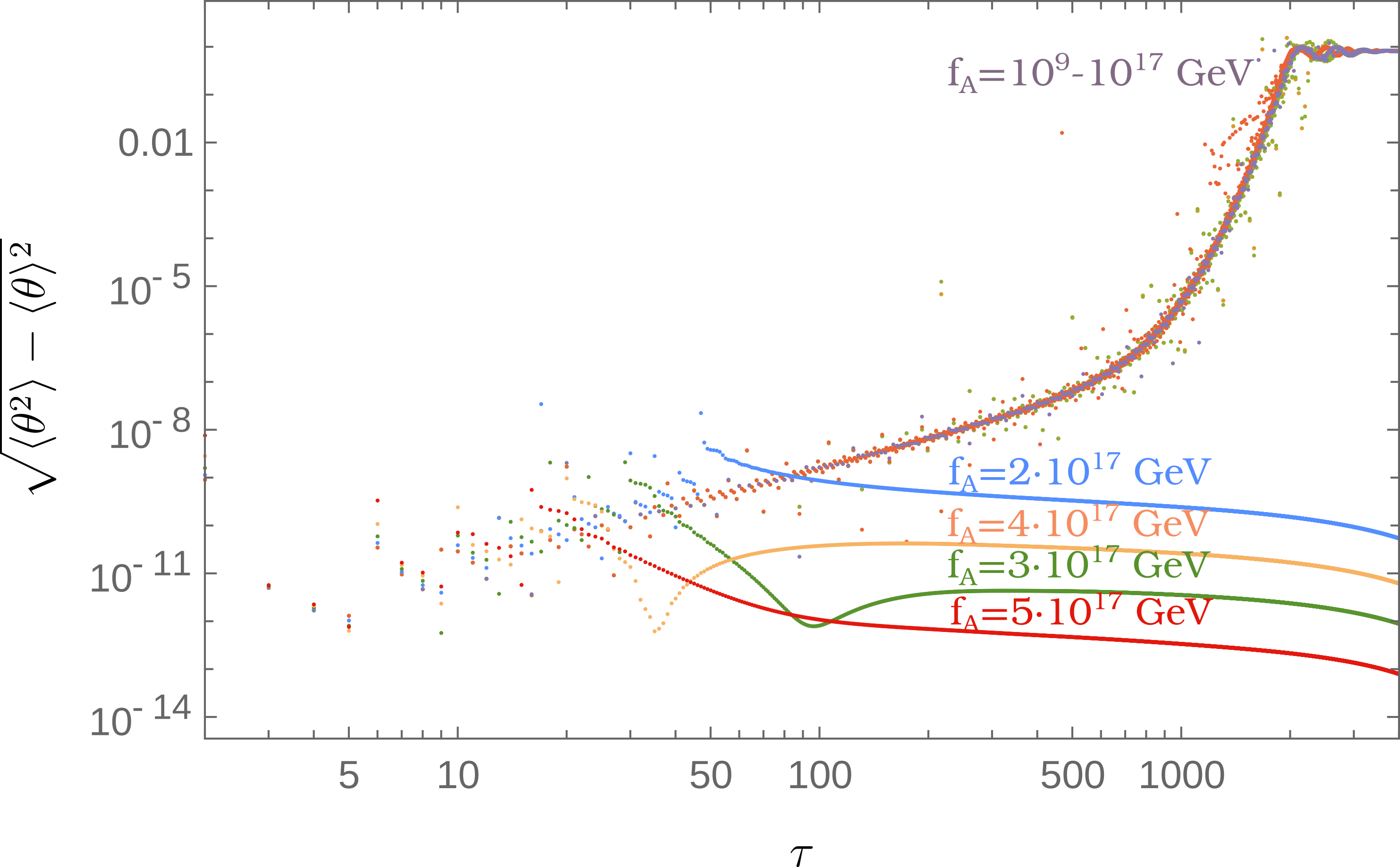}\\
\vskip.3cm
\,\hskip.2cm\includegraphics[width=.57\textwidth]{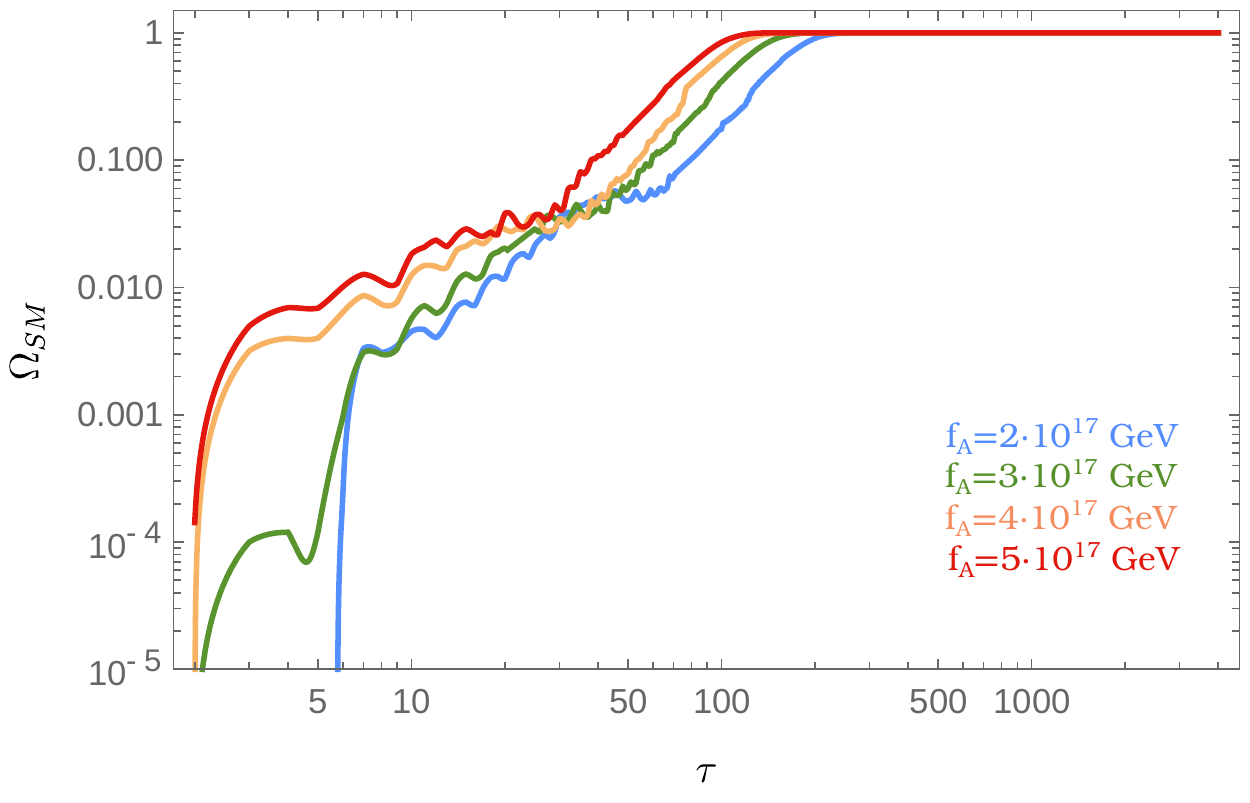}
\caption{{Upper plot: Variance of the angular variable in Model 1 as a function of conformal time $\tau$ for 9 different values of $f_A$, and with $\lambda=1.27\times10^{-11}$, $\lambda_{H\sigma}=10^{-6}$. The simulations were done for  $L=14848$, $N=64$ and a timestep $\Delta\tau=0.0005$. Lower plot: Growth of the relative energy density in the SM bath at early times, $\Omega_{\rm SM}=\rho_{\rm SM}/\rho$, for the same choices of parameters. For $f_A\lesssim10^{16}$ GeV the values of $\Omega_{\rm SM}$ remain well below the range of the vertical axis of the plot.} } 
\label{fig:variance}
\end{figure}

We thus conclude that avoiding the restoration of the PQ symmetry in Model 1 requires $f_A\gtrsim2\times 10^{17}$ GeV. This is one order of magnitude above the value estimated in Ref.~\cite{Ballesteros:2016xej}.

\begin{figure}[h]
\centering     
\includegraphics[width=.57\textwidth]{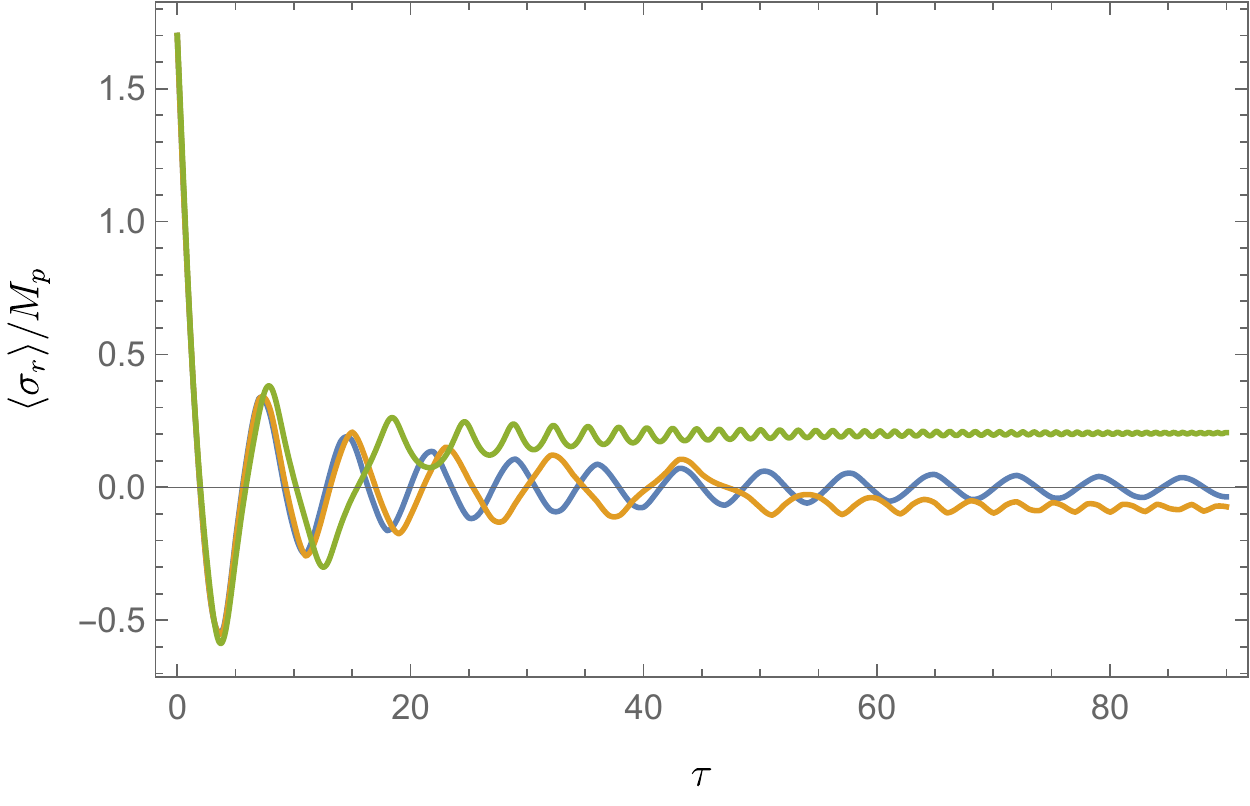}
\caption{{Evolution of the spatial average of $\sigma_r$ as a function of time, for $f_A=10^{15}$ GeV (blue), $f_A=2\times10^{17}$ GeV (orange), and $f_A=5\times10^{17}$ GeV (green), for simulations with the same parameters as in Fig.~\ref{fig:variance}. Note how for the latter two curves the field settles early on around a minimum of the potential (given by $\pm f_a$).} } 
\label{fig:background_fa}
\end{figure}

\paragraph{\bf{Isocurvature power spectrum at superhorizon scales for \boldmath$ f_A=5\times10^{17}$ GeV.}}\mbox{}\vspace{-0.3cm}\\

After having identified $f_A=5\times10^{17}$ GeV as a value for which the restoration of the PQ symmetry is avoided and the fluctuations in the angular variable $\theta$ are minimized,
we proceed to carry out more simulations for this value of $f_A$, and compute the power spectrum of isocurvature fluctuations as a function of time using Eq.~\eqref{eq:Sagammalattice}.

The results show that the isocurvature power spectrum goes over several periods of exponential amplification when the average background crosses the origin --as expected from the tachyonic axion mass-- and then settles into a phase in which it oscillates with an amplitude decaying as $1/a^2 \sim 1/\tau$. We illustrate these findings in Fig.~\ref{fig:Sagammatime} for a simulation with $L=14848, N=64$. On the left plot we show the evolution of the ratio 
{of}
the power spectrum at a given time 
{to} its value at the end of inflation, for 53 frequency bins  with a rainbow hue going from red for the lowest frequencies to blue for the highest frequencies. 
We {also} plot the evolution of $|\langle \varphi_r\rangle|$, rescaled so as to fit in the {figure}. Notice how when $|\langle \varphi_r\rangle|$ crosses zero there is a strong growth in the power spectrum, as expected from the parametric resonance in the Lam\'{e} equation alluded to in the previous section. Once the background {field} stops crossing the origin the power spectrum decays, while experiencing some oscillations. On the right plot we show the same ratio of power spectra, multiplied by the square of the scale factor, {which tends to a constant value}.
As is clear from the plot, the lowest frequencies (in red) experience a larger amplification with respect to the power spectrum at the end of inflation. This fits the naive expectation from the tachyonic mass of the axion near the origin of field space, which plays a bigger role for lower frequencies.

\begin{figure}[h]
\centering   
\includegraphics[width=.49\textwidth]{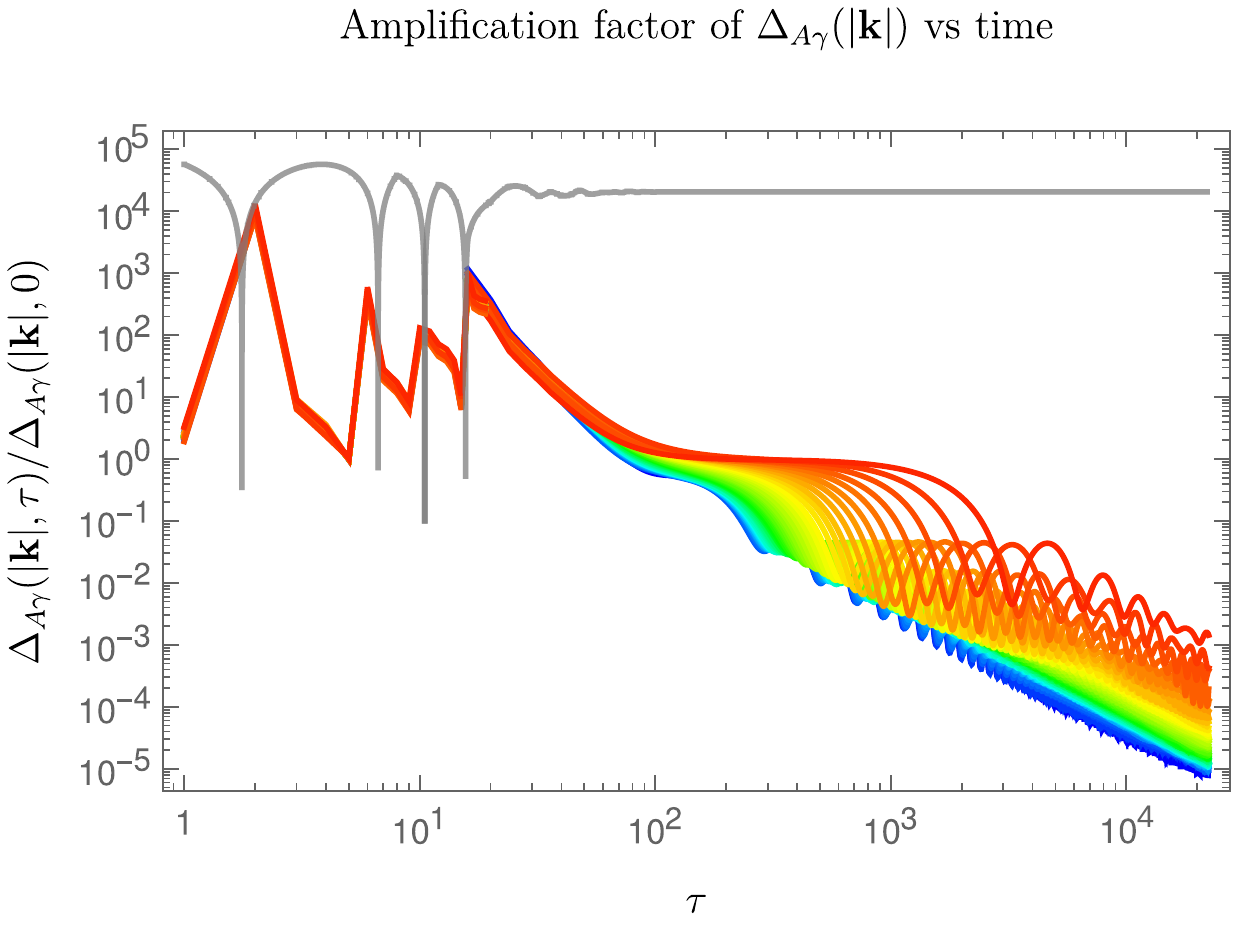}
\includegraphics[width=.49\textwidth]{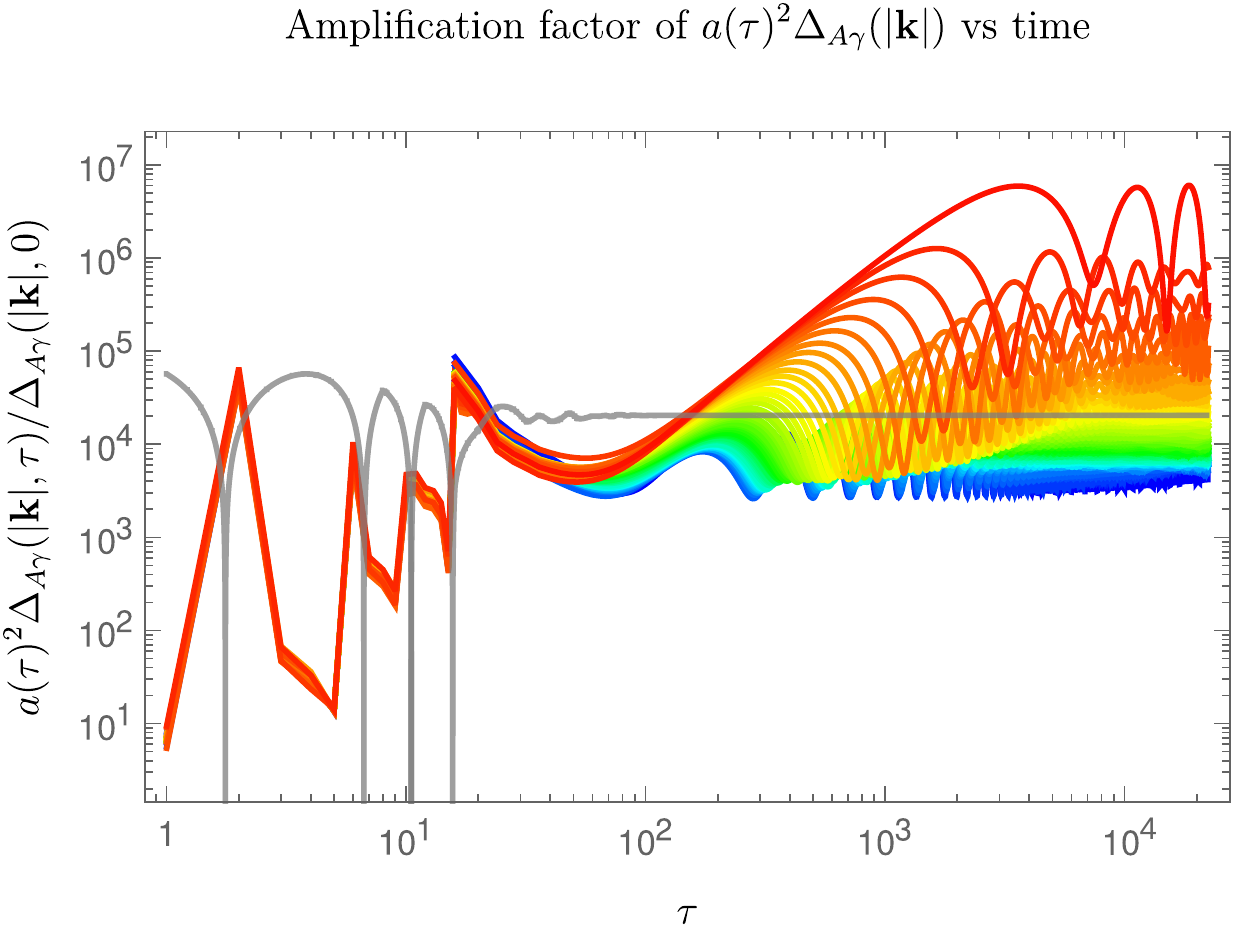}
\caption{Left plot: Ratio of the isocurvature power spectrum with respect to its value at the end of inflation in Model 1, as a function of $\tau$ for a simulation with $f_A=5\times10^{17}$ GeV, $L=14848, N=64$, $\Delta \tau=0.0005$, for 53 frequency bins from the lowest frequency (red) to the highest (blue). The spectra are overlaid over a rescaled plot of $|\langle\varphi_r\rangle|(\tau)$, shown by a gray line. Right plot: Same power spectra, multiplied by $a(\tau)^2$. In all plots, the choices of quartic couplings are as in Fig.~\ref{fig:variance}.} 
\label{fig:Sagammatime}
\end{figure}
\begin{figure}[h]
\centering   
\includegraphics[width=.49\textwidth]{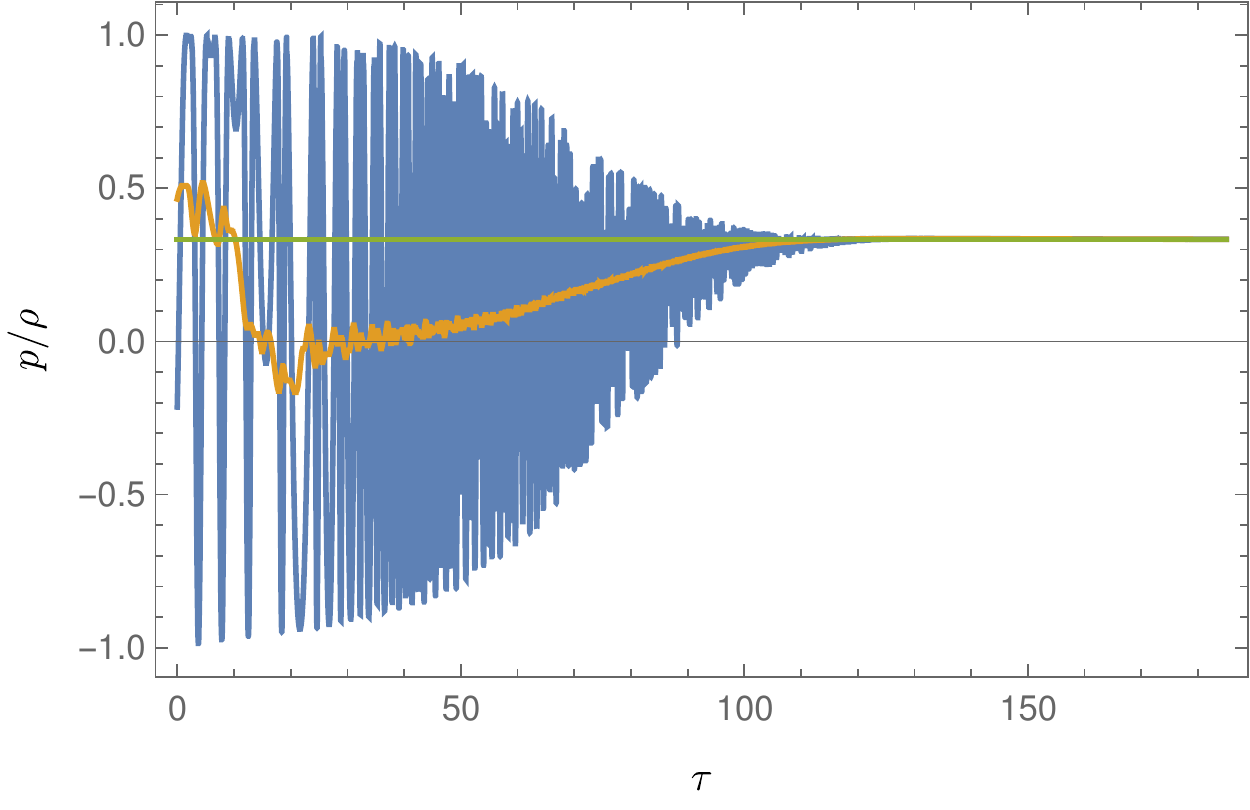}
\caption{{Equation {of state} $\omega=p/\rho$ obtained for the same simulation as in Fig.~\ref{fig:Sagammatime} (blue). The orange line gives a rolling time average over an interval of ten units of $\tau$, while the horizontal green line {corresponds to} {radiation} ($p=\rho/3$). Notice a brief period of matter domination ($p\approx0$) around $\tau\sim20$.}} 
\label{fig:raddom}
\end{figure}
The $1/a(\tau)^2$ behaviour of the isocurvature power spectrum for large $\tau$ can be understood as follows. As seen in Fig.~\ref{fig:background_fa}, at late times the {background} fields get trapped in a potential well with small oscillations whose amplitude decreases with time. Eventually, when the amplitude of oscillations can be neglected and $\sigma_r$ sits at $f_A$, the perturbations in $\sigma_\theta$ have zero mass and their Fourier components $\delta\sigma_{\theta,k}$ satisfy the equation
\begin{equation}\label{eq:eqdeltasigmamassless}
\ddot\delta\sigma_{\theta,k} + 3\frac{\dot a}{a}
\delta\dot\sigma_{\theta,k}+ \frac{k^2}{a^2}\, \delta\sigma_{\theta,k} = 0 
\end{equation}
The simulations confirm that the scale factor behaves as in radiation domination, {with only a very brief period of matter domination for large $f_A$ (see Fig.~\ref{fig:raddom})}. This happens because, even if the background scalar field oscillating around the minimum would behave as a matter component, the Higgs production and decay into the radiation field $\rho_{\rm SM}$ are fast enough such that $\rho_{\rm SM}$ quickly dominates the total energy density. In this case one has $a(t) = b \sqrt{t}$
for some constant $b$, {and} 
$\sqrt{t}\delta\sigma_{\theta,k}(t) = c_1(k) e^{2 i k\sqrt{t}/b} +c_2(k) e^{-2 i k  \sqrt{t}/b}$
with $c_1(k)$, $c_2(k)$ additional constants. {Therefore} the power spectrum {is} $\Delta_{\delta\sigma_\theta}\propto |\delta\sigma_{\theta,k}(t)|^2\propto1/a(t)^2$. From Eq.~\eqref{eq:Sagammaapprox} with $\bar\sigma_r=f_A$, one concludes that $\Delta_{{\cal S}_{A\gamma}}\propto 1/a(t)^2$ as observed in our results. Moreover, given that during radiation domination 
the dimensionless conformal time of Eq.~\eqref{eq:tauBdef} goes as $\tau\propto\sqrt{t}$, 
the power spectrum {oscillates} in $\tau$ with a frequency proportional to $k$, as 
{shown} in Fig.~\ref{fig:Sagammatime}. Though we can understand analytically the $1/a^2$ behaviour of the isocurvature power spectrum ,
we need to resort to the lattice simulations in order to estimate the momentum-dependent normalization amplitude, which is sensitive to the initial nonperturbative amplification.  For each momentum $|\bf k|$, we may look at the amplification factor times $a(t)^2$ --which, as follows from the above arguments, should become  an oscillating function with a constant momentum-dependent amplitude at late times-- and estimate the maximum value reached during the oscillations. In this way one can define an approximate upper bound for subsequent times, 
\begin{align}\label{eq:hatkappamax}
 \Delta_{A\gamma}(|{\bf k}|,\tau)\lesssim  \frac{\Delta_{A\gamma}(|{\bf k}|,0)}{a(\tau)^2}
 \,\hat{\kappa}_{\rm max}(|{\bf k}|),\quad \hat{\kappa}_{\rm max}(|{\bf k}|)\equiv{\rm Max} \left(\frac{\Delta_{A\gamma}(|{\bf k}|,\tau)a(\tau)^2}{\Delta_{A\gamma}(|{\bf k}|,0)}\right),
\end{align}
where again $\tau=0$ corresponds to the end of inflation. The quantity $\hat{\kappa}$ can be estimated directly from the lattice simulations. Alternatively, rather than an upper bound we can carry out a direct estimate by performing time averages of the amplification factor times $a(t)^2$ in the oscillating phase:
\begin{align}\label{eq:hatkappamean}
 \Delta_{A\gamma}(|{\bf k}|,\tau)\approx \frac{\Delta_{A\gamma}(|{\bf k}|,0)}{a(\tau)^2}
 \,\hat{\kappa}_{\rm mean}(|{\bf k}|),\quad \hat{\kappa}_{\rm mean}(|{\bf k}|)\equiv\overline{\left(\frac{\Delta_{A\gamma}(|{\bf k}|,\tau)a(\tau)^2}{\Delta_{A\gamma}(|{\bf k}|,0)}\right)},
\end{align}
where $\overline{x}$ denotes a time average of the quantity $x$ over several oscillations. Note that estimating $\hat\kappa_{\rm mean}$ requires simulating for long enough times such that the power spectrum for the lowest nonzero frequency $k_{\rm min}$ goes over at least one full oscillation. 
This requires very long computing times, and we 
have prioritized longer intervals in $\tau$ over simulations with more lattice points.  We fixed $N=64$ and, in order to cover a wider range of frequencies and avoid systematic effects for frequencies near the infrared or ultraviolet cutoff, we run 11 different simulations with $L=29\times 2^k$, $k=1,2,\dots,11$. From the different simulations we extracted estimates of $\hat\kappa_{\rm max/mean}$ for the whole range of frequencies by proceeding as follows. First, we checked that the simulations of different boxes give compatible results for the mid and low frequencies, while the high frequencies seem to be affected by systematic cutoff effects. This can be seen for example  in  the upper plot in Fig.~\ref{fig:hatkappas}, which shows results for $\hat\kappa_{\rm max}$ as a function of the physical momentum $|{\bf k}|/a_0$ in ${\rm Mpc}^{-1}$ units for the different simulations. The range of frequencies of each simulation is indicated with horizontal arrows between values of $k_{\rm min}$ (vertical dotted, orange lines) and $k_{\rm max}$ (vertical dashed, purple lines). The faint dots give the results for all the simulations superimposed together, and it can be seen that the values of $\hat\kappa_{\rm max}$  for the highest frequencies of each simulation deviate systematically from the results for the same frequencies of a simulation with a smaller box. To avoid these discretization effects, we drop the highest frequencies in each simulation. We also drop the lowest frequency in each simulation, as it has poor statistics and is more susceptible to fluctuations (only 3 modes in the lattice have $|{\bf k}|=k_{\rm min}$). The result of this procedure, supplemented by averaging the results of  $\hat\kappa_{\rm max}$  from different simulations in overlapping windows of $|{\bf k}|$, give the blue points in the upper plot of Fig.~\ref{fig:hatkappas}. 
\begin{figure}[t]
\centering   
\includegraphics[width=.7\textwidth]{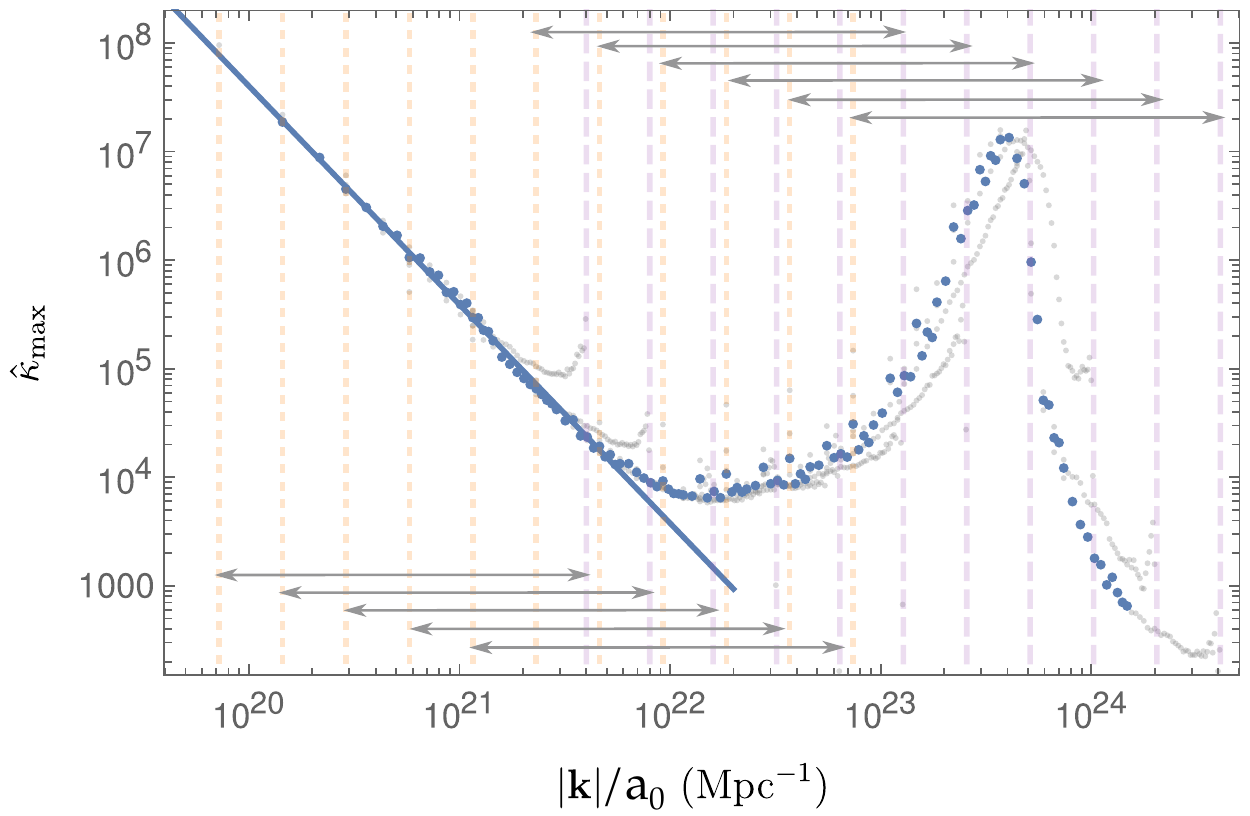}
\vskip.2cm
\includegraphics[width=.7\textwidth]{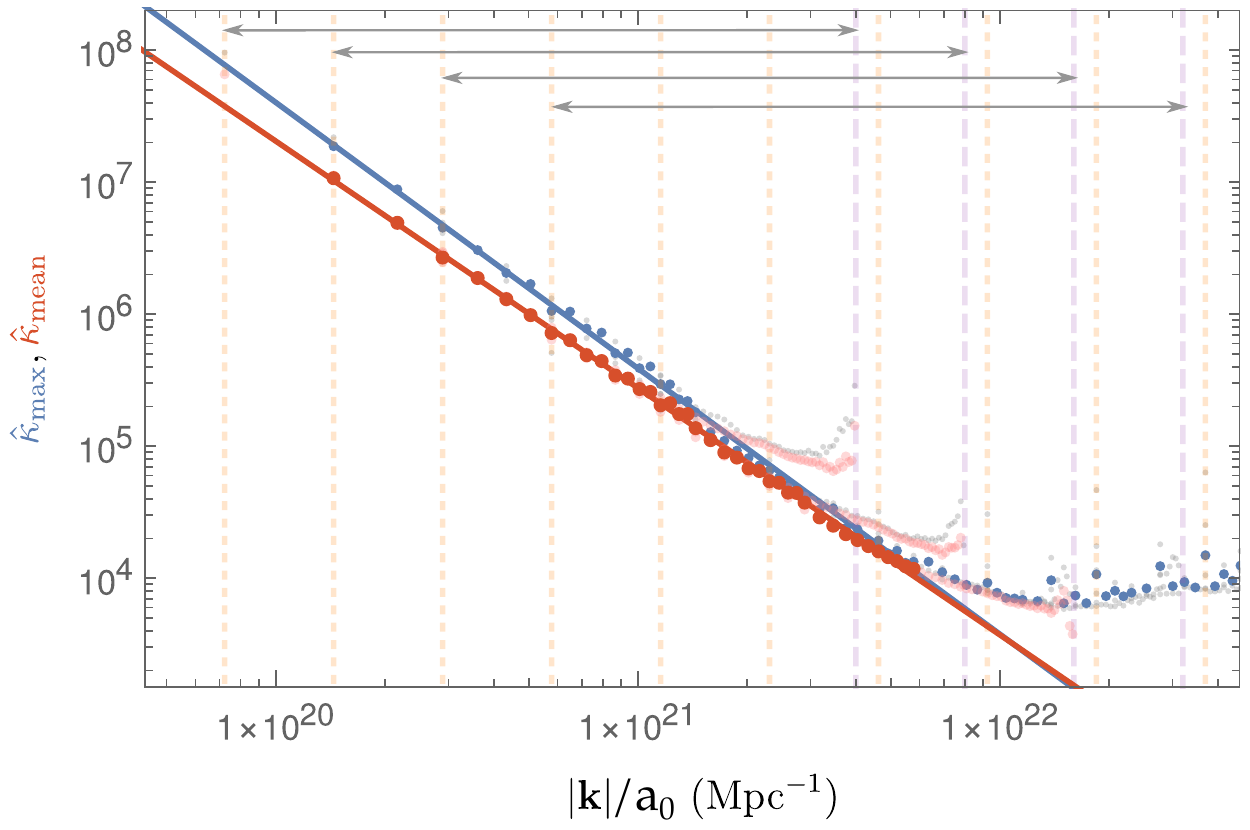}
\caption{Upper plot: Estimate of the amplification factor $\hat\kappa_{\rm max}$ in Model 1 for 11 different lattice simulationswith $N=64$, $\Delta\tau=0.0005$ and $L=29\times 2^{k}$ with $k=1,2,\dots11$, $f_A=5\times10^{17}$ GeV and the rest of the model parameters fixed as mentioned in the main text. The dotted orange vertical lines give the minimum nonzero frequencies covered by each simulation, while the dashed purple lines give the maximum frequencies. The horizontal arrows denote the frequency range of the different simulations. The faint gray dots show the full data of the simulations, while the colored blue dots give the results when dropping the lowest and highest frequencies of each simulation, and averaging the spectra for frequencies covered by more than one simulation. The diagonal blue line is the power-law fit of Eq.~\eqref{eq:kappamaxfit} for low frequencies. Lower plot: Estimates for both $\hat\kappa_{\rm max}$ (gray and blue dots, blue line) and $\hat\kappa_{\rm mean}$ (faint pink and red dots) focusing on the lower frequencies. The pink and red dots correspond to the results before/after dropping high frequencies and averaging over simulations. The red line is the power-law fit of Eq.~\eqref{eq:kappameanfit}.} 
\label{fig:hatkappas}
\end{figure}
One can clearly see a resonant peak for $|{\bf k}|\sim 4\times10^{23}\,{\rm Mpc}^{-1}$, which is of the order of the frequency of oscillation of {the} background at the end of inflation. Actually, this value of the peak frequency corresponds to 
\begin{align}
 |{\bf k}|_{\rm peak}\approx 0.6 B= 0.6 \sqrt{\lambda_{\rm inf}}\,\rho_{\rm inf,end},
\end{align}
which in lattice units corresponds to  $|{\bf \hat{k}}|_{\rm peak}\approx 0.6$. This is of the order of the upper limit of the resonant frequency band $|\hat{\bf k}|\leq 1/2$ expected from  the Lam\'{e} equation obeyed by the $\sigma_\theta$ perturbations in the linear regime obtained when the quadratic interactions of the fields are ignored, as discussed earlier. For lower frequencies the value of $\hat\kappa_{\rm max}$ reaches a simple power law behaviour, illustrated by the solid blue lines in Fig.~\ref{fig:hatkappas}, and given by
\begin{align}\label{eq:kappamaxfit}
 \hat\kappa_{\rm max}(|{\bf k}|)\approx7.57\times10^{47}\left(\frac{{\rm Mpc}^{-1}}{|{\bf k}|}\right)^{2.01}.
\end{align}
A similar procedure can be applied to estimate $\hat\kappa_{\rm mean}$, for which we centered on the three simulations with larger boxes/lower frequencies. The results  are illustrated in the lower plot in Fig.~\ref{fig:hatkappas}, where the estimates of $\hat\kappa_{\rm mean}$ before/after dropping the high frequencies and averaging over simulations are shown by pink and red dots, respectively. Again, for low frequencies one gets a simple power law behaviour, illustrated by the red line in Fig.~\ref{fig:hatkappas} and given by
\begin{align}\label{eq:kappameanfit}
 \hat\kappa_{\rm mean}(|{\bf k}|)\approx7.12\times10^{44}\left(\frac{{\rm Mpc}^{-1}}{|{\bf k}|}\right)^{1.87}.
\end{align}
From the previous estimates of $\hat\kappa_{\rm max/mean}$, taking a leap of faith and extrapolating over 24 orders of magnitude to the CMB pivot scale $|{\bf k}|_{\star}=0.002{\,\rm Mpc}^{-1}$, we can use Eqs.~\eqref{eq:hatkappamax} and \eqref{eq:hatkappamean} to obtain the following upper bound and direct estimate for the amplification factor $\kappa$ of the isocurvature power spectrum at CMB scales (see Eq.~\eqref{eq:kappadef}):
\begin{align}\label{eq:finalamplification}
 \kappa=\frac{\Delta_{{\cal S}_{A\gamma}}(\tau_{\rm CMB},k_\star)}{\Delta_{{\cal S}_{A\gamma}}(0,k_\star)}\lesssim {\cal O}(10^4), \quad  \kappa\approx {\cal O}(10).
\end{align}
In the equations above, we used the fact that the CMB is generated at redshift $z_{
\rm CMB}=a_0/a(\tau_{\rm CMB})-1\approx1100$, with $a_0$ in our units given by Eq.~\eqref{eq:a0end} with $a_{\rm end}=1$. The above estimates of $\kappa$ give a net amplification of the isocurvature power spectrum between the end of inflation and the CMB time, and are incompatible with the suppression needed to {satisfy}
the CMB isocurvature bound, given by Eq.~\eqref{eq:kappamax}. We see that the latter equation is violated by 6-7 orders of magnitude. Even when the large extrapolation to CMB scales can be questioned, the fact that we find a very sizable violation of the isocurvature bound {suggests a potentially significant overabundance of}
isocurvature fluctuations. As we saw before, pre-inflationary axion dark matter requires $f_A\gtrsim10^{17}$~GeV, yet the viability of these scenarios 
{is threatened by this apparent}
overproduction of isocurvature fluctuations. This {conclusion diminishes the likeliness of the validity of the naive isocurvature expectation derived from the linear analysis of inflationary perturbations} 
of Model 1 {in the context of a} pre-inflationary dark matter {scenario}.

\subsection{Model 2}

For Model 2 we use the initial conditions associated with the power spectra computed in Section~\ref{subsec:model2:inflation},  and follow an analogous implementation in {\tt{CLUSTEREASY}} as the one detailed in Section~\ref{subsec:lattice_simulations} for Model 1, adding two additional equations for the new real scalar fields contained in the complex field $\phi$. For the parameter choices we fix the quartic couplings as $\lambda_\phi=4.01\times 10^{-9},\lambda_\sigma=10^{-5},\lambda_{\phi\sigma}=0$,  $-\hat\lambda_{\phi\sigma}=\lambda_{H\phi}=\lambda_{H\sigma }=10^{-7}$. The above values give the same effective inflationary quartic coupling as was chosen for Model 1. We further choose $m^2_\phi=m^2_{\phi_r,\rm vac}-2\hat\lambda_{\phi\sigma}f_A^2$ where $m^2_{\phi_r,\rm vac}=(6.31\times10^{7}\,{\rm GeV})^2$ corresponds, up to subleading effects that depend on the Higgs VEV, to the square of the mass of the $\phi_r$ excitation at the vacuum $\sigma_r=f_A$, $\phi_r=\phi_\theta=\sigma_\theta=0$.

As was done for Model 1, we start by investigating the potential restoration of the PQ symmetry. Surprisingly, despite the suppression of the power spectra at the end of inflation, we do find a restoration of the PQ symmetry for $f_A\lesssim 10^{16}$ GeV, as shown in the upper plot of Fig.~\ref{fig:variance_2scalars}. As shown further in the lower plot, in this case we do not find a large production of SM radiation.

\begin{figure}[h]
\centering     
\includegraphics[width=.57\textwidth]{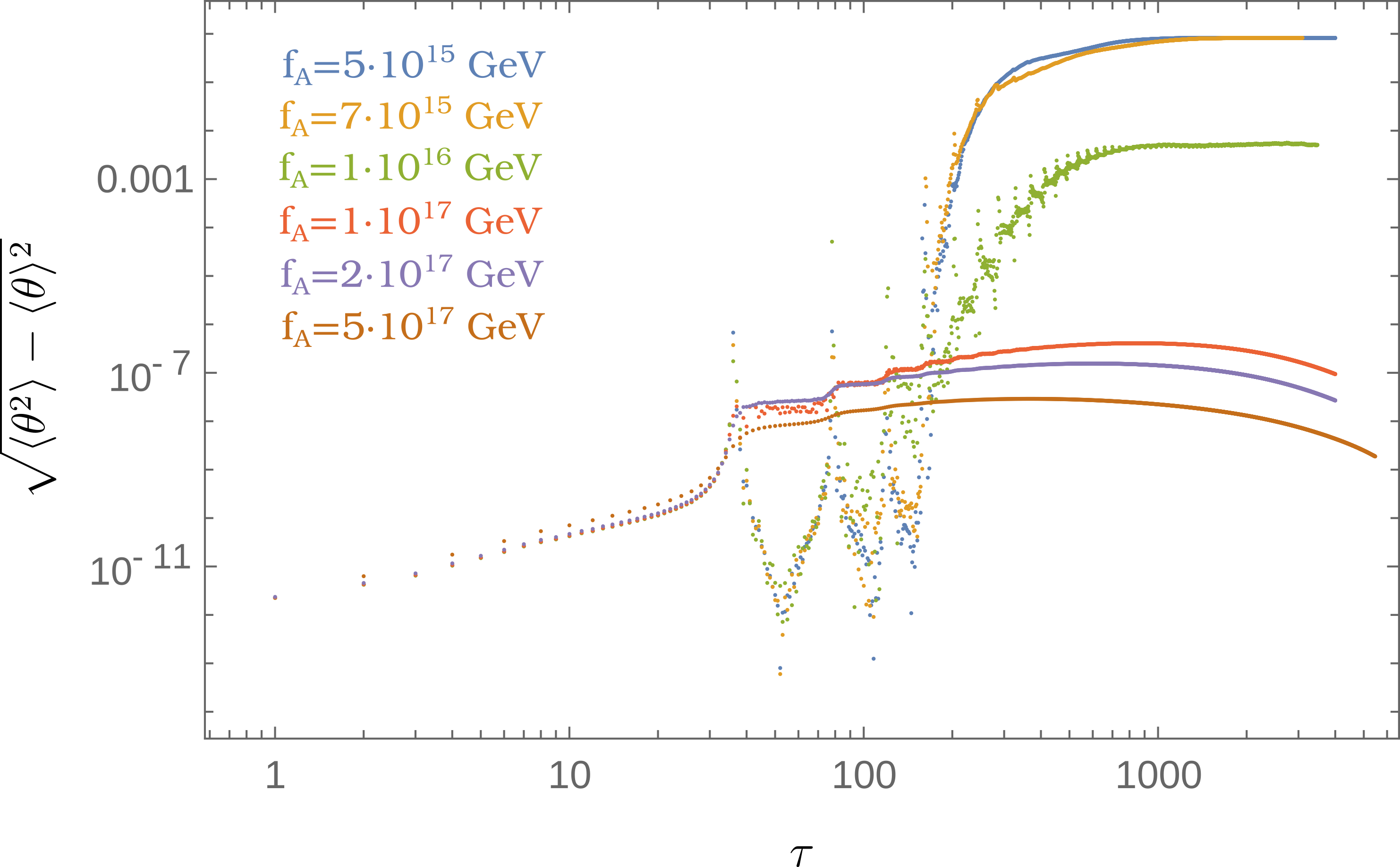}\\
\vskip.3cm
\,\hskip.2cm\includegraphics[width=.57\textwidth]{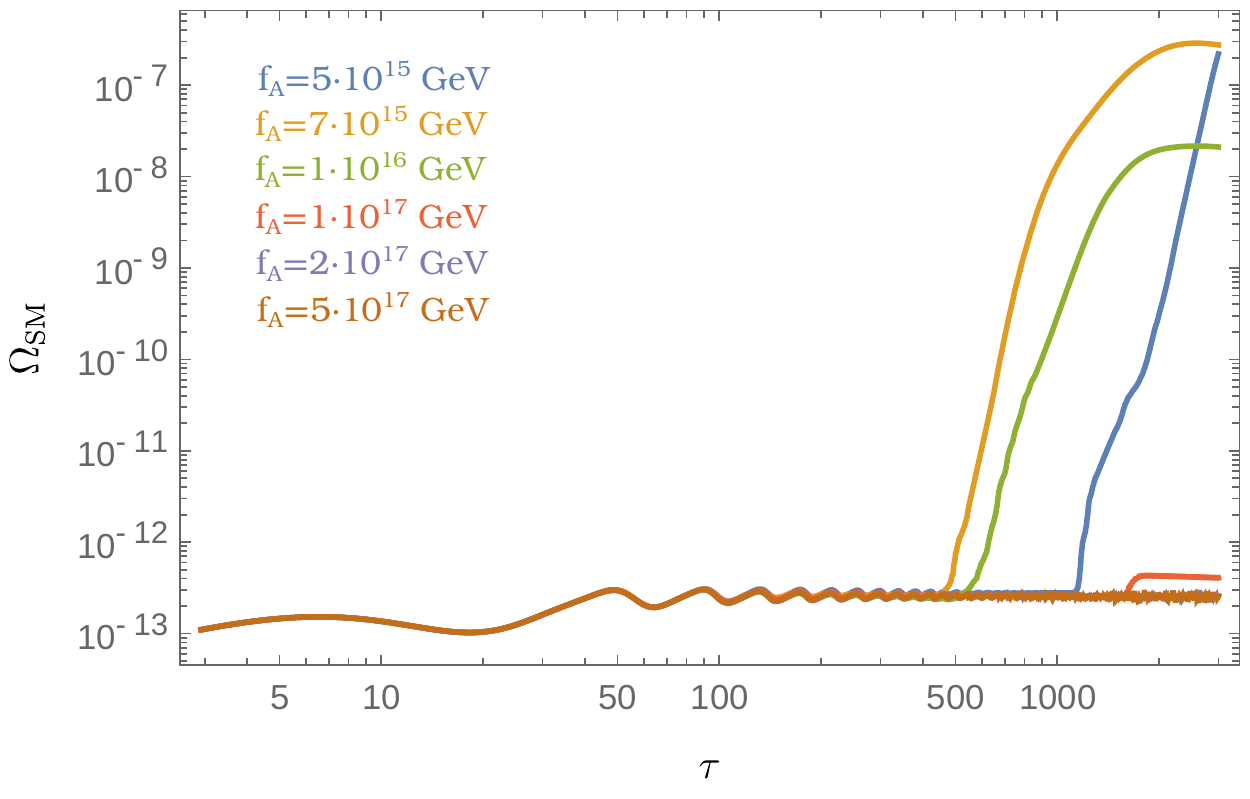}
\caption{ Upper plot: Variance of the angular variable in Model 2 as a function of conformal time $\tau$ for 6 different values of $f_A$, and with the parameters fixed as mentioned in the main text. The simulations were done for  $L=14848$, $N=64$ and a timestep $\Delta\tau=0.0005$. Lower plot: Growth of the relative energy density in the SM bath at early times, $\Omega_{\rm SM}=\rho_{\rm SM}/\rho$, for the same choices of parameters.} 
\label{fig:variance_2scalars}
\end{figure}

\begin{figure}[h]
\centering     
\includegraphics[width=.57\textwidth]{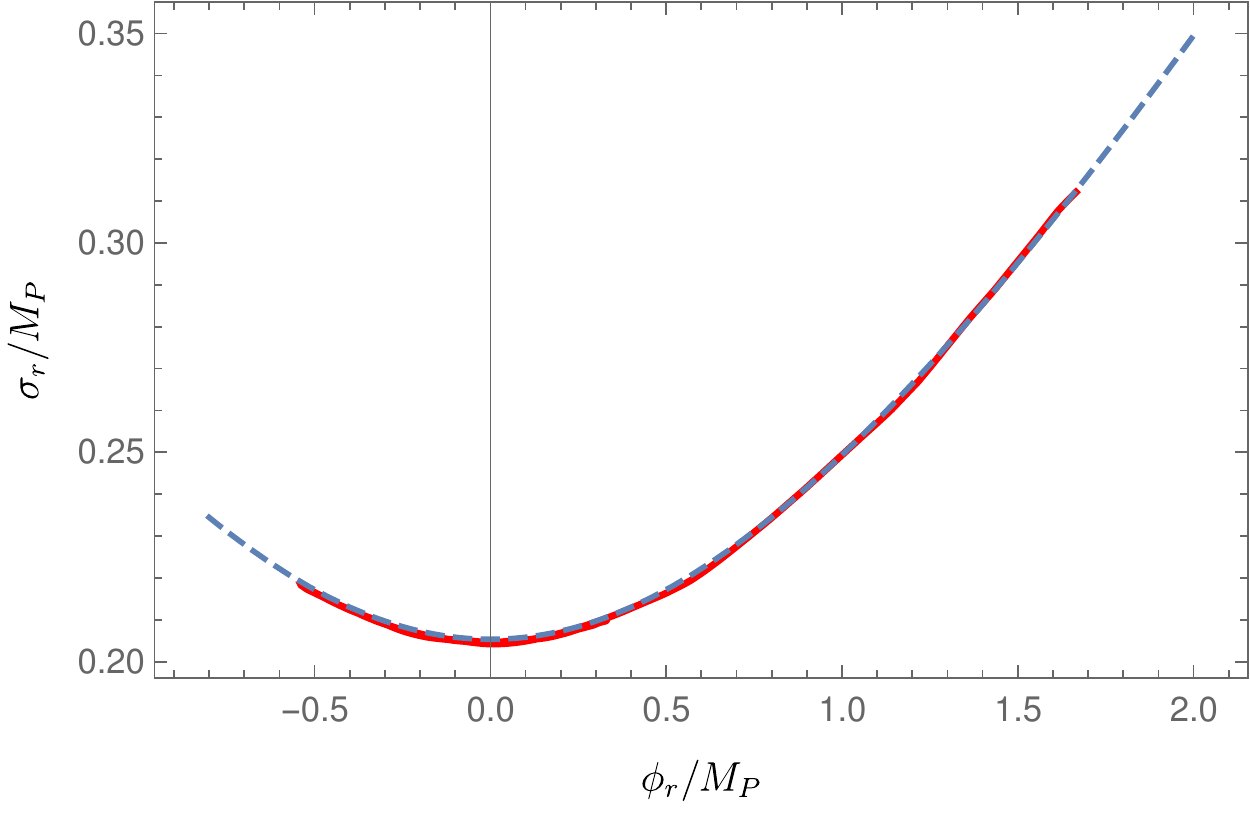}
\caption{Trajectory of the lattice averages of the fields $\phi_r$,$\sigma_r$ in Model 2  with $f_A=5\times10^{17}$ GeV during the simulation (red), versus the valley trajectory of Eq.~\eqref{eq:valley2} (dashed blue). The simulation parameters are as in Fig.~\ref{fig:variance_2scalars}, and the couplings were fixed as mentioned in the main text. } 
\label{fig:2fieldvalley}
\end{figure}

For large values of $f_A$, in which the PQ restoration is avoided, we find as in the previous section an initial exponential amplification of the isocurvature power spectrum, followed by a decay going as $1/a(t)^2$. This is illustrated for $f_A=5\times10^{17}$ GeV in Fig.~\ref{fig:Sagammatime_model2}. As is clear when comparing with Fig.~\ref{fig:Sagammatime}, the amplification is in fact much larger than in Model 1. As will be seen below, such large amplification can overcome the suppression of the power spectrum during inflation and again puts into question the compatibility of the model with current isocurvature bounds. 

As in the previous section, the results can be understood qualitatively from the early-time parametric resonance effects in a background oscillating in a quartic potential, as well as the late time oscillations around a quadratic potential well. The enhanced parametric amplification {in Model 2} with respect to Model 1 can be attributed to the fact that the average background is approximately confined to a potential energy valley which, in contrast to Model 1, has much smaller quadratic terms (which are positive, and not set by $f_A$), so that one naturally expects more oscillations in the quartic regime in which perturbations are generated efficiently. Assuming $\lambda_{\phi\sigma}=0$, $\sigma_\theta,\phi_\theta\approx0$ and keeping the terms involving $f_A$ that were ignored in Eq.~\eqref{eq:valley}, the potential energy valley is approximately given by
\begin{align}\label{eq:valley2}
 \sigma_r^2\approx\, f_A^2-2\frac{\hat\lambda_{\phi\sigma}}{\lambda_\sigma}\phi_r^2,
\end{align}
and the potential along the valley, as a function of $\phi_r$, is
\begin{align}V_{\rm valley}\approx\frac{1}{2}m^2_{\phi_r,\rm vac} \phi _r^2 + \frac{1}{4} \left(\lambda_\phi-4 \frac{\hat\lambda_{\phi\sigma
}^2}{\lambda_\sigma}\right)\,\phi_r^4.
\end{align}
As advertised before, the previous potential has a positive quadratic term fixed by the effective mass $m^2_{\phi_r,\rm vac}$, rather than $f_A$.  
In Fig.~\ref{fig:2fieldvalley} we show how  the trajectories of the lattice averages of the fields remain indeed close to the above valley. The value of $\langle\phi_r\rangle$ as a function of time can also be seen in the gray curves of Fig.~\ref{fig:Sagammatime_model2}. One can see that $\phi_r$ remains oscillating with a large amplitude (and thus in the quartic regime) for longer time than the field $\sigma_r$ did in Model 1 (see Figs.~\ref{fig:background_fa} and \ref{fig:Sagammatime}). The large oscillations of the background and the enhanced fluctuations are expected to lead to a larger effective Higgs mass and block the production of SM radiation, which is confirmed by the lower plot in Fig.~\ref{fig:variance_2scalars}. Another reason for the larger amplification than observed in Model 1 is that we empirically find that the final amplitude is not very sensitive to the initial amplitude of perturbations. By rerunning lattice codes with an $\mathcal{O}(10)$ rescaled initial amplitude we see no relevant effect on the final spectra. Therefore, by starting with a more suppressed spectrum, the enhancement factor will be expected to be larger on these grounds as well.

\begin{figure}[h]
\centering   
\includegraphics[width=.49\textwidth]{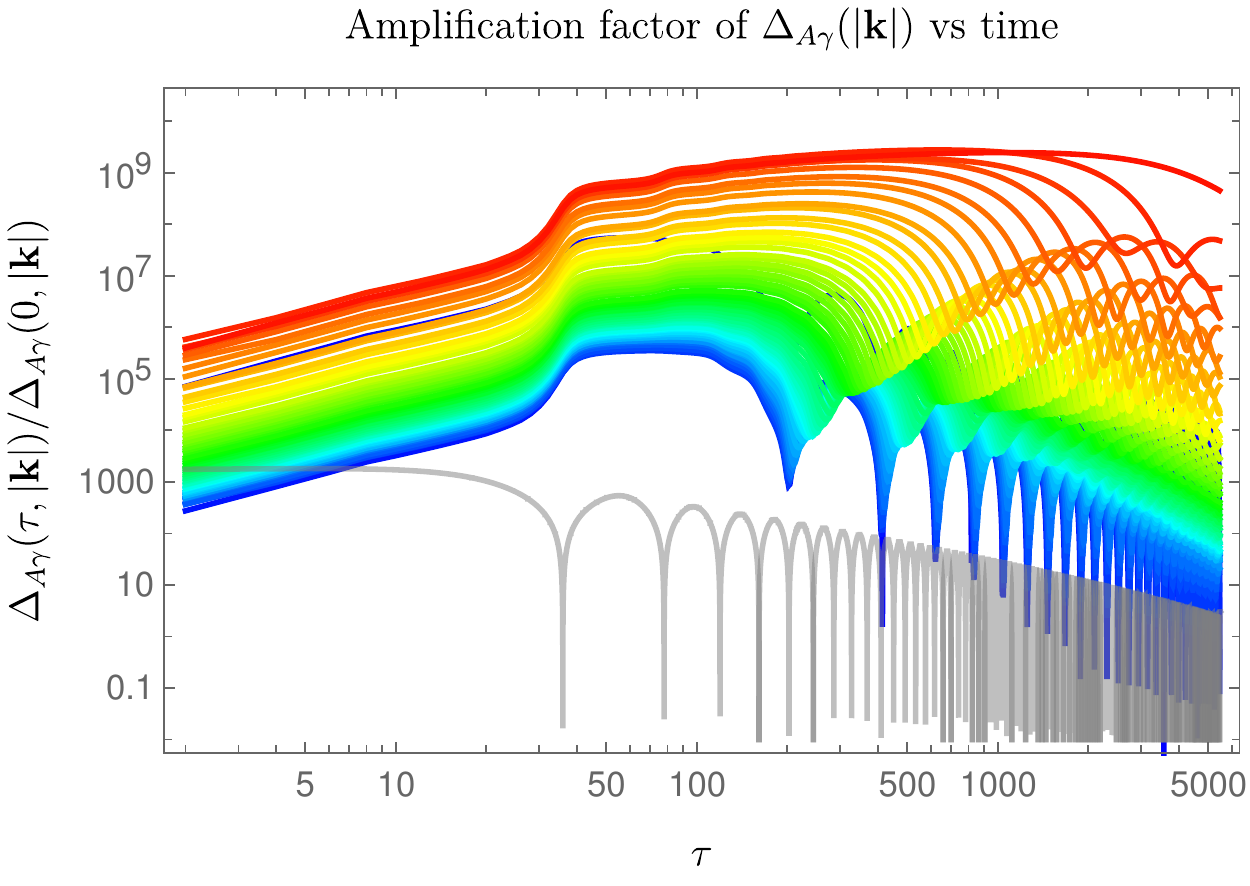}
\includegraphics[width=.49\textwidth]{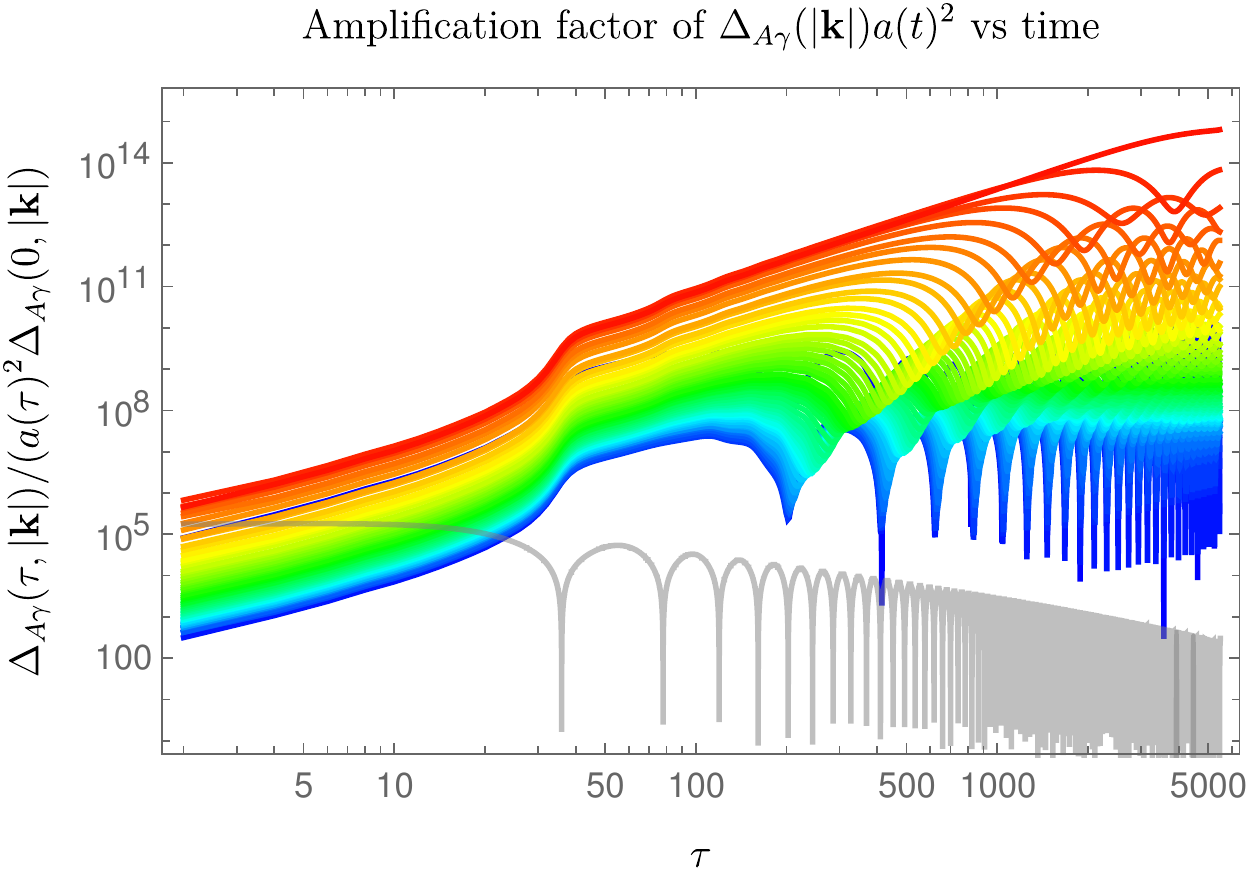}
\caption{Left plot: Ratio of the isocurvature power spectrum with respect to its value at the end of inflation in Model 2, as a function of $\tau$ for a simulation with $f_A=5\times10^{17}$ GeV, $L=14848, N=64$, for 55 frequency bins from the lowest frequency (red) to the highest (blue). The spectra are overlaid over a rescaled plot of $|\langle\phi_r\rangle|(\tau)$, shown by a gray line. Right plot: Same power spectra, multiplied by $a(\tau)^2$. In both plots, the rest of the model parameters where fixed as specified in the main text.} 
\label{fig:Sagammatime_model2}
\end{figure}

Finally, we can attempt a quantitative estimate of the power spectrum at the CMB pivot scale using fits of the amplification factors $\hat \kappa_{\rm max},\hat \kappa_{\rm mean}$ of Eqs.~\eqref{eq:hatkappamax} and \eqref{eq:hatkappamean}. Proceeding as in the previous section leads to the results of Fig.~\ref{fig:hatkappas2scalars}. For late times, the self-consistency checks of { \tt{LATTICEEASY}} indicate a loss of numerical precision, which are not surprising given the large amplification factors at late times in the right plot of Fig.~\ref{fig:Sagammatime_model2}. For this reason, when performing numerical fits of the amplification factors we drop the lowest frequencies, which only achieve the first peak and subsequent oscillating regime at late times. Doing so we find
\begin{align}\label{eq:kappafits2scalars}
 \hat\kappa_{\rm max}(|{\bf k}|)\approx1.93\times10^{121}\left(\frac{{\rm Mpc}^{-1}}{|{\bf k}|}\right)^{5.25},\quad
 \hat\kappa_{\rm mean}(|{\bf k}|)\approx9.53\times10^{117}\left(\frac{{\rm Mpc}^{-1}}{|{\bf k}|}\right)^{5.1}.
\end{align}
To extrapolate to CMB frequencies using Eqs.~\eqref{eq:hatkappamax} and \eqref{eq:hatkappamean} we need an estimate of $\Delta_{A\gamma}(\tau,k_\star)$. In Section~\ref{subsec:model2:inflation} we saw that the power spectra for the massive canonical fields is accurately captured by using the Minkowski modes of Eq.~\eqref{eq:bc}. Using this in Eqs.~\eqref{eq:Sagammaapprox} and \eqref{eq:power_spectra_from_modes} gives
\begin{align}
 \Delta_{A\gamma}(0,k_\star)\approx \frac{k_\star^3}{\theta_i^2 \pi^2\bar\sigma_r(0)^2 m_{A}(0)}.
\end{align}
Upon substitution into Eqs.~\eqref{eq:hatkappamax} and \eqref{eq:hatkappamean}, using the fits of Eq.~\eqref{eq:kappafits2scalars} and the value of $\theta_i$ corresponding to Eq.~\eqref{eq:Omegah2preinf} with $f_A=5\times10^{17}$ GeV, one gets values of $\Delta_{A\gamma}(\tau_{\rm CMB},k_\star)$ which are more than a hundred orders of magnitude above the maximum value of $\Delta_{A\gamma}(0,k_\star)$ that would be compatible with the CMB bound.}

\begin{figure}[h]
\centering   
\includegraphics[width=.7\textwidth]{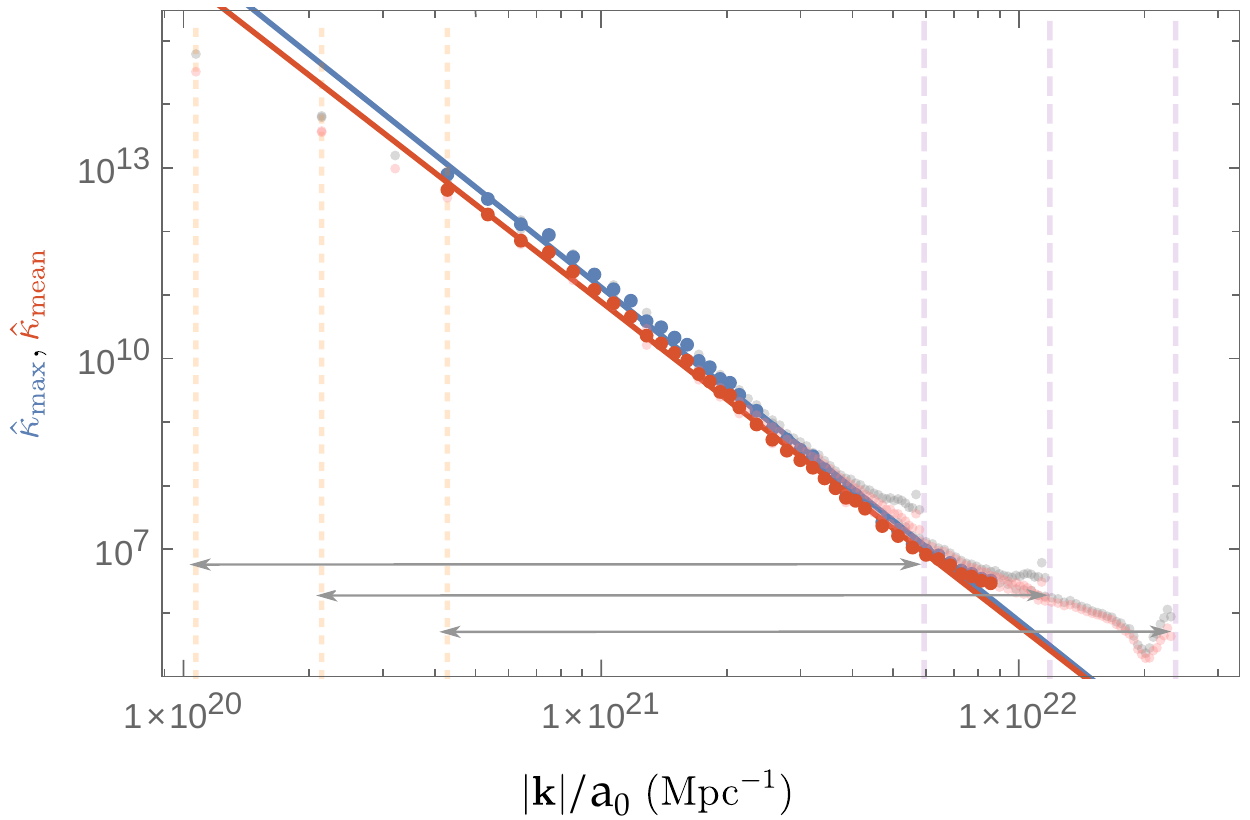}
\caption{Estimate of the amplification factor $\hat\kappa_{\rm max}$  in Model 2 for $f_A=5\times10^{17}$ GeV and 3 different lattice simulations with $N=64$, $\Delta\tau=0.0005$ and $L=29\times 2^{k}$ with $k=7,8,9$. The colour coding, vertical lines and arrows follow the conventions of  Fig.~\eqref{fig:hatkappas}. See the main text for the choice of model parameters.} 
\label{fig:hatkappas2scalars}
\end{figure}

\section{\label{sec:conclusions}Discussion}

In this paper we have shown that the common expectation that axion perturbations freeze at horizon crossing during inflation is violated in pre-inflationary axion models featuring axion-inflaton interactions, which {have been motivated in previous works \cite{Fairbairn:2014zta, Ballesteros:2016euj, Nakayama:2015pba} as a way to} suppress isocurvature fluctuations. This expectation, however, can be far from the truth, as the axion-inflaton interactions can give rise to a large nonperturbative growth of axion perturbations during reheating. The latter can 
lead to a complete restoration of the PQ symmetry {(invalidating the premise of pre-inflationary scenario and, leading to an overabundance of dark matter) or give rise to a large isocurvature power spectrum at CMB scales, in conflict with current bounds from Planck.}

To illustrate these effects we have considered two models. In the first one, the axion {arises from a complex scalar which also contains the inflaton}.  
This {implies} a large effective axion decay constant during inflation, which {has been argued to suppress (linear)} 
axion perturbations during inflation \cite{Fairbairn:2014zta, Ballesteros:2016euj}. The second model involves 2 scalar fields in which axion-inflaton interactions make the axion field very massive during inflation, suppressing its fluctuations \cite{Nakayama:2015pba}. For both models we have followed the evolution of axion perturbations during inflation and reheating. We solved the differential equations in the linear regime during inflation (fully accounting for the fact that the axion is also massive in the first model) while for reheating we performed lattice simulations accounting for couplings to the Higgs and {its} decays into relativistic SM degrees of freedom. Our results show that the PQ symmetry is restored for values of $f_A$ below {$10^{17}$ GeV}, while for larger values {of $f_A$} the growth of {low momentum} axion perturbations 
{is substantial, and a naive extrapolation} to CMB scales {suggests}
a violation of current isocurvature bounds by several orders of magnitude. Though we cannot claim that such extrapolation is under control {due to huge range of scales involved,} the results {sow doubts about the}
viability of these models {as pre-inflationary axion dark matter scenarios.} {As a consequence, axion dark matter in both of the models that we have studied can only be compatible 
with the post-inflationary scenario, which in turn implies that $f_A$ is bounded from above by $2 \times 10^{11}$~GeV (see Eq.\ \eq{abundance}) and that, correspondingly, $m_A$ is bounded from below by  $~30$~$\mu$eV (see Eq.\ \eq{eq:ma})}. With SMASH being a variant of Model 1, our findings support
the focus on post-inflationary scenarios in Refs.~\cite{Ballesteros:2016euj,Ballesteros:2016xej}, with the present work providing an improved justification on the following grounds. First, the above references 
assumed the freezing of perturbations  after horizon crossing during inflation. Second, we have now obtained a direct estimate of the value of $f_A$ beyond which the PQ symmetry stops being restored by nonperturbative effects during reheating. The latter was estimated in Ref.~\cite{Ballesteros:2016xej} to lie around $f_A\approx 10^{16}$ GeV, one order of magnitude below the value found in this paper.

{In our lattice simulations we have not included perturbations of the spacetime metric. This is usually the case in studies of preheating, where subhorizon scales are commonly considered and for which metric fluctuations of wavenumber $k$ are expected to be suppressed by $\sim(k/H)^2$ (or higher powers) with respect to density ones. However, for superhorizon perturbations such as the ones we have considered in this paper metric fluctuations should, in principle, be included. Without a full analysis which takes them into account we do not know to which extent they can affect our results. For an incomplete list of references on the topic see \cite{Bassett:1998wg,Bassett:1999mt,Bassett:1999cg,Bassett:1999ta,Zibin:2000uw,Finelli:2000ya,Huang:2011gf}.}

\subsection*{Acknowledgments}

The authors thank L.~Di~Luzio and J.~Redondo for discussions. The work of GB is funded by a Contrato de Atracci\'on de Talento (Modalidad 1) de la Comunidad de Madrid (Spain), number 2017-T1/TIC-5520, by MCIU (Spain) through contract PGC2018-096646-A-I00 and the IFT Centro de Excelencia Severo Ochoa Grant SEV-2016-0597. AR acknowledges support by the Deutsche Forschungsgemeinschaft (DFG, German Research Foundation) under Germany's Excellence Strategy – EXC 2121 ``Quantum Universe". 
CT acknowledges financial support by SFB 1258 and the ORIGINS cluster of DFG.  YW is supported by the ERC Consolidator Grant STRINGFLATION under the HORIZON 2020 grant agreement no. 647995.

\appendix

\section{Adiabatic and isocurvature fluctuations} \label{app:iso}

The simplest way to understand  isocurvature fluctuations is by defining first the concept of {\it adiabatic fluctuations}. Let us consider $n$ cosmological species living in an approximately homogeneous and isotropic universe, each of which has an equation of state $w_j$ ($j=1,\ldots,n$) which defines the relation between the background (i.e.\ homogeneous) energy density $\bar \rho_j(t)$ and pressure $\bar p_j(t)$ for each species: $\bar p_j = w_j \bar \rho_i$. The energy density (and the pressure) of each species feature small, space dependent, fluctuations around their background values: $\rho_j(t,{\bf x})=\bar\rho_j(t)+\delta\rho_j(t,{\bf x})$. It is customary to write this as $\rho_j(t,{\bf x})=\bar\rho_j(t)(1+\delta_j(t,{\bf x}))$, defining $\delta_j(t,{\bf x})$ as the ratio between the density perturbation and its background value. Let us now assume that, for each species, we can write the energy density as $\rho_j(t,{\bf x})=\bar\rho_j(t+\delta t(t,{\bf x}))$, where $\delta t(t,{\bf x})$ can be interpreted as a coordinate dependent shift in time that is, crucially, the same for all species. An analogous expression can be assumed for the pressures, with the same $\delta t(t,{\bf x})$. Obviously, this is a very specific restriction on the properties of the energy densities and pressure perturbations.  Expanding at linear order in $\delta t(t,{\bf x})$, we can write $\delta\rho_j(t,{\bf x}) = \dot{\bar\rho}_j(t)\,\delta t(t,{\bf x}) = -3(1+w_j)H(t){\bar\rho}_j(t)\,\delta t(t,{\bf x})$, where we use a dot to indicate a time derivative and, in the second equality we have assumed that {the energy densities of each species evolve independently} at the background level. From the last expression we obtain
\begin{align}  \label{adiabic}
\frac{\delta_1(t,{\bf x})}{1+w_1} = \ldots = \frac{\delta_n(t,{\bf x})}{1+w_n}\,,
\end{align}
so that the relative energy density fluctuations of the species are related among them through weight factors that depend on their equations of state. An excellent fit to the CMB is obtained assuming the relation \eq{adiabic} holds among all cosmological species (photons, cold dark matter, baryons, neutrinos) as an initial condition deep in the radiation era, i.e.\ for the superhorizon Fourier modes that satisfy $k\tau \ll 1$, where  $\tau$ is conformal time and $k$ is a comoving wavelength. If the condition \eq{adiabic} holds for all species but one, say cold dark matter, then we talk about isocurvature initial conditions  for that particular species. The initial conditions in the radiation era are set by inflation and the subsequent reheating process. The CMB sets stringent bounds on the maximum amount of isocurvature initial conditions --i.e.\ on the deviation from eq.\ \eq{adiabic}-- for all cosmological species \cite{Akrami:2018odb}. These can therefore be used to constrain the physics of inflation and reheating. Scalar fields that are light during inflation (meaning that their mass is much smaller than $H$) can generate large isocurvature fluctuations that get imprinted in the initial conditions for radiation after reheating. This is a powerful way of setting bounds on axion models in which the PQ symmetry breaks before or during inflation. 

The observant reader may be wondering whether the above definition of adiabatic initial conditions (and, by extension, of isocurvature) depends on the choice of coordinates $(t, {\bf x})$, and more specifically on the gauge choice for the metric. It can be checked that at linear order in fluctuations this is not a concern, due to the way in which $\delta_j$ transforms from one gauge to another. Under a general coordinate transformation of the form $\tilde t = t +T(t,{\bf x})$ and $\tilde {\bf x} = {\bf x} + L(t,{\bf x})$ the energy density of any species, being a scalar (i.e.\ bearing no spatial indexes), transforms at linear order as $\tilde\delta\rho_j=\delta\rho_j - \dot{\bar\rho}_j(t)T(t,{\bf x})$. Dividing this expression by $\bar\rho_j(t)$ and using the (energy {density evolution}) 
equation $\dot{\bar\rho}_j(t)+3(1+w_j)H(t)\bar\rho_j(t)=0$, we obtain $\tilde\delta_j=\delta_j + 3(1+w_j)H(t)\, T(t,{\bf x})$, so that if the condition \eq{adiabic} holds for the variables $\delta_j$, it is also true for the variables $\tilde\delta_j$. 
Alternatively, we can work directly with the (gauge invariant) quantities that in any gauge read $\zeta_j=-\psi- H\,\delta\rho_j/\dot{\bar\rho}_j$, where $\psi(t,x)$ is the spatial part of the (FLRW) metric fluctuation $ds^2 \ni a^2 (1-2\psi)dx^2$ and $a(t)$ denotes the (time-dependent) scale factor of the Universe. If the condition \eq{adiabic} is satisfied for the variables $\delta_j$, then $\zeta_1=\ldots=\zeta_n$. Notice that in reality the condition that each species satisfies $\dot{\bar\rho}_j(t)+3(1+w_j)H(t)\bar\rho_j(t)=0$ is accessory. We can completely forgo this assumption and define adiabaticity directly from $\delta\rho_j(t,{\bf x}) = \dot{\bar\rho}_j(t)\,\delta t(t,{\bf x})$, which leads to a more general condition than \eq{adiabic}:
\begin{align}  \label{adiabic2}
\frac{\delta {\rho_1}(t,{\bf x})}{ \dot{\bar\rho}_1}= \ldots = \frac{\delta {\rho_n}(t,{\bf x})}{{ \dot{\bar\rho}_n}}\,.
\end{align}
However, the condition \eq{adiabic} is commonly used to define adiabatic initial conditions, because photons, baryons, neutrinos and cold dark matter are assumed to have no interactions (other than gravity) during radiation domination and therefore they are thought to {evolve independently}. 

Another objection that can be made is whether there actually exists a solution of the equations of motion for the {metric} and matter fluctuations satisfying $\delta\rho_j(t,{\bf x}) = \dot{\bar\rho}_j(t)\,\delta t(t,{\bf x})$. This was solved by S.\ Weinberg, who proved that in the limit of large wavelength (that is, for superhorizon modes) there are two different such solutions, finding the coordinate dependence of $\delta t(t,{\bf x})$ for both of them \cite{Weinberg:2003sw}. One of these solutions decays with the expansion of the Universe as $\delta t(t,{\bf x})\propto 1/a(t)$ and the other is $\delta t(t,{\bf x})=-(\zeta/a(t))\int_{\tilde t}^t a(t')dt'$, where  $\tilde t$ is an arbitrary reference time. $\zeta$ {is a constant that corresponds to the long wavelength limit of the} curvature perturbation on uniform total density hypersurfaces \cite{Bardeen:1983qw}, defined as $\zeta=-\psi- H\,\delta\rho/\dot{\bar\rho}$, where $\delta\rho$ and $\dot{\bar\rho}$ refer to the total energy-momentum tensor, in analogy to the quantities $\zeta_j$ introduced above. The name {\it adiabatic} for these modes {is not related to the absence of changes in the entropy of the cosmological fluid, but is rather associated to the fact that perturbations can be captured by time-shifts:}  any scalar quantity $s(t, {\bf x})$ in such a mode satisfies $\delta s/ \dot{\bar{s}} = \delta t(t,{\bf x})$, where $\delta t(t,{\bf x})$ is of any of the two forms above.  In particular, for any energy-momentum tensor the fluctuations in the energy density and pressure are related by $\delta \rho/ \dot{\bar{\rho}} = \delta p/ \dot{\bar{p}}$, hence the name. 

If the only mode produced during inflation is the adiabatic one related to $\dot\zeta =0$ (with $\zeta \neq 0$), the fluctuations continue in this mode also after inflation, provided that they remain superhorizon. This applies in particular during reheating. 

The amount of isocurvature is conventionally defined with respect to photons through the relation \cite{Kodama:1985bj}
\begin{align}\label{eq:calSj}
\mathcal{S}_{j} = 3(\zeta_j -\zeta_\gamma)\,,
\end{align}
where the factor 3 is included so that {{in the gauge $\psi=0$} $\mathcal{S}_{\rm baryons}$ equals the relative fluctuation in the baryon to photon number {density} ratio, {$\delta(n_B/n_\gamma)/(n_B/n_\gamma)$}, see e.g.\ \cite{Bassett:2005xm}.

\section{Axion isocurvature fluctuations} \label{app:isoax}

Next we consider the isocurvature perturbation related to the axion field, $S_{A\gamma}$. As we are interested in how they impact the CMB, we want to estimate the perturbation at the time of matter-radiation decoupling. This happens after the QCD phase transition, during which the axion acquires a mass. Thus the axion component of the plasma behaves as a pressureless fluid with energy density $\rho_A= m_A n_A$ --where $n_A$ is the axion number density-- whose average $\bar\rho_A$ satisfies $\dot{\bar\rho}_A=-3H\bar\rho_A$. Then one has $3\zeta_A=-3\psi-3H\delta\rho_A/\dot{\bar\rho}_A=-3\psi+\delta\rho_A/\bar\rho_A=-3\psi+\delta n_A/\bar n_A$. On the other hand, for the photon radiation field with $\rho_\gamma\propto T^4$, $\dot{\bar\rho}_\gamma=-4H\bar{\rho}_\gamma$, and $n_\gamma\propto T^3$,  one has $3\zeta_\gamma=-3\psi+3/4\delta\rho_\gamma/\bar{\rho}_\gamma=-3\psi+\delta n_\gamma/\bar{n}_\gamma.$ Hence one can write
\begin{align}
 S_{A\gamma}=\frac{\delta n_A}{\bar{n}_A}-\frac{\delta n_\gamma}{\bar{n}_\gamma}=\frac{\delta n_A}{\bar{n}_A}-3\frac{\delta T}{T}.
\end{align}
With the axion perturbations beeing seeded during inflation while the axion was massless, their corresponding initial density perturbation is only due to subleading gradient effects; thus, we might assume that the axion isocurature perturbations have an initial value of $\delta_\rho=0$. It is usually assumed that under cosmological evolution $\delta_\rho$ stays suppressed \cite{Hertzberg:2008wr}. With $\delta{\rho}=m_A\delta n_A+\sum_{i\neq a}m_i\delta n_i+\delta\rho_r\approx0$ and $\delta \rho_r=4\rho_r \delta T/T$, then if the energy density in axions is subdominant, $m_A n_A\ll \rho_r$, one concludes that $\delta T/T \ll \delta n_A/n_A.$ The same conclusion can be arrived without imposing $\delta{\rho}=0$, but rather as a consequence of the fact that for the axion isocurvature perturbation one initially has $\delta n_{\gamma}/n_{\gamma}=3\delta T/T=0$, and this quantity is expected to remain small during the subsequent evolution, again leading to  $\delta T/T \ll \delta n_A/n_A$ \cite{Beltran:2006sq}. Either way one can then approximate $S_{A\gamma}$ as
\begin{align}
 S_{A\gamma}\approx\frac{\delta n_A}{n_A}.
\end{align}

}

{Using that in the misalignment mechanism the amount of dark matter scales as $n_A \sim \theta^2$, this gives us
\begin{equation}
 \mathcal{S}_{A\gamma} = \frac{\theta^2-\langle \theta^2 \rangle }{\langle \theta^2 \rangle}\,.
 \label{eq:isospectrumCMB}
\end{equation}
The fractional isocurvature component is defined as
\begin{equation}
\beta_\text{iso} \equiv\frac{\Delta_{S_{A\gamma}}(k_\star)}{\Delta_{S_{A\gamma}}(k_\star)+\Delta_{{\cal R}}(k_\star)}\,.
\end{equation}
In the above equation, $\Delta_X(k_\star)$ denotes the power spectrum of the fluctuations of the quantity $X$ evaluated at the pivot scale $k_\star$ of CMB measurements. The power spectrum is related to correlators of fluctuations as in Eq.~\eqref{eq:Deltadef}.
In the limit of small angle fluctuations, writing $\theta= \theta_i+\delta \theta,$ with $\theta_i$ the initial misalignment angle and $\delta \theta_i\ll1$, one has
\begin{align}
 \langle S_{A\gamma}({\bf k}) S_{A\gamma}({\bf k}')\rangle\approx\frac{4}{\theta_i^2}\,\langle \delta\theta({\bf k})\delta\theta({\bf k}')\rangle,
\end{align}
so that we may write
\begin{align}
 \Delta_{S_{A\gamma}}(k_\star)\approx\frac{4}{\theta_i^2}\Delta_{\delta\theta}(k_\star).
\end{align}
Further assuming that the isocurvature power spectrum is subdominant with respect to $\Delta_{\cal R}(k_\star)$, one arrives at
\begin{align}
\beta_\text{iso}\approx\frac{4}{\theta_i^2}\frac{\Delta_{\delta\theta}(k_\star)}{\Delta_{\cal R}(k_\star)}.
\end{align}

\section{{Mass of the axion fluctuations during and after inflation}}\label{app:isomass}

\subsection{Model 1}

To study the evolution of cosmological perturbations in two-field inflationary models it is convenient to define the fluctuations along the directions parallel and orthogonal to the trajectory of the fields. In the case of Model 1, assuming an inflationary trajectory aligned with the radial direction (as can always be enforced by an appropriate redefinition of the phase of $\sigma$), the axion fluctuation ${\cal F}_A$ can be identified with a canonically normalized gauge invariant (isocurvature) perturbation associated to the direction orthogonal to the inflationary trajectory.\footnote{In the literature, such orthogonal fluctuations are also referred to as {\it isocurvature perturbations}, as they vanish when Eq.~\eqref{adiabic2} holds for the two scalars. However, they should not be confused with the late time isocurvature perturbations  ${\cal S}_i$ of Eq.~\eqref{eq:calSj}.} In general, the latter variable obeys the equation \eq{eq:iso_orthogonal_evolution} of Section \ref{sec:model1}:
\begin{equation}\label{eq:iso_orthogonal_evolution_rep}
 \frac{\partial^2 {\mathcal{F}}}{\partial N^2} +  (3-\epsilon) \frac{\partial {\mathcal{F}}}{\partial N} + \frac{m^2_\text{iso}}{H^2}\mathcal{F} = 0\ ,
\end{equation}
modified with the addition of a source term (not shown here) which depends on the time derivative of the comoving curvature perturbation and which vanishes if the trajectory does not have turns. The full expression for the  mass is given by
\begin{equation}\label{eq:totmass}
m^2_\text{iso} = N^i N^j \nabla_i \nabla_j V + \epsilon \mathbb{R} H^2 + 3\omega^2\ .
\end{equation}
In this expression 
\begin{equation}
m^2_A = N_A^i N_A^j \nabla_i \nabla_j V,
\label{eq:isomass}
\end{equation}
can be identified in our case with the 'static' axion perturbation mass, where $N_A^i$ are the components of the unit-norm vector in the direction of the axion field and $\nabla_i$ denotes the the field covariant derivative. Both of them are defined  with respect to the field metric $G_{ij}$ in the Einstein frame (see Eq.~\eqref{eq:Einstein_frame_action}), which is also used to raise and lower the indices $(i,j,\ldots)$. The last term in \eq{eq:totmass} is related to the angular velocity of the background trajectory in field space; being $\omega$ the `turn rate' of the trajectory, which measures the deviation from a geodesic in the (curved) field space. Thanks to the $U(1)$ symmetry of Model 1, we can assume that the background solution proceeds only in a radial direction\footnote{In the scenario of \cite{Co:2019wyp} this term might constitute an important correction.} (any deviations from it will be quickly damped during inflation), and hence this term evaluates to zero. Finally, the second term of \eq{eq:totmass} contains the Ricci scalar of field space, $\mathbb{R}$, and is proportional to the first slow-roll parameter, which suppresses the contribution of this term during inflation. 

For concreteness, let us consider now the field basis of Eq.~\eqref{eq:Adef} and assume $\xi=0$, so that $N^i =\{0,1/\sqrt{G_{AA}}\} = \{0,1\}$, and ${\cal F}_A=A$. Then the static axion fluctuation mass is
\begin{equation}\label{eq:m2Astatic}
m^2_A =  \left(\partial_A^2 - \Gamma^{\rho}_{AA} \partial_\rho\right) V = \frac{\partial_\rho V}{\rho} =\lambda(\rho^2-f_A^2)\,.
\end{equation}
As discussed in Section \ref{sec:model1}, it differs from zero if the inflaton is displaced away from the minimum of the potential. For non-zero $\xi$ the isocurvature mass evaluates to 
\begin{equation}
m^2_A = \frac{\rho \partial_\rho G_{\theta\theta} }{2G_{\theta\theta} G_{\rho\rho}}\frac{\partial_\rho V}{\rho}=\frac{ 1-\xi f_A^2/M_p^2}{\Omega^2(\Omega^2+6\xi\rho^2/M_p^2)}\left(\frac{\partial_\rho V_J}{\rho}-\frac{4\xi}{\Omega^2}V_J\right)=\frac{ 1-\xi f_A^2/M_p^2}{\Omega^4(\Omega^2+6\xi\rho^2/M_p^2)}
\lambda  \left(\rho^2 - f_A^2\right)\,,
\label{eq:isomassxi}
\end{equation}
which is the same expression as Eq. \eqref{eq:m2Astatic} times a multiplicative correction which becomes unity for $\xi=0$. For $\xi =0.1$ --which we use in this paper-- this correction is small during inflation. In addition, 
\begin{equation}
 \frac{m^2_A}{H^2}  \lesssim \frac{3 M_p}{\rho}\sqrt{2\epsilon} \ll 1,
\end{equation}
assuming slow-roll and regardless of possible higher dimension (Planck suppressed) operators that may appear in the scalar potential. In the first inequality we used that the correction factor  is smaller than unity. 

In our lattice simulations of reheating, described in Section \ref{sec:results}, we  take $\xi =0$. There are two possibly relevant $\xi$-dependent corrections to the isocurvature mass in this period, both of which we can ignore.  The first is the correction proportional to the Ricci scalar of field space, cf. Eq. \eqref{eq:isomass}. After inflation this term redshifts as $H^2$ and for $1 \gtrsim \xi \gtrsim 0.005$ it 
might substantially reduce the first tachyonic spike. However, in our simulations we saw that the tachyonic growth dominates the dynamics until it hits a saturation point where non-linear effects become important. We expect to reach this saturation point regardless of whether the first tachyonic burst can effectively be suppressed. To confirm this intuition we rerun a simulation with an $\mathcal{O}(0.1)$ smaller initial amplitude of perturbations and find that the final isocurvature amplitude is unaffected. The second possible correction may in principle arise if the condition $\xi \rho^2/M_p^2 \ll 1 $ is violated. However, tachyonic amplification of the isocurvature perturbations happens when $\rho \in [0, f_A]$, hence in this regime we can safely neglect it. Similarly, the background solution will also be affected by a non-zero $\xi$. But again, for the same reason as above, this effect is small in the region where tachyonic enhancement takes place. 
Hence we do not expect the $\xi$-dependent corrections to the isocurvature mass and the background evolution to affect the final isocurvature fraction substantially and hence we can safely use Eq. \eqref{eq:m2Astatic}.

\subsection{Model 2}
We can proceed similarly for Model 2, for which the axion is now defined as in Eq.~\eqref{eq:AdefModel2}. The static contribution to the axion perturbation mass can be obtained from Eq.~\eqref{eq:isomass}; evaluating the result in the inflationary background of Eq.~\eqref{eq:valley} one gets the result of Eq.~\eqref{eq:ma2}.

\section{Higgs decays\label{app:Higgsdecays}}
Here we collect the relevant formulae for the Higgs decay rate in Eqs.~\eqref{eq:eomlattice}.
\begin{align}
\label{eq:Gammah}\begin{aligned}
 \Gamma_h=& \,\Gamma_{h\rightarrow t\bar t}+\Gamma_{h\rightarrow b\bar b}+\Gamma_{h\rightarrow W^+W^-}+\Gamma_{h\rightarrow ZZ},\\
 %%%%%
 \Gamma_{h\rightarrow t\bar t}=&\,\frac{3y_t^2}{16\pi}m_h\left(1-\frac{4m^2_t}{m^2_h}\right)^{3/2},\\
 %%%%
  \Gamma_{h\rightarrow b\bar b}=&\,\frac{3y_b^2}{16\pi}m_h\left(1-\frac{4m^2_b}{m^2_h}\right)^{3/2},\\
  %%%
 \Gamma_{h\rightarrow ZZ}=&\,\frac{g^2}{128\pi}\frac{m^3_h}{m^2_W}\sqrt{1-x_Z}\left(1-x_Z+\frac{3}{4}x
 ^2_Z\right),\\
 %%%%
 \Gamma_{h\rightarrow W^+W^-}=&\,\frac{g^2}{64\pi}\frac{m^3_h}{m^2_W}\sqrt{1-x_W}\left(1-x_W+\frac{3}{4}x
 ^2_W\right),
\end{aligned}\end{align}
where
\begin{align}\label{eq:xs}
 x_{Z/W}=\frac{4m^2_{Z/W}}{m^2_h},\quad m^2_W=\frac{g^2 \langle h^2\rangle}{4}, \quad m^2_Z=\frac{(g^2+{g'}^2)\langle h^2\rangle}{4}.
\end{align}
For the lattice implementation,  we interpret $\langle h^2\rangle$ as an average of $h^2$ over the lattice, while we determine $m^2_h$ from  the second derivative of the potential with respect to the Higgs field, again averaged over the lattice. For the decay to be allowed we  demand a positive value of $m^2_h$. We implement this by using the following value of $m_h$ in Eqs.~\eqref{eq:Gammah} and \eqref{eq:xs}:
\begin{align}
m_h=\Theta\left(\frac{\partial ^2V(\varphi_n)}{\partial^2 h}\right) \left|\frac{\partial ^2V(\phi_n)}{\partial^2h}\right|^{1/2}
\end{align}
Note that the decay rates into $W,Z$ diverge for $\langle h^2\rangle\rightarrow0$. This follows from the fact that in this case there is no spontaneous symmetry breaking and the computation of the decay rates assuming three massive gauge boson polarizations is no longer valid. During the lattice evolution one quickly gets $\langle h^2\rangle\neq0$ when the fluctuations grow at early times; nevertheless, for numerical stability we only consider these decay channels for 
$\langle h^2\rangle$ above a certain cutoff which we implement by requiring $x_W>10^{-3}$.}

\section{Mean field approximation}
\label{app:meanfieldapproximation}

\begin{figure}
\centering     
\includegraphics[width=0.49\textwidth]{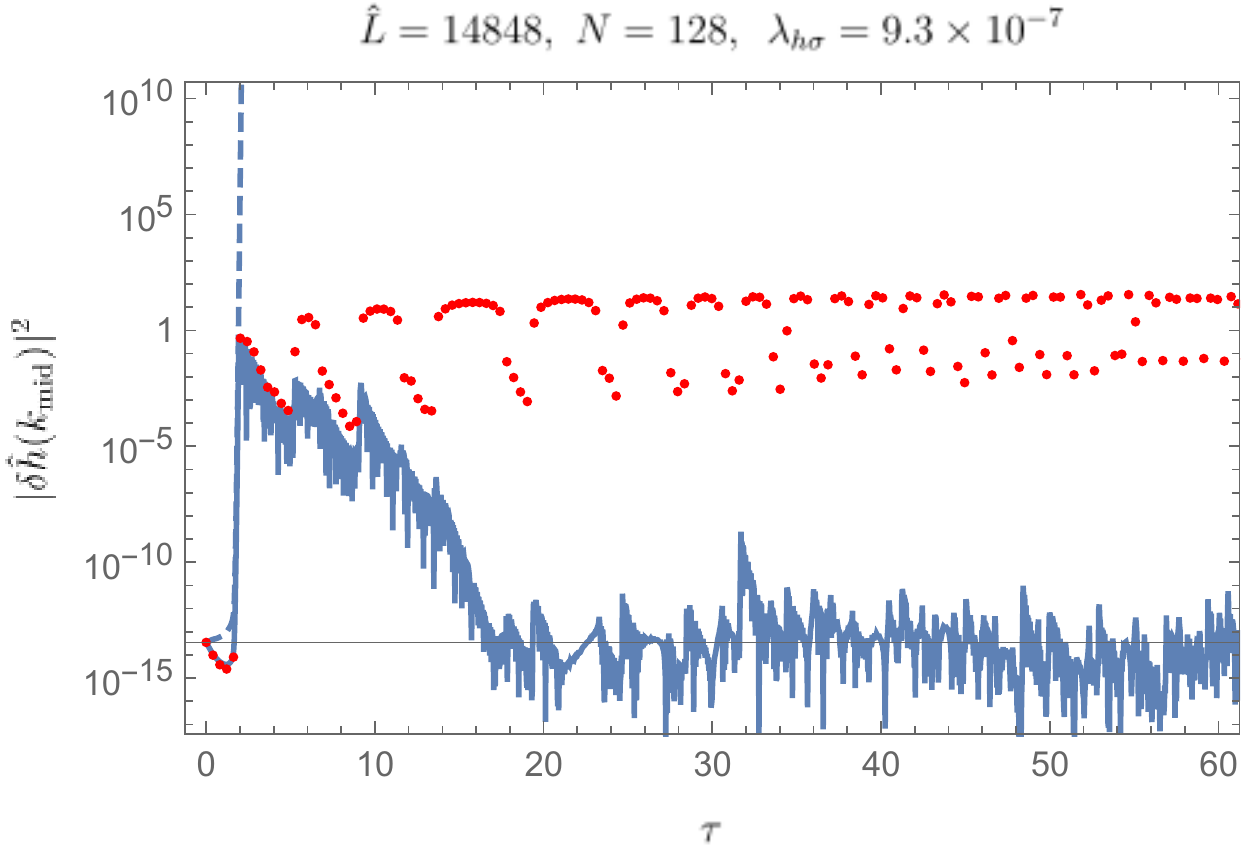}
\hfill
\includegraphics[width=0.49\textwidth]{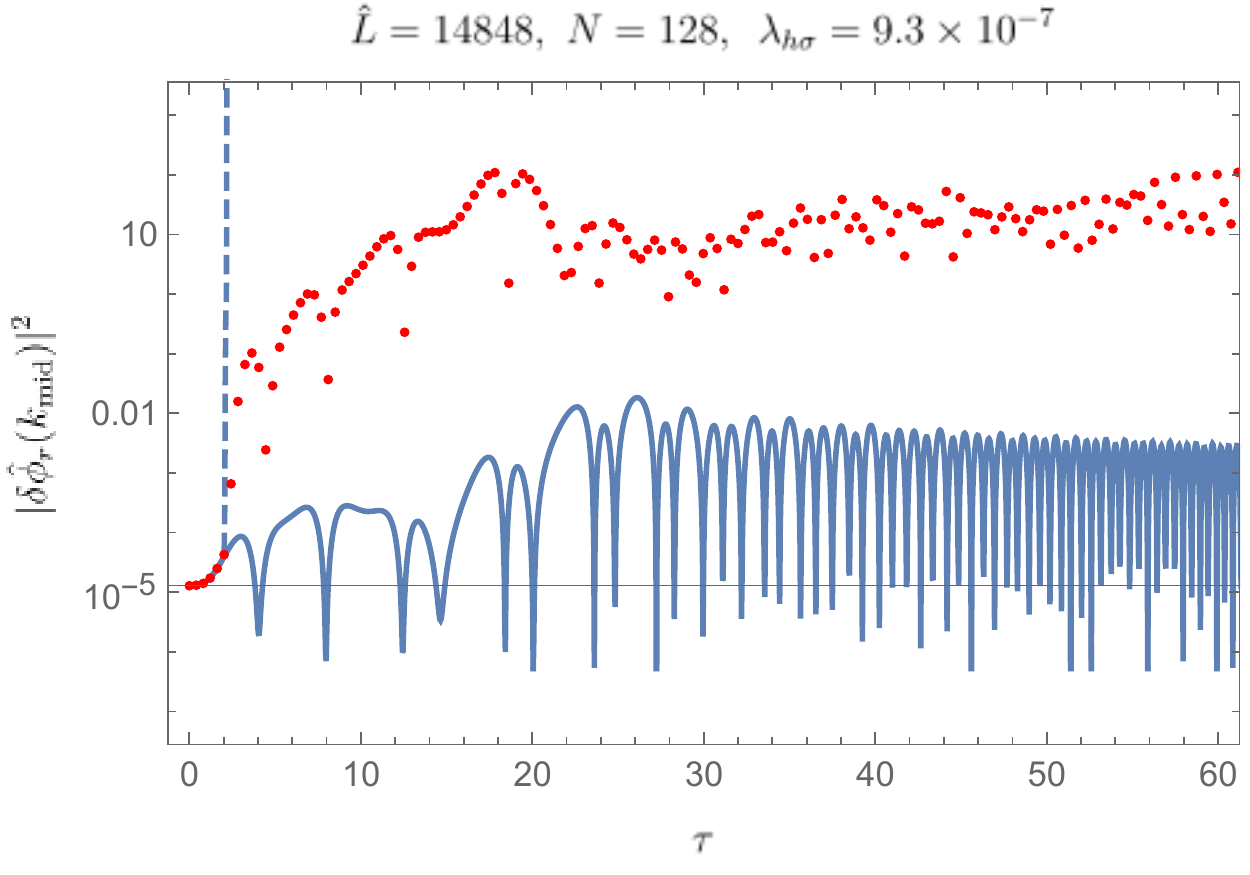}
\includegraphics[width=0.49\textwidth]{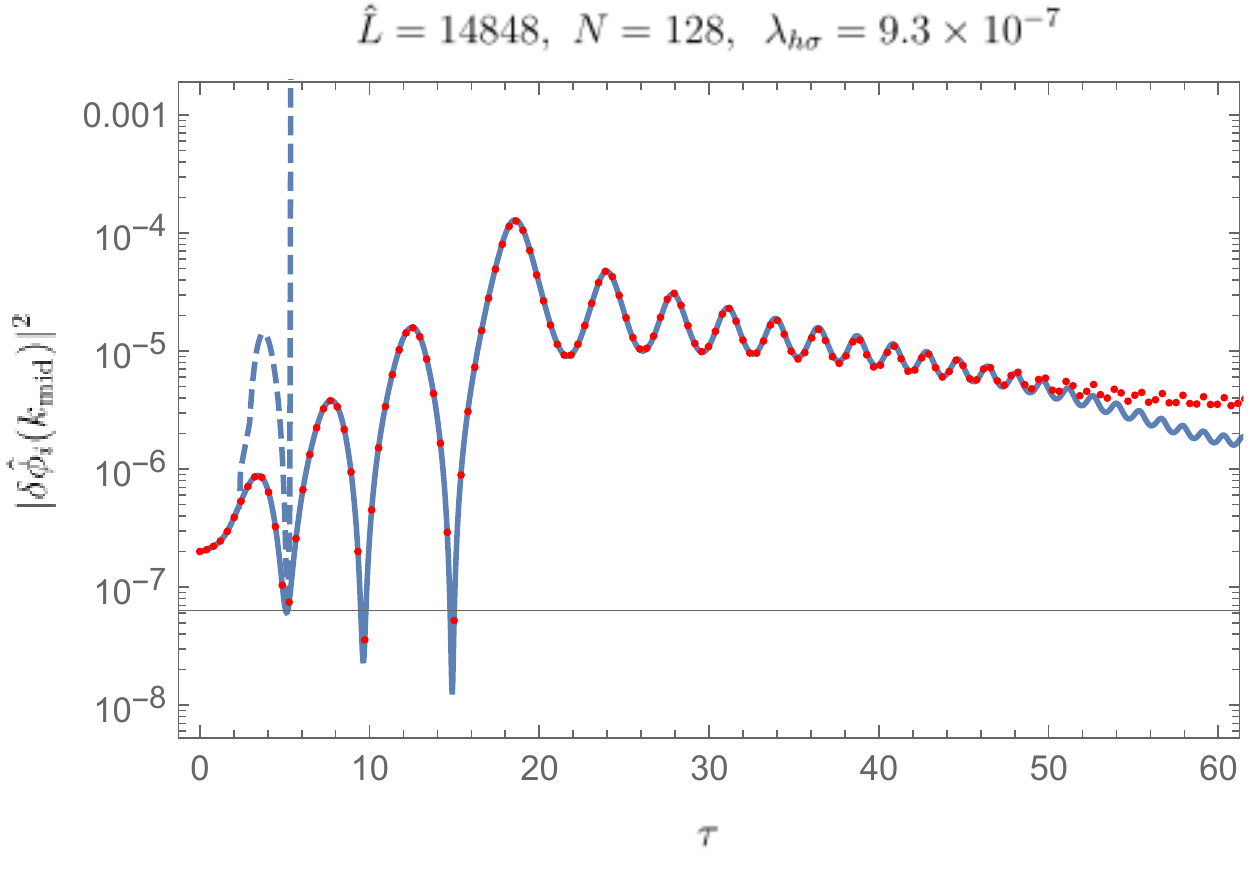}
\caption{{Comparison between the power spectra computed in the lattice (red points) and the results obtained from solving the linearized mean-field equations using the means and averages of the lattice simulation (blue curves). The dashed blue lines corresponds to using a simplified version of Eqs.~ \eqref{eq:linearized_eqs} in which one makes the following two approximations: first, one ignores the variances $\sigma^2_n$ in the the relations \eqref{eq:avsigma}, and second, in the equation for every perturbation $\delta\varphi_n$ one  ignores the terms sourced by $\delta\varphi_m,\,m\neq n$. The solid blue curves correspond to solving the full coupled Eqs.~\eqref{eq:linearized_eqs}. Note how the mean field approximation captures with excellent accuracy the full non-linear evolution of the perturbation of $\phi_2$, which is tied to the axion perturbation, up to $\tau\approx 50$.}} 
\label{fig:meanfieldapproximation_hphi1}
\end{figure}

As a cross check of our lattice simulations for {Model 1} we compare {the power spectra obtained on the lattice}
with the {result of applying the} mean field approximation {to the non-linear equation for the field fluctuations, which corrects} various terms in the linearized equations of motion {for the modes in Fourier space} with the spatial {means and} variances of the field{s}. {Considering the real fields $\varphi_n=\{\sigma_r,\sigma_\theta,h\}$ with perturbations $\delta\varphi_{n,k}$  in momentum space (where we assume rotational invariance and a dependence on $k\equiv|\bf k|$),} the mean field approximation of the {linearized equations of motion on top of a homogeneous time-dependent background} for $\xi = 0$ {can be obtained  as follows: starting from  Eqs.~\eqref{eq:eomlattice}, the fields are expanded around a homogeneous background that satisfies the equations; then only the terms in the fluctuations are kept, and a Fourier transform of the spatial coordinates is performed. Next, the  time-dependent background fields or their products/powers are reinterpreted as spatial averages over the lattice. The resulting equations are:}
{
\begin{align}
\label{eq:linearized_eqs}
\begin{aligned}
\left(\frac{d^2}{dt^2}+ 3H \frac{d}{dt}+{k^2}\right)&\,\delta\sigma_{r,k} + \left[ \lambda(3\langle\sigma_{r}^2\rangle+\langle\sigma_{\theta}^2\rangle-f_A^2)+ \lambda_{H\sigma} (\langle h^2\rangle -v^2) \right] \delta\sigma_{r,k} \\
%%%
&+2\lambda \langle \sigma_r\sigma_\theta\rangle\delta\sigma_{\theta,k} + 2\lambda_{H\sigma} \langle \sigma_r h\rangle\delta h_k= 0,\\
\left(\frac{d^2}{dt^2}+ 3H \frac{d}{dt}+{k^2}\right)&\,\delta\sigma_{\theta,k} + \left[ \lambda(3\langle\sigma_{\theta}^2\rangle+\langle\sigma_{r}^2\rangle-f_A^2)+ \lambda_{H\sigma} (\langle h^2\rangle-v^2) \right] \delta\sigma_{\theta,k} \\
%%%
&+2\lambda \langle \sigma_r\sigma_\theta\rangle\delta\sigma_{r,k} + 2\lambda_{H\sigma} \langle \sigma_\theta h\rangle\delta h_k= 0,\end{aligned}\end{align}\begin{align}\begin{aligned}
%%
%% full non-
\left(\frac{d^2}{dt^2}+ (3H+\Gamma_h) \frac{d}{dt}+{k^2}\right)&\, {\delta}h_k +\left[ \lambda_H(3\langle h^2\rangle-v^2)+ \lambda_{H\sigma} (\langle \sigma_r^2\rangle+\langle \sigma_\theta^2\rangle-f_A^2) \right] \delta h_{k}\\
%%%
&+2\lambda_{H\sigma} \langle \sigma_r h\rangle\delta\sigma_{r,k} + 2\lambda_{H\sigma} \langle  \sigma_\theta h\rangle\delta \sigma_{\theta,k}=0\,,
\end{aligned}
\end{align}}
{where $\langle\quad\rangle$ denotes a spatial (time-dependent) average on the lattice. In practice,} 
{we compute} spatial means $\bar\varphi_n(t)=\langle\varphi_n(t)\rangle$ and spatial variances $\delta^2_n(t)\equiv\langle\varphi_n(t)^2\rangle-\langle\varphi_n(t)\rangle^2$ {on the lattice.} Then in Eqs.~\eqref{eq:linearized_eqs} we 
{substitute}
\begin{align}\label{eq:avsigma}
 \langle\varphi^2_n\rangle\rightarrow \bar\varphi_n(t)^2+\delta^2_n(t), \quad {\rm and} \quad  \langle\varphi_m\varphi_n\rangle\rightarrow \bar\phi_m(t)\bar\phi_n(t),\quad {\rm for}\quad  m\neq n,
\end{align}
{neglecting cross-correlations among different fields for simplicity. For the value of the Hubble constant $H$ we also use the result of lattice simulations, in which the evolution of the scale factor is computed self-consistently.}

} We expect the mean field approximation above to capture the evolution of the {fluctuations} initially, until the perturbations grow large enough such that the non-linear interactions between the $\delta\varphi_n$ cannot be ignored. 
 This provides us with {a} 
 consistency check of our numerics. In Figure \ref{fig:meanfieldapproximation_hphi1} we show the results for Model 1 with lattice size $\hat L=14848$, number of grid points per dimension $N=128$, 
 {$\sigma_r=1.7 M_P$ at the end of inflation,}
{$\lambda=1.27\times10^{-11}$ and 
$\lambda_{H\sigma}=9.3\times 10^{-7}$.}  The time evolution of the power spectra of the modes with comoving  momentum {$k_\text{mid}  \equiv \sqrt{\lambda}\phi_0(126{\pi}/{\hat L})$} ({corresponding to roughly half of the largest momentum resolved by the lattice simulation)}
 is shown. 
The first few oscillations in  ${\delta}h$ and $\delta \sigma_r$ are captured, while the solution for ${\delta}\sigma_\theta$ (which is directly related to the axion isocurvature perturbation {(}$\delta\theta\sim\delta\sigma_\theta/\langle\sigma_r\rangle$) agrees with remarkable accuracy with the full lattice results all the way up to $\tau\sim50$.  We also show the {result} of solving a simplified version of Eqs.~\eqref{eq:linearized_eqs} {setting to zero variances and cross-interactions}.
{This shows} that the variances (which {capture} averages of non-linear effects in the lattice) are 
responsible for curbing the initial exponential growth of perturbations. {The fact that the mean field approximation captures the evolution of the $\sigma_\theta$ power spectrum  better than that of $\sigma_r,\delta h$ can be understood from the less explosive growth of $\delta\sigma_\theta$ at early times, and the fact that the source terms in the equation for $\delta\sigma_\theta$ proportional to $\delta h_k, \delta\sigma_r$ remain suppressed as one has $\langle\sigma_\theta\rangle\approx0$ as long as the perturbations in $\sigma_\theta$ remain small. This protects the evolution of $\delta\sigma_\theta$ from the influence of the large values of $\delta h,\delta\sigma_r$ at early timems.}

%%%%%%%%%%%%%%%%%%%%%%%%%%%%
%%%%%%%%%%%%%%%%%%%%%%%%%%%%
\bibliographystyle{hunsrtm}
\bibliography{ISObiblio}

\begin{thebibliography}{10}

\bibitem{Afach:2015sja}
J.~M. Pendlebury et~al.
\newblock {\em Phys. Rev. D}, 92(9):092003, 2015, 1509.04411.

\bibitem{Ai:2020mzh}
W.-Y. Ai, J.~S. Cruz, B.~Garbrecht, and C.~Tamarit.
\newblock 1 2020, 2001.07152.

\bibitem{Peccei:1977hh}
R.~Peccei and H.~R. Quinn.
\newblock {\em Phys. Rev. Lett.}, 38:1440--1443, 1977.

\bibitem{Vafa:1984xg}
C.~Vafa and E.~Witten.
\newblock {\em Phys. Rev. Lett.}, 53:535, 1984.

\bibitem{Weinberg:1977ma}
S.~Weinberg.
\newblock {\em Phys. Rev. Lett.}, 40:223--226, 1978.

\bibitem{Wilczek:1977pj}
F.~Wilczek.
\newblock {\em Phys. Rev. Lett.}, 40:279--282, 1978.

\bibitem{diCortona:2015ldu}
G.~Grilli~di Cortona, E.~Hardy, J.~Pardo~Vega, and G.~Villadoro.
\newblock {\em JHEP}, 01:034, 2016, 1511.02867.

\bibitem{Abbott:1982af}
L.~F. Abbott and P.~Sikivie.
\newblock {\em Phys. Lett. B}, 120:133--136, 1983.

\bibitem{Dine:1982ah}
M.~Dine and W.~Fischler.
\newblock {\em Phys. Lett. B}, 120:137--141, 1983.

\bibitem{Preskill:1982cy}
J.~Preskill, M.~B. Wise, and F.~Wilczek.
\newblock {\em Phys. Lett. B}, 120:127--132, 1983.

\bibitem{Borsanyi:2016ksw}
S.~Borsanyi et~al.
\newblock {\em Nature}, 539(7627):69--71, 2016, 1606.07494.

\bibitem{Aghanim:2018eyx}
N.~Aghanim et~al.
\newblock {\em Astron. Astrophys.}, 641:A6, 2020, 1807.06209.

\bibitem{Klaer:2019fxc}
V.~B. Klaer and G.~D. Moore.
\newblock {\em JCAP}, 06:021, 2020, 1912.08058.

\bibitem{Gorghetto:2018myk}
M.~Gorghetto, E.~Hardy, and G.~Villadoro.
\newblock {\em JHEP}, 07:151, 2018, 1806.04677.

\bibitem{Hindmarsh:2019csc}
M.~Hindmarsh, J.~Lizarraga, A.~Lopez-Eiguren, and J.~Urrestilla.
\newblock {\em Phys. Rev. Lett.}, 124(2):021301, 2020, 1908.03522.

\bibitem{Gorghetto:2020qws}
M.~Gorghetto, E.~Hardy, and G.~Villadoro.
\newblock {\em SciPost Phys.}, 10:050, 2021, 2007.04990.

\bibitem{Wise:1981ry}
M.~B. Wise, H.~Georgi, and S.~L. Glashow.
\newblock {\em Phys. Rev. Lett.}, 47:402, 1981.

\bibitem{Reiss:1981nd}
D.~B. Reiss.
\newblock {\em Phys. Lett.}, 109B:365--368, 1982.

\bibitem{Ernst:2018bib}
A.~Ernst, A.~Ringwald, and C.~Tamarit.
\newblock {\em JHEP}, 02:103, 2018, 1801.04906.

\bibitem{DiLuzio:2018gqe}
L.~Di~Luzio, A.~Ringwald, and C.~Tamarit.
\newblock {\em Phys. Rev. D}, 98(9):095011, 2018, 1807.09769.

\bibitem{Ernst:2018rod}
A.~Ernst, L.~Di~Luzio, A.~Ringwald, and C.~Tamarit.
\newblock {\em PoS}, CORFU2018:054, 2019, 1811.11860.

\bibitem{Boucenna:2018wjc}
S.~M. Boucenna, T.~Ohlsson, and M.~Pernow.
\newblock {\em Phys. Lett. B}, 792:251--257, 2019, 1812.10548.
\newblock [Erratum: Phys.Lett.B 797, 134902 (2019)].

\bibitem{Babu:2018qca}
K.~S. Babu, T.~Fukuyama, S.~Khan, and S.~Saad.
\newblock {\em JHEP}, 06:045, 2019, 1812.11695.

\bibitem{Chakrabortty:2019fov}
J.~Chakrabortty, R.~Maji, and S.~F. King.
\newblock {\em Phys. Rev. D}, 99(9):095008, 2019, 1901.05867.

\bibitem{Ballesteros:2019tvf}
G.~Ballesteros, J.~Redondo, A.~Ringwald, and C.~Tamarit.
\newblock {\em Front. Astron. Space Sci.}, 6:55, 2019, 1904.05594.

\bibitem{FileviezPerez:2019ssf}
P.~Fileviez~P\'erez, C.~Murgui, and A.~D. Plascencia.
\newblock {\em JHEP}, 01:091, 2020, 1911.05738.

\bibitem{DiLuzio:2020qio}
L.~Di~Luzio.
\newblock {\em JHEP}, 11:074, 2020, 2008.09119.

\bibitem{Lazarides:2020frf}
G.~Lazarides and Q.~Shafi.
\newblock {\em Phys. Lett. B}, 807:135603, 2020, 2004.11560.

\bibitem{Bahre:2013ywa}
R.~B\"ahre et~al.
\newblock {\em JINST}, 8:T09001, 2013, 1302.5647.

\bibitem{Anastassopoulos:2017ftl}
V.~Anastassopoulos et~al.
\newblock {\em Nature Phys.}, 13:584--590, 2017, 1705.02290.

\bibitem{Ouellet:2018beu}
J.~L. Ouellet et~al.
\newblock {\em Phys. Rev. Lett.}, 122(12):121802, 2019, 1810.12257.

\bibitem{Braine:2019fqb}
T.~Braine et~al.
\newblock {\em Phys. Rev. Lett.}, 124(10):101303, 2020, 1910.08638.

\bibitem{Chung:2016ysi}
W.~Chung.
\newblock {\em PoS}, CORFU2015:047, 2016.

\bibitem{Gramolin:2020ict}
A.~V. Gramolin, D.~Aybas, D.~Johnson, J.~Adam, and A.~O. Sushkov.
\newblock {\em Nature Phys.}, 17(1):79--84, 2021, 2003.03348.

\bibitem{Kwon:2020sav}
O.~Kwon et~al.
\newblock 12 2020, 2012.10764.

\bibitem{Brubaker:2016ktl}
B.~M. Brubaker et~al.
\newblock {\em Phys. Rev. Lett.}, 118(6):061302, 2017, 1610.02580.

\bibitem{McAllister:2017lkb}
B.~T. McAllister, G.~Flower, E.~N. Ivanov, M.~Goryachev, J.~Bourhill, and M.~E.
  Tobar.
\newblock {\em Phys. Dark Univ.}, 18:67--72, 2017, 1706.00209.

\bibitem{Alesini:2020vny}
D.~Alesini et~al.
\newblock 12 2020, 2012.09498.

\bibitem{Geraci:2017bmq}
A.~A. Geraci et~al.
\newblock {\em Springer Proc. Phys.}, 211:151--161, 2018, 1710.05413.

\bibitem{brass01}
\url{http://wwwiexp.desy.de/groups/astroparticle/brass/brassweb.htm}.

\bibitem{Budker:2013hfa}
D.~Budker, P.~W. Graham, M.~Ledbetter, S.~Rajendran, and A.~Sushkov.
\newblock {\em Phys. Rev. X}, 4(2):021030, 2014, 1306.6089.

\bibitem{Alesini:2017ifp}
D.~Alesini, D.~Babusci, D.~Di~Gioacchino, C.~Gatti, G.~Lamanna, and C.~Ligi.
\newblock 7 2017, 1707.06010.

\bibitem{Brun:2019lyf}
P.~Brun et~al.
\newblock {\em Eur. Phys. J. C}, 79(3):186, 2019, 1901.07401.

\bibitem{Armengaud:2014gea}
E.~Armengaud et~al.
\newblock {\em JINST}, 9:T05002, 2014, 1401.3233.

\bibitem{Irastorza:2018dyq}
I.~G. Irastorza and J.~Redondo.
\newblock {\em Prog. Part. Nucl. Phys.}, 102:89--159, 2018, 1801.08127.

\bibitem{Raffelt:2006cw}
G.~G. Raffelt.
\newblock {\em Lect. Notes Phys.}, 741:51--71, 2008, hep-ph/0611350.

\bibitem{Carenza:2020cis}
P.~Carenza, B.~Fore, M.~Giannotti, A.~Mirizzi, and S.~Reddy.
\newblock {\em Phys. Rev. Lett.}, 126(7):071102, 2021, 2010.02943.

\bibitem{Arvanitaki:2010sy}
A.~Arvanitaki and S.~Dubovsky.
\newblock {\em Phys. Rev. D}, 83:044026, 2011, 1004.3558.

\bibitem{Bar:2019ifz}
N.~Bar, K.~Blum, and G.~D'Amico.
\newblock {\em Phys. Rev. D}, 101(12):123025, 2020, 1907.05020.

\bibitem{Akrami:2018odb}
Y.~Akrami et~al.
\newblock {\em Astron. Astrophys.}, 641:A10, 2020, 1807.06211.

\bibitem{Bezrukov:2007ep}
F.~L. Bezrukov and M.~Shaposhnikov.
\newblock {\em Phys. Lett. B}, 659:703--706, 2008, 0710.3755.

\bibitem{Fairbairn:2014zta}
M.~Fairbairn, R.~Hogan, and D.~J.~E. Marsh.
\newblock {\em Phys. Rev. D}, 91(2):023509, 2015, 1410.1752.

\bibitem{Ballesteros:2016xej}
G.~Ballesteros, J.~Redondo, A.~Ringwald, and C.~Tamarit.
\newblock {\em JCAP}, 08:001, 2017, 1610.01639.

\bibitem{Boucenna:2017fna}
S.~M. Boucenna and Q.~Shafi.
\newblock {\em Phys. Rev. D}, 97(7):075012, 2018, 1712.06526.

\bibitem{Ballesteros:2016euj}
G.~Ballesteros, J.~Redondo, A.~Ringwald, and C.~Tamarit.
\newblock {\em Phys. Rev. Lett.}, 118(7):071802, 2017, 1608.05414.

\bibitem{Nakayama:2015pba}
K.~Nakayama and M.~Takimoto.
\newblock {\em Phys. Lett. B}, 748:108--112, 2015, 1505.02119.

\bibitem{Ringwald:2020vei}
A.~Ringwald, K.~Saikawa, and C.~Tamarit.
\newblock {\em JCAP}, 02:046, 2021, 2009.02050.

\bibitem{Barbon:2009ya}
J.~L.~F. Barbon and J.~R. Espinosa.
\newblock {\em Phys. Rev. D}, 79:081302, 2009, 0903.0355.

\bibitem{Burgess:2009ea}
C.~P. Burgess, H.~M. Lee, and M.~Trott.
\newblock {\em JHEP}, 09:103, 2009, 0902.4465.

\bibitem{Ema:2017rqn}
Y.~Ema.
\newblock {\em Phys. Lett. B}, 770:403--411, 2017, 1701.07665.

\bibitem{Gorbunov:2018llf}
D.~Gorbunov and A.~Tokareva.
\newblock {\em Phys. Lett. B}, 788:37--41, 2019, 1807.02392.

\bibitem{Ema:2019fdd}
Y.~Ema.
\newblock {\em JCAP}, 09:027, 2019, 1907.00993.

\bibitem{Cespedes:2012hu}
S.~Cespedes, V.~Atal, and G.~A. Palma.
\newblock {\em JCAP}, 05:008, 2012, 1201.4848.

\bibitem{Beltran:2006sq}
M.~Beltran, J.~Garcia-Bellido, and J.~Lesgourgues.
\newblock {\em Phys. Rev. D}, 75:103507, 2007, hep-ph/0606107.

\bibitem{Linde:1985yf}
A.~D. Linde.
\newblock {\em Phys. Lett. B}, 158:375--380, 1985.

\bibitem{Stompor:1995py}
R.~Stompor, A.~J. Banday, and K.~M. Gorski.
\newblock {\em Astrophys. J.}, 463:8, 1996, astro-ph/9511087.

\bibitem{Hertzberg:2008wr}
M.~P. Hertzberg, M.~Tegmark, and F.~Wilczek.
\newblock {\em Phys. Rev. D}, 78:083507, 2008, 0807.1726.

\bibitem{Wantz:2009it}
O.~Wantz and E.~P.~S. Shellard.
\newblock {\em Phys. Rev. D}, 82:123508, 2010, 0910.1066.

\bibitem{Linde:1991km}
A.~D. Linde.
\newblock {\em Phys. Lett. B}, 259:38--47, 1991.

\bibitem{Gordon:2000hv}
C.~Gordon, D.~Wands, B.~A. Bassett, and R.~Maartens.
\newblock {\em Phys. Rev. D}, 63:023506, 2000, astro-ph/0009131.

\bibitem{Achucarro:2012sm}
A.~Achucarro, J.-O. Gong, S.~Hardeman, G.~A. Palma, and S.~P. Patil.
\newblock {\em JHEP}, 05:066, 2012, 1201.6342.

\bibitem{Hertzberg:2016tal}
M.~P. Hertzberg.
\newblock {\em JCAP}, 11:037, 2016, 1609.01342.

\bibitem{Felder:2000hq}
G.~N. Felder and I.~Tkachev.
\newblock {\em Comput. Phys. Commun.}, 178:929--932, 2008, hep-ph/0011159.

\bibitem{Felder:2008zz}
G.~N. Felder.
\newblock {\em Comput. Phys. Commun.}, 179:604--606, 2008, 0712.0813.

\bibitem{Greene:1997fu}
P.~B. Greene, L.~Kofman, A.~D. Linde, and A.~A. Starobinsky.
\newblock {\em Phys. Rev. D}, 56:6175--6192, 1997, hep-ph/9705347.

\bibitem{Kofman:1997yn}
L.~Kofman, A.~D. Linde, and A.~A. Starobinsky.
\newblock {\em Phys. Rev. D}, 56:3258--3295, 1997, hep-ph/9704452.

\bibitem{Bassett:1998wg}
B.~A. Bassett, D.~I. Kaiser, and R.~Maartens.
\newblock {\em Phys. Lett. B}, 455:84--89, 1999, hep-ph/9808404.

\bibitem{Bassett:1999mt}
B.~A. Bassett, F.~Tamburini, D.~I. Kaiser, and R.~Maartens.
\newblock {\em Nucl. Phys. B}, 561:188--240, 1999, hep-ph/9901319.

\bibitem{Bassett:1999cg}
B.~A. Bassett and F.~Viniegra.
\newblock {\em Phys. Rev. D}, 62:043507, 2000, hep-ph/9909353.

\bibitem{Bassett:1999ta}
B.~A. Bassett, C.~Gordon, R.~Maartens, and D.~I. Kaiser.
\newblock {\em Phys. Rev. D}, 61:061302, 2000, hep-ph/9909482.

\bibitem{Zibin:2000uw}
J.~P. Zibin, R.~H. Brandenberger, and D.~Scott.
\newblock {\em Phys. Rev. D}, 63:043511, 2001, hep-ph/0007219.

\bibitem{Finelli:2000ya}
F.~Finelli and R.~H. Brandenberger.
\newblock {\em Phys. Rev. D}, 62:083502, 2000, hep-ph/0003172.

\bibitem{Huang:2011gf}
Z.~Huang.
\newblock {\em Phys. Rev. D}, 83:123509, 2011, 1102.0227.

\bibitem{Weinberg:2003sw}
S.~Weinberg.
\newblock {\em Phys. Rev. D}, 67:123504, 2003, astro-ph/0302326.

\bibitem{Bardeen:1983qw}
J.~M. Bardeen, P.~J. Steinhardt, and M.~S. Turner.
\newblock {\em Phys. Rev. D}, 28:679, 1983.

\bibitem{Kodama:1985bj}
H.~Kodama and M.~Sasaki.
\newblock {\em Prog. Theor. Phys. Suppl.}, 78:1--166, 1984.

\bibitem{Bassett:2005xm}
B.~A. Bassett, S.~Tsujikawa, and D.~Wands.
\newblock {\em Rev. Mod. Phys.}, 78:537--589, 2006, astro-ph/0507632.

\bibitem{Co:2019wyp}
R.~T. Co and K.~Harigaya.
\newblock {\em Phys. Rev. Lett.}, 124(11):111602, 2020, 1910.02080.

\end{thebibliography}
 
 \end{document}